\documentclass[12pt]{article}
\pdfoutput=1

\usepackage{amsmath}
\usepackage{slashed}
\usepackage{amssymb}
\usepackage{epsfig}
\usepackage{graphicx}
\usepackage{multirow}
\usepackage{color}
\usepackage[normal]{subfigure}
\usepackage{rotating}
\usepackage{hyperref}
\usepackage[margin=0.9in]{geometry}
\usepackage[table]{xcolor}
\usepackage{enumitem}
\usepackage[utf8x]{inputenc}
\usepackage[compress,numbers,sort]{natbib}
\usepackage{authblk}
\usepackage{colortbl}
\usepackage{pdflscape}
\definecolor{LightCyan}{rgb}{0.88,1,1}
\definecolor{piggypink}{rgb}{0.99, 0.87, 0.9}

\usepackage{ucs}
\usepackage{amsmath}
\usepackage{amsfonts}
\usepackage{amssymb}
\usepackage{eucal}               % Modifica i caratteri in stile {\cal M} e \mathcal{M}. I vecchi caratteri sono disponibili con il comando \CMcal{M}.
\usepackage{mathrsfs}          % Alfabeto di lettere "corsive" maiuscole per le Lagrangiane e gli eleMenti di Matrice. \mathscr{M}
\usepackage[sans]{dsfont}     % Questo serve per fare l'1 "doppio" ovvero l'operatore identita' con \mathds{1}
\usepackage[Symbol]{upgreek}  % Lettere greche

\usepackage{array}             % Ulteriori comandi per le tabelle. E' necessario per colortbl
\usepackage{booktabs}       % Miglior layout per le tabelle
\usepackage{colortbl}          % Serve per colorare le tabelle. Necessita array e color
\usepackage{multirow}
\usepackage{longtable}

\usepackage{fancyhdr}

\usepackage[normalem]{ulem}             % Underline and strike text etc.
\usepackage{cancel}           % Per cancellare il testo con linee oblique

\usepackage{slashed}
\usepackage{simplewick}     % Wick contractions
\usepackage{feynmf}           % processare con le seguenti istruzioni da riga di comando:
\usepackage{color}
\definecolor{grigio}{cmyk}{0,0,0,0.1}
\definecolor{rosa}{cmyk}{0,0.1,0.1,0.02}
\definecolor{rosino}{cmyk}{0,0.05,0.05,0.02}
\definecolor{rosas}{cmyk}{0,0.3,0.25,0.05}
\definecolor{celeste}{cmyk}{0.1,0,0,0.02}
\definecolor{giallino}{cmyk}{0,0,0.1,0.02}
\definecolor{rosso}{cmyk}{0,1,1,0.4}
\definecolor{rossos}{cmyk}{0,1,1,0.55}
\definecolor{rossoc}{cmyk}{0,1,1,0.2}
\definecolor{blu}{cmyk}{1,1,0,0.3}
\definecolor{blus}{cmyk}{1,1,0,0.5}
\definecolor{bluc}{cmyk}{1,1,0,0.1}
\definecolor{blucc}{cmyk}{0.7,0.5,0,0}
\definecolor{viola}{cmyk}{0,1,0,0.6}
\definecolor{viola2}{cmyk}{0,1,0.2,0.6}
\definecolor{verde}{cmyk}{0.92,0,0.59,0.25}
\definecolor{verdec}{cmyk}{0.92,0,0.59,0.15}
\definecolor{verdes}{cmyk}{0.92,0,0.59,0.4}
\definecolor{verdino}{cmyk}{0.12,0,0.09,0.02}
\definecolor{giallo}{cmyk}{0,0,1,0}
\definecolor{gialloverde}{cmyk}{0.44,0,0.74,0}
\definecolor{Titolo}{rgb}{0.752941176,0.576470588,0.992156863}% #C093FD
\definecolor{altro}{rgb}{0.094117647,0.650980392,0.643137255}% #24A6A4
\definecolor{Peanuts}{rgb}{0.2, 0.4, 0.6}% #336699
\definecolor{Pean1}{rgb}{0.6, 0.8, 0.4}% #99cc66
\definecolor{BHO}{rgb}{0.2, 0.8, 1}% #33CCFF
\definecolor{Daria}{rgb}{0, 0.9412, 0}% #00F000 e' uguale al verde u_u
\definecolor{UniPi}{rgb}{0.2549, 0.4627, 0.6275}% #4176a0
\definecolor{UniPidue}{rgb}{0.3216, 0.5804, 0.7882}% #5294c9
\definecolor{rossoCP3}{cmyk}{0,.88,.77,.40}
\definecolor{verdeCP3}{rgb}{0.09765625, 0.57421875, 0.1015625}
\definecolor{bluCP3}{rgb}{0, 0.23, 0.67}

\newcommand{\Eq}[1]{Eq.~\eqref{#1}}
\newcommand{\Fig}[1]{Fig.~\ref{#1}}

\newcommand{\Ref}[1]{Ref.~\cite{#1}}           % non viene letto dal motore di SPIRES che genera automaticamente la bibliografia

\newcommand{\GeV}{\,{\rm GeV}}

\newcommand{\ZZ}{\mathbb{Z}}

\newcommand{\beq}{\begin{equation}}
\newcommand{\eeq}{\end{equation}}

\newcommand{\FERMI}{{\sf Fermi}}

\newcommand{\HESS}{{\sf H.E.S.S.}}

\newcommand{\Planck}{{\sf Planck}}
\newcommand{\LHC}{{\sf LHC}}
\newcommand{\LUX}{{\sf LUX}}
\newcommand{\LZ}{{\sf LZ}}

\newcommand{\squared}[1]{\left| #1 \right|^2}
\newcommand{\bma}{\begin{pmatrix}}
\newcommand{\ema}{\end{pmatrix}}
\newcommand{\nn}{\nonumber}

\ifx\pdfoutput\undefined
\usepackage[dvips,bookmarks]{hyperref}    % This is for arXiv.org
\else
\usepackage{hyperref}    % This is for pdftex
\fi
\hypersetup{colorlinks, bookmarksopen, bookmarksnumbered,
citecolor=verdeCP3, linkcolor=bluCP3, pdfstartview=FitH, urlcolor=rossoCP3}

\def\hhref#1{\href{http://arxiv.org/abs/#1}{#1}} % in bibliography
\begin{document}

\begin{flushright}
{\footnotesize
{\sc SCIPP 16/02\\
\sc NIKHEF 2016/006}
%\\{\sc }
}
\end{flushright}
\color{black}
%\vspace{0.3cm}

\begin{center}

%{\Huge\bf Dark Matter Interactions Mediated \\ \vspace{0.5cm} by a 750 GeV Scalar in the Long Run}

{\Huge\bf A 750 GeV Portal: \\ [2mm] LHC Phenomenology \\ [4mm] and Dark Matter Candidates}

\medskip
\bigskip\color{black}\vspace{0.6cm}

{
{\large\bf Francesco D'Eramo}$\, ^{a, b}$,
{\large\bf Jordy de Vries}$\, ^c$,
{\large\bf Paolo Panci}$\, ^d$
}
\\[7mm]
{\it $^a$ \href{http://www.physics.ucsc.edu/}{Department of Physics, University of California Santa Cruz}, \\ 1156 High St., Santa Cruz, CA 95064, USA} \\ [3mm]
{\it $^b$ \href{http://scipp.ucsc.edu/}{Santa Cruz Institute for Particle Physics}, \\ 1156 High St., Santa Cruz, CA 95064, USA} \\ [3mm]
{\it $^c$ \href{http://www.nikhef.nl/pub/theory/general.html}{Nikhef, Theory Group}, \\ Science Park 105, 1098 XG, Amsterdam, The Netherlands} \\ [3mm]
{\it $^d$ \href{http://www.iap.fr}{Institut d'Astrophysique de Paris}, UMR 7095 CNRS, Universit\'e Pierre et Marie Curie, \\ 98 bis Boulevard Arago, Paris 75014, France}\\[3mm]
\end{center}

\bigskip

\centerline{\large\bf Abstract}
\begin{quote}
%\color{black}
\large

We study the effective field theory obtained by extending the Standard Model field content with two singlets:
 a $750 \, {\rm GeV}$ (pseudo-)scalar and a stable fermion. Accounting for collider productions initiated by both gluon and photon fusion, we investigate where the theory is consistent with both the \LHC\ diphoton excess and bounds from Run 1. We analyze dark matter phenomenology in such regions, including relic density constraints as well as collider, direct, and indirect bounds. Scalar portal dark matter models are very close to limits from direct detection and mono-jet searches if gluon fusion dominates, and not constrained at all otherwise. Pseudo-scalar models are challenged by photon line limits and mono-jet searches in most of the parameter space.
   
\end{quote}

\newpage
%\tableofcontents

%%%%%%%%%%%%%%%%%%%%%%%%%%%%%%%%%%%%%%%%%%%%%%%%
%%%%%%%%%%%%%%%%%% INTRO %%%%%%%%%%%%%%%%%%%%%%%%%%
%%%%%%%%%%%%%%%%%%%%%%%%%%%%%%%%%%%%%%%%%%%%%%%%

\section{Introduction}

The ATLAS ~\cite{ATLAS:2015abc} and CMS collaborations~\cite{CMS:2015dxe} recently pointed out an excess in diphoton events with a peak in the invariant mass distribution around $m_{\gamma\gamma} \simeq 750 \, {\rm GeV}$. Upon interpreting the events as the production and two-body decay of a new $750 \, {\rm GeV}$ particle, current data cannot discriminate between a narrow or broad (up to $\sim 45 \, {\rm GeV}$) resonance. Although the evidence is far from conclusive, if it is confirmed with more luminosity it would be a monumental discovery after decades of undisputed success for the Standard Model (SM). Furthermore, it is natural to believe that such a hypothetical particle is linked to a bigger framework addressing, for instance, the gauge hierarchy problem and would be the herald of additional discoveries.

A more robust and elderly motivation for physics beyond the SM is the evidence for dark matter (DM)~\cite{Bertone:2004pz}. Among several candidates~\cite{Feng:2010gw}, a weakly interacting massing particle (WIMP) produced through thermal freeze-out~\cite{Lee:1977ua,Scherrer:1985zt,Srednicki:1988ce} is undeniably one of the most appealing. Thus it is tempting to investigate whether the potentially new $750 \, {\rm GeV}$ degree of freedom could act as a portal field, allowing DM and the SM to communicate beyond gravitational interactions. 

This work focuses on (pseudo-)scalar portals and fermion DM candidates, both SM singlets. New (peudo-)scalars are ubiquitous in well-motivated frameworks for physics beyond the SM. At the same time, fermion singlets are DM candidates begging for new weak scale degrees of freedom, as gauge invariance forbids renormalizable interactions with SM particles~\cite{Kim:2008pp}. We work within an Effective Field Theory (EFT) framework and write down the minimal theory for the \LHC\ diphoton excess with a DM candidate. We define the theory at a cutoff scale $\Lambda$ interpreted as the scale where heavy degrees of freedom are integrated out and we apply EFT methods to connect the interactions at different scales. While we present our analysis and results for the case of a Dirac fermion DM, it is straightforward to generalize to the Majorana case.

We start our phenomenological study with a comprehensive analysis of \LHC\ results. Two different mechanisms, gluon and photon fusion, can be responsible for the (pseudo-)scalar production at colliders. In spite of being mediated by strong interactions, gluon fusion does not necessarily have to be the dominant production mechanism at the \LHC\ since we have no actual evidence that the new particle couples to gluons at all. From the diphoton excess, we do know that the resonance must couple to photons. This implies that there exists an irreducible photon-fusion contribution to the resonance production, which can be dominant one or not depending on the relative sizes of the couplings to photons and gluons. We therefore include both production mechanisms in our study, and we identify where the EFT is capable of accounting for the diphoton events while at the same time being consistent with $\sqrt{s} = 8 \, {\rm TeV}$ data. 

The presence of a DM candidate in the EFT impacts our analysis even before discussing any DM phenomenology. Once produced at the \LHC, the (pseudo-)scalar could be allowed to decay to invisible final states, altering the width and diphoton rate. For this reason, we find it convenient to divide our \LHC\ study into two scenarios:
\begin{itemize}
\item[$\diamond$] {\bf{SM Dominated Resonance:}} The DM mass is above the critical value $ \simeq 375 \, {\rm GeV}$. The resonance only decays to SM final states and it is typically narrow. The ATLAS preferred value of $\Gamma_S\simeq 45 \, {\rm GeV}$ can be obtained only for large couplings to SM fields which are inconsistent with searches in other decay channels such as $Z\gamma$.
\item[$\diamond$] {\bf{DM Dominated Resonance:}} The DM mass is below $\simeq 375 \, {\rm GeV}$ such that decays to DM pairs are kinematically allowed. This invisible channel is very likely to dominate the total width and the resonance is now quite broad.
\end{itemize}
 In each of the above we perform a thorough exploration of the parameter space. The presence of a sizeable coupling to gluons utterly drives \LHC\ phenomenology, as gluon fusion is clearly the leading candidate for the (pseudo-)scalar production. However, photon fusion can still dominate the production if the coupling to gluons is small enough. For example, this would be the case for UV-complete theories where any  heavy particles that are integrated out at the cutoff scale $\Lambda$ do not carry color charge. Despite the apparently large parameter space, we identify two main EFT regimes where the production is dominated by a single partonic process and where the couplings of the new particle to SM gauge bosons are quite constrained. We emphasize that every specific UV completion with no additional degrees of freedom below the cutoff $\Lambda$ must satisfy the constraints of our EFT analysis.

In the second part of our study we incorporate the DM phenomenology. For parameters favored by \LHC\ data, we further impose constraints from DM searches and also identify regions where the DM has a correct thermal relic density. Collider searches for mono-jet events turn out to be relevant only in the DM dominated scenario, as the associated cross section falls rapidly as the DM mass increases above the resonant value $ \simeq 375 \, {\rm GeV}$. Direct Detection (DD) experiments constrain only the scalar portal case, and the coupling to gluons is again crucial. If such a coupling is present, DD rates are dominated by the gluon content of the nucleons. If not, both the coupling to photons and the loop-induced couplings to gluons and light quarks contribute to the signal. In each case we evaluate DD rates through a rigorous Renormalization Group (RG) procedure, which is mandatory as the scale separation between DD and \LHC\ experiments is large. On the contrary, indirect detection (ID) experiments put bounds only on pseudo-scalar mediators because annihilations mediated by a scalar portal are p-wave suppressed. This difference also explains why larger couplings to the scalar are necessary to reproduce the observed DM abundance through thermal freeze-out.

The paper is structured as follows. In Section~\ref{sec:EFT} we introduce the EFT that will be the basis of our study. Section~\ref{sec:RGE} deals with the connection between energy scales. In Section~\ref{sec:LHCscenarios} we introduce the two different \LHC\ scenarios and identify the parameter space region allowed by collider searches in both cases.  Finally, we present our DM analysis in Section~\ref{sec:DM} and summarize our main findings in Section~\ref{sec:con}. We provide appendices with explicit expressions for decay rates and cross sections, details of the RG procedure, and methods for the relic density calculation.

%%%%%%%%%%%%%%%%%%%%%%%%%%%%%%%%%%%%%%%%%%%%%%%%
%%%%%%%%%%%%%%%%%%% EFT %%%%%%%%%%%%%%%%%%%%%%%%%%
%%%%%%%%%%%%%%%%%%%%%%%%%%%%%%%%%%%%%%%%%%%%%%%%

\section{The EFT for Dark Matter and the Diphoton Excess}
\label{sec:EFT}

We introduce the minimal EFT necessary to describe the diphoton excess at the \LHC, while simultaneously providing a stable DM candidate. We augment the SM by two singlet fields: a real scalar $S$ with mass $m_S = 750 \, {\rm GeV}$ and a fermion $\chi$. The formalism developed in this Section is valid for both Dirac and Majorana $\chi$. 
 Although we give the details of the EFT for the case of a scalar $S$, the generalization to the case of a pseudo-scalar $P$ is straightforward as shown at the end of this Section. 

Within our framework, the \LHC\ excess is accounted for by the production of S and its subsequent decay to photons. At the same time, $S$ also acts as a portal to the DM particle $\chi$ assumed to be a stable field as a consequence of a $\ZZ_2$ symmetry. The EFT Lagrangian reads
\beq
\mathcal{L}_{\rm EFT} = \mathcal{L}_{\rm SM} + \frac{1}{2} \partial_\mu S \partial^\mu S - \frac{1}{2} m_S^2 S^2 + 
\overline{\chi} \left( i \slashed{\partial} - m_\chi \right) \chi + \mathcal{L}_{\rm int} \ ,
\eeq
with $ \mathcal{L}_{\rm SM}$ the SM Lagrangian. We organize the interactions in $ \mathcal{L}_{\rm int}$ by distinguishing between renormalizable and non-renormalizable operators, and we further classify the latter according to their mass dimension
\beq
 \mathcal{L}_{\rm int} = \mathcal{L}_{\rm int}^{(\rm ren)} + \sum_{d > 4} \sum_{\alpha_d} \frac{c_{\alpha_d}}{\Lambda^{d-4}} \mathcal{O}_{\alpha_d}^{(d)} \ .
\label{eq:Lintdef}
\eeq
The sum on the right-hand side of the above equation runs over all SM gauge invariant operators for each mass dimension $d$. The higher-dimensional operators $\mathcal{O}_{\alpha_d}^{(d)}$ are suppressed by powers of the EFT cutoff $\Lambda$, understood as the mass scale where heavy degrees of freedom generating the interactions are integrated out.  The dimensionless and renormalization-scale, $\mu$, dependent Wilson coefficients, $c_{\alpha_d}$, encode unresolved dynamics above the EFT cutoff.

The renormalizable piece contains the portal interaction between the two singlets 
\beq
\mathcal{L}_{\rm int}^{(\rm ren)} =  c_{\chi S} \, S \, \overline{\chi} \chi  \ .
\label{eq:Lintren}
\eeq
Additional renormalizable interactions in the scalar potential are not forbidden by any symmetry. In particular, operators involving both $S$ and the SM Higgs doublet $H$ would induce a mixing between $S$ and the SM Higgs boson $h$, affecting production and decays of both. This scenario is quite constrained by Higgs coupling measurements and it has been recently studied in Refs.~\cite{Falkowski:2015swt,Berthier:2015vbb}. In this work, we assume these scalar potential interactions to be absent, as realized in several UV completions (see e.g. Ref.~\cite{Backovic:2015fnp}). 

Moving on to higher-dimensional operators, we consider the $d = 5$ contact interactions
\beq
 \mathcal{L}^{\rm d = 5 }_{(\rm int)} = \frac{S}{\Lambda} \left[ c_{GG} \, G^{A\,\mu\nu} G^A_{\mu\nu} + c_{WW} \, W^{I\,\mu\nu} W^I_{\mu\nu} + c_{BB} \, B^{\mu\nu} B_{\mu\nu} \right]  \ .
 \label{eq:Lintdim5}
\eeq
Here, we assume the EFT cutoff to be much higher than the weak scale, $\Lambda \gg m_Z$, and therefore we couple our degrees of freedom in a $SU(3)_c \times SU(2)_L \times U(1)_Y$ gauge invariant way. Also in this case, other operators in \Eq{eq:Lintdim5} are in principle allowed by symmetry considerations: the Higgs portal operator $\overline{\chi} \chi H^\dag H$, the coupling to SM fermions $S H \overline{f_L} f_R$, and additional $d=5$ scalar potential interactions. We assume again that these couplings are not present at the EFT cutoff, as it would be the case in several UV completions (also discussed in Ref.~\cite{Backovic:2015fnp}). However, assuming that they vanish at the cutoff does not save us from having them at other scales: the absence of a symmetry protection allows the RG evolution to switch them on through radiative corrections, and it the next Section we quantify how this happens. These radiative contributions play no role for \LHC\ physics, and we can safely use the Lagrangian in \Eq{eq:Lintdim5} to study \LHC\ phenomenology. The situation is rather different for DD, since we evolve the EFT all the way down to the scale of nuclear physics. To summarize, the EFT obtained by adding $S$ and $\chi$ to the SM and with the interactions in \Eq{eq:Lintren} and \Eq{eq:Lintdim5} will be the basis for this work. The EFT where the new bosonic degree of freedom is a pseudo-scalar $P$ is very similar
\begin{align}
\mathcal{L}_{\rm int}^{(\rm ren)} = & \, c_{\chi P} \, P \, \overline{\chi} \, i \gamma^5 \chi  \ , \\
 \mathcal{L}^{\rm d = 5 }_{(\rm int)} = & \, \frac{P}{\Lambda} \left[ \tilde{c}_{GG} \, G^{A\,\mu\nu} \widetilde{G}^A_{\mu\nu} + \tilde{c}_{WW} \, W^{I\,\mu\nu} \widetilde{W}^I_{\mu\nu} + \tilde{c}_{BB} \, B^{\mu\nu} \widetilde{B}_{\mu\nu} \right]  \ .
\end{align}

The analysis of a specific UV realization for the above EFT goes beyond the scope of this work. Nevertheless, we make sure that the use of the EFT is consistent with the energy scales we consider. For resonant (pseudo-)scalar production at the \LHC\ we have a partonic center of mass energy of the order $m_{S,P} = 750 \, {\rm GeV}$. We assume the EFT cutoff, understood as the mass of the particles we integrate-out to give the above contact interactions, to be larger than the energy scale for resonant productions. Thus we restrict our analysis to $\Lambda \gtrsim 1 \, {\rm TeV}$. For simple UV completions where the operators are generated by integrating-out heavy vector-like fermions with mass $M_f$, the use of EFT is justified up to $10 \%$ for $M_f \simeq 1 \, {\rm TeV}$, and the accuracy rapidly improves for larger $M_f$. Consequently, we truncate the sum in \Eq{eq:Lintdef} at dimension 5 and we do not consider operators of higher dimensions, since their effects are power suppressed.

Barring substantial CP violation, we have either the scalar $S$ or the pseudo-scalar $P$. RG effects turn out to be negligible for the pseudo-scalar because DD constraints are very weak, so the results in Section~\ref{sec:RGE} are relevant for the scalar only. Furthermore, \LHC\ rates are identical for the two cases, so the analysis performed in Section~\ref{sec:LHCscenarios} is valid for both. The DM phenomenology is drastically different between the two cases, since DM annihilations mediated by a (pseudo-)scalar are (s-)p-wave processes, and DD constraints are negligible for the pseudo-scalar. For this reason, we keep the DM discussion in Section~\ref{sec:DM} separated for the two cases.

%%%%%%%%%%%%%%%%%%%%%%%%%%%%%%%%%%%%%%%%%%%%%%%%
%%%%%%%%%%%%%%%%%%% RGE %%%%%%%%%%%%%%%%%%%%%%%%%%
%%%%%%%%%%%%%%%%%%%%%%%%%%%%%%%%%%%%%%%%%%%%%%%%

\section{RGE Scale Connection and Direct Detection Rates}
\label{sec:RGE}

The Wilson coefficients in \Eq{eq:Lintdim5} are generated at the cutoff $\Lambda$ by integrating out heavy degrees of freedom, while \LHC\ data bound their values at the typical collider scale. In order to perform a consistent EFT analysis, we would have to RG evolve the interactions down to \LHC\ scales before putting limits. As we will show shortly, these corrections turn out to be inconsequential. Nevertheless, our EFT looks very different at energy scale of the order of $\sim 1 \, {\rm GeV}$, where nuclear matrix elements are evaluated to compute DD rates. This procedure can significantly affect DD rates, as pointed out for several cases in Refs.~\cite{Kopp:2009et,Hill:2011be,Frandsen:2012db,Haisch:2013uaa,Hill:2013hoa,Vecchi:2013iza,Kopp:2014tsa,Crivellin:2014qxa,Crivellin:2014gpa,D'Eramo:2014aba,Hill:2014yxa,Berlin:2015njh}.

The only SM fields accessible at such a low-energy scale and relevant for DD observables are light quarks, gluons, and photons. We define a different EFT for DD in terms of these light degrees of freedom, and the relevant interactions for our study are the following
\beq
\mathcal{L}_{\rm EFT}^{\rm DD} = \sum_{q = u, d, s} \mathcal{C}_q m_q \, \overline{\chi} \chi \,\overline{q} q +
 \mathcal{C}_G \overline{\chi} \chi \, G^{A\,\mu\nu} G^A_{\mu\nu} + \mathcal{C}_F\, \overline{\chi} \chi \, F^{\mu\nu} F_{\mu\nu} \ ,
 \label{eq:EFTDD}
\eeq
with Wilson coefficients evaluated at the nuclear scale $\mu_N \simeq 1 \, {\rm GeV}$. Our goal here is to connect the Wilson coefficients at the cutoff scale appearing in Eqs.~(\ref{eq:Lintren}) and (\ref{eq:Lintdim5}) with the ones at the nuclear scale in \Eq{eq:EFTDD}. This is achieved by performing the RG evolution (RGE)
\beq
\left( c_{\chi S}, c_{GG}, c_{WW}, c_{BB} \right)_{\mu = \Lambda} \quad \xrightarrow{RGE} \quad 
\left( \mathcal{C}_q, \mathcal{C}_G, \mathcal{C}_F \right)_{\mu = \mu_N} \ ,
\eeq
obtained via the following steps
\begin{itemize}
\item perform the RGE from $\mu = \Lambda$ down to the scalar mass $\mu = m_S$;
\item integrate out the scalar field $S$ at the scale $\mu = m_S$;
\item perform the RGE from $\mu = m_S$ down to the weak scale $\mu \simeq m_Z$;
\item integrate out the heavy SM degrees of freedom (top, W, Z, Higgs);
\item perform the RGE from $\mu = m_Z$ down to the nuclear scale $\mu \simeq \mu_N$, and in the process integrate out the intermediate heavy quarks (bottom and charm) at their mass threshold.
\end{itemize}
Here, we present the main RGE results with details of calculations deferred to Appendix~\ref{app:RG}.

\subsection{Running from $\Lambda$ to $m_S$?}
\label{sec:RunningLambdatomS}

The RG evolution to lower scales has two main effects: multiplying couplings by overall constants (self-renormalization) and inducing new interactions (operator mixing). Here, we inspect if the latter is ever relevant for \LHC\ physics, as it is the only process which could induce a different phenomenology. Self-renormalization can always be taken care of by considering the Wilson coefficients at $\mu = m_S$.

If the scalar couples to gluons, QCD running induces couplings to quarks at the scale $m_S$. This can be phenomenologically relevant only for the top quark, since the effect is proportional to the Yukawa coupling. If sizeable, this coupling can contribute to the total width of the scalar and open the $\bar{t} t$ production channel at the \LHC. This effect is quantified by the induced partial width to $\bar{t} t$ in units of the one to gluons
\beq
\frac{\Gamma_{S \rightarrow \bar{t} t} }{\Gamma_{S \rightarrow G G} } = 
\frac{3}{16} \frac{m_t^2}{m_S^2} \frac{c_{\bar{t} t}(m_S)^2}{c_{GG}(m_S)^2} \left( 1 - \frac{4 m_t^2}{m_S^2} \right)^{3/2} \simeq 6.3 \times 10^{-3} \frac{c_{\bar{t} t}(m_S)^2}{c_{GG}(m_S)^2}  \ ,
\eeq
where the Wilson coefficient $c_{\bar{t} t}(m_S)$ can be obtained from the results in Appendix~\ref{app:RG} 
\beq
\frac{c_{\bar{t} t}(m_S)}{c_{GG}(m_S)} = \frac{8}{\pi}  \int_\Lambda^{m_S} \frac{\alpha^2_s(\mu)}{\alpha_s(m_S)} d \ln \mu \simeq 0.23 \ln(m_S / \Lambda) \ .  
\eeq
This effect is too small to play any role at the \LHC. The operator mixing and the radiatively induced interactions for the case of no coupling to gluons are even more suppressed as a consequence of the weak fine structure constant and smaller anomalous dimensions. 

Other potentially relevant RG effects arise from inducing operators involving both the new resonance and the the SM Higgs doublet. In our case we would induce the dimension 5 operator $S (H^\dagger H)^2$, which has two main effects. First, it opens up the new $S$ decay channel into two or more Higgs bosons. Second, it induces a mixing between $S$ and $h$ with consequent change of production and decay rates for both scalars. In particular, the couplings between $h$ and other SM fields would be different with respect to their SM value. The Wilson coefficient for this dimension 5 operator as induced at the scale $m_S$ can be calculated using the analysis in Appendix~\ref{app:RG}, and it results in
\beq
c_H(m_S) \simeq  \left[0.0027 \, c_{BB}(\Lambda)+0.023\,c_{WW}(\Lambda)\right]  \ln(m_S / \Lambda) \ .
\label{cHmS}
\eeq
Contributions from the gluonic coupling $c_{GG}$ only appear at the two-loop level and we neglect them. The decay width into Higgs bosons, relative to the one into photons, is then given by
\beq
\frac{\Gamma_{S \rightarrow hh }}{\Gamma_{S \rightarrow \gamma \gamma} } = 
\frac{9}{16} \frac{v^4}{m_S^4} \frac{c_{H}(m_S)^2}{c_{\gamma\gamma}(m_S)^2} \left( 1 - \frac{4 m_h^2}{m_S^2} \right)^{1/2} \simeq 6.1 \times 10^{-3} \frac{c_{H}(m_S)^2}{c_{\gamma\gamma}(m_S)^2}  \ ,
\eeq
where $c_{\gamma\gamma}(m_S) \simeq s_w^2 c_{WW}(\Lambda) + c_w^2 c_{BB}(\Lambda)$ (see Appendix~\ref{app:DecayAndXS}). This RG induced decay width is too small to play any role in our analysis. The induce mixing between $S$ and $h$ is quantified by the mixing angle
\beq
\tan 2 \alpha \simeq \frac{c_H(m_S) \, v^3}{\Lambda \, m_S^2} \ ,
\eeq
with $c_H(m_S)$ still given in the expression in \Eq{cHmS}. As discussed extensively in Section~\ref{sec:LHCscenarios}, the coupling of the new resonance to electroweak gauge bosons can be at most $c_{BB}/\Lambda \simeq c_{WW}/\Lambda = 0.3 \, {\rm TeV}^{-1}$ in the photon fusion regime. Such couplings give rise to a very small mixing angle, $\alpha \simeq 10^{-4}$, well below any experimental constraints \cite{Falkowski:2015swt}. 
 
\subsection{Connecting $m_S$ to $\mu_N$ I: Scalar coupled to gluons}
\label{sec:DD1}

 If $c_{GG}$ does not vanish at the cutoff, the couplings to electroweak gauge bosons provides a negligible contribution to DD rates, thus we ignore their effects and only consider QCD running. This has been extensively studied in the literature (see e.g. Refs.~\cite{Vecchi:2013iza,Hill:2014yxa}). We repeat the leading order (LO) analysis for completeness in Appendix~\ref{app:RG}, where we argue that next-to-LO corrections only modify the final result by a few percent.
 
We summarize here the main results. As discussed in Section~\ref{sec:RunningLambdatomS}, we should only be concerned about the RG from $m_S$ to $\mu_N$. Thus we start our analysis at the scale $m_S$, where we integrate out the scalar and write down the effective Lagrangian 
\begin{equation}
\label{eq:EFTabovetopbelowS}
\mathcal{L}_{\rm EFT}^{m_t < \mu < m_S} = \mathcal{C}_{GG} \, \bar \chi \chi\, G^{A\,\mu\nu} G^A_{\mu\nu}\ ,
\end{equation}
valid for energy scales above the top mass. The coupling is obtained via a tree-level matching
\beq
\mathcal {C}_{GG}(m_S)  = \frac{c_{\chi S}}{\Lambda m_S^2}c_{GG}(m_S)\ .
\eeq
The connection between the Wilson coefficient in \Eq{eq:EFTabovetopbelowS} evaluated at the renormalization scale $\mu = m_S$ and the ones of the effective Lagrangian for DD in \Eq{eq:EFTDD} evaluated at the nuclear scale is achieved as follows
\begin{align}
\label{eq:RGgluonFINALa} \mathcal C_q (\mu_N) \simeq & \, -5.86 \, \mathcal C_{GG}(m_S)\ ,\\
\label{eq:RGgluonFINALb} \mathcal {C}_{GG}(\mu_N) \simeq & \, 4.01 \,\mathcal C_{GG}(m_S) \ .
\end{align}

\subsection{Connecting $m_S$ to $\mu_N$ II: Scalar coupled EW gauge bosons}
\label{sec:DD2}

On the other hand, if $S$ does not couple to gluons at the scale $\Lambda$ the running driven by electroweak gauge bosons turns out to be relevant for the rate calculation~\cite{Frandsen:2012db,Crivellin:2014gpa}. We start at the scale $\mu = m_S$, where we integrate out the scalar and end up with the effective Lagrangian
\beq
\label{eq:EFTabovetopbelowSEW}
\mathcal{L}_{\rm EFT}^{m_t < \mu < m_S}  = \mathcal{C}_{WW}\, \bar \chi \chi\, W^{I\,\mu\nu} W^I_{\mu\nu} +
\mathcal{C}_{BB}\, \bar \chi \chi\, B^{\mu\nu} B_{\mu\nu} \ .
\eeq
The tree-level matching in this case is analogous 
\begin{align}
 \mathcal{C}_{BB}(m_S) = & \, \frac{c_{\chi S}}{\Lambda m_S^2} c_{BB}(m_S)\ ,\\
  \mathcal{C}_{WW}(m_S) = & \, \frac{c_{\chi S}}{\Lambda m_S^2} c_{WW}(m_S) \ .
\end{align}
The connection between the couplings in \Eq{eq:EFTabovetopbelowSEW} and the ones for DD in \Eq{eq:EFTDD} reads
\begin{align}
\label{eq:RGEWFINALa} \mathcal C_u (\mu_N)  \simeq & \, - 0.046\, \mathcal C_{BB}(m_S) + 0.15 \, \mathcal C_{WW}(m_S) \ , \\
\label{eq:RGEWFINALb} \mathcal C_{d,s} (\mu_N) \simeq & \, - 0.021\, \mathcal C_{BB}(m_S) + 0.14\, \mathcal C_{WW}(m_S) \ , \\
\label{eq:RGEWFINALc} \mathcal {C}_{GG}(\mu_N) \simeq & \, 5.5 \times 10^{-4} \mathcal C_{BB}(m_S) + 2.5 \times 10^{-3} \,\mathcal C_{WW}(m_S)  \ , \\
\label{eq:RGEWFINALd} \mathcal {C}_{FF}(\mu_N)  \simeq & \, 0.77\,\mathcal C_{BB}(m_S) + 0.23\, \mathcal C_{WW}(m_S)\ .
\end{align}

\subsection{RGE Analysis: Summary}

If it useful to summarize the main results of this Section. The RGE from $\Lambda$ to $m_S$ does not affect the \LHC\ phenomenology, thus we use the EFT defined at the scale $\mu = m_S$ to perform the \LHC\ phenomenological analysis. Then we have to connect the couplings at $\mu = m_S$ with DD rates. For coupling to gluons we use the results in Eqs.~(\ref{eq:RGgluonFINALa})-(\ref{eq:RGgluonFINALb}), whereas for interactions with electroweak gauge bosons we have the low-energy couplings given in Eqs.~(\ref{eq:RGEWFINALa})-(\ref{eq:RGEWFINALd}).

%%%%%%%%%%%%%%%%%%%%%%%%%%%%%%%%%%%%%%%%%%%%%%%%
%%%%%%%%%%%%%%%%%%% LHC %%%%%%%%%%%%%%%%%%%%%%%%%%
%%%%%%%%%%%%%%%%%%%%%%%%%%%%%%%%%%%%%%%%%%%%%%%%

\section{Two Different Scenarios for \LHC}
\label{sec:LHCscenarios}

The Lagrangians introduced in Section~\ref{sec:EFT} contain seven free parameters: three mass scales $(m_S, m_\chi, \Lambda)$ and four dimensionless couplings $(c_{GG}, c_{WW}, c_{BB}, c_{\chi S})$. We set $m_S = 750 \, {\rm GeV}$ motivated by the diphoton excess. As justified in Section~\ref{sec:RunningLambdatomS}, we start with the EFT defined at $\mu = m_S$ and present the results of our \LHC\ analysis in terms of $c_{XX} / \Lambda$, where $X = \{G, W, B\}$. It is worth recalling that we always have in mind values $\Lambda \gtrsim {\rm few \; TeV}$ to safely satisfy the EFT hypothesis. DD rates are computed through the RGE from $m_S$ down to the nuclear scale as discussed in Section~\ref{sec:RGE}, thus they also depend on the combination $c_{XX} / \Lambda$ only.

This leaves us with the DM mass and four couplings. The DM mass value $m_S / 2 \simeq 375 \, {\rm GeV}$ is quite special, as for masses smaller (greater) than this critical value the scalar $S$ is (not) allowed to decay to DM pairs. In view of this, we divide our study into these two main cases. They correspond to quite different scenarios at the \LHC, since the scalar resonance is typically narrow unless we open the decay to DM. The origin of this can be traced back to the fact that decays to DM are the only ones mediated by a renormalizable interaction. Before introducing and studying the two scenarios, we discuss the cross sections for gluon and photon fusion. Related \LHC\ studies considering gluon fusion partonic processes have been recently performed in Refs.~\cite{Knapen:2015dap,Buttazzo:2015txu,Franceschini:2015kwy,DiChiara:2015vdm,McDermott:2015sck,Ellis:2015oso,Low:2015qep,Gupta:2015zzs,Dutta:2015wqh,Chakrabortty:2015hff,Agrawal:2015dbf,Falkowski:2015swt,Kim:2015ksf,Alves:2015jgx,Berthier:2015vbb,Craig:2015lra,Altmannshofer:2015xfo}, whereas photon fusion was considered in Refs.~\cite{Csaki:2015vek,Fichet:2015vvy,Altmannshofer:2015xfo,Csaki:2016raa,Martin:2016byf}. If the coupling $c_{GG}$ is non vanishing at the cutoff then gluon fusion certainly dominates partonic productions for every channel. In the absence of such a coupling at the cutoff scale, one may wonder about the main production mechanism. The two-loop induced coupling $c_{GG}$ at the scale $m_S$ turns out to induce a gluon fusion rate that it is subdominant compared to the vector boson fusion (VBF) contribution. Strictly speaking, all possible VBFs contribute to the cross section in our EFT, namely partonic processes with initial state $ZZ$, $WW$, $WZ$, $Z\gamma$, $W\gamma$ and $\gamma\gamma$. In particular, we cannot have only the $\gamma\gamma$ process since this would require to have the three effective vertices vanishing ($S WW$, $S ZZ$, and $S Z\gamma$) with the only freedom of tuning the two Wilson coefficients $c_{BB}$ and $c_{WW}$, as pointed out in \Ref{Altmannshofer:2015xfo}. Photon fusion diagrams dominate at LO, and the next relevant contribution is the interference between photon and weak boson processes. We neglect this correction since it is approximately only a $10 \%$ modification of the total cross section~\cite{Fichet:2015vvy}. However, these couplings to weak gauge bosons lead to proton-proton collisions with $WW$, $ZZ$, and $Z \gamma$ final states that are bounded from \LHC\ Run 1 searches.

We include both gluon and photon fusion in our analysis and identify for what parameters each one is dominant. As already stressed in Section~\ref{sec:EFT}, the \LHC\ analysis performed here is valid for both scalar and pseudo-scalar mediators.

\subsection{\LHC\ production cross sections}
\label{sec:LHCXS}

Whether the resonance is broad or not, the question about the partonic production mechanism is still open. The general formalism for \LHC\ cross sections can be found in Appendix~\ref{app:DecayAndXS}. Here, we specialize the general expression given in \Eq{eq:sxLHCschannelNarrow2} to the two cases of our interest.  

Gluon fusion is a natural candidate, as long as the coupling to gluons $c_{GG}$ is switched on. If this is the case, the production cross section results in
\beq
\sigma_{p p \rightarrow ij}(\sqrt{s}) = \mathcal{K}_{GG} \, \frac{\pi^2 \, C_{GG}^{\sqrt{s}}}{8 m_S s} \frac{\Gamma_{S \rightarrow G G} \, \Gamma_{S \rightarrow i j}}{\Gamma_S}  \ .
\label{eq:XSgg}
\eeq
The overall multiplicative factor $\mathcal{K}_{GG}$ accounts for QCD higher-order contributions. This K-factor correction does not depend on the CM energy of the proton-proton collision, as we are always interested in resonant processes, and in what follows we use the value $\mathcal{K}_{GG} = 1.48$~\cite{Franceschini:2015kwy}.  The quantity $ C_{GG}^{\sqrt{s}}$ is defined in \Eq{eq:Cabdef} as an integral over the gluon parton distribution function (pdf) in the proton. We adopt the gluon pdf from \Ref{Martin:2009iq}, and using their public code\footnote{\href{https://mstwpdf.hepforge.org/}{https://mstwpdf.hepforge.org/}} 
we find the following numbers
\begin{align}
C_{GG}^{8 \, {\rm TeV}} = & \, 140.097 \ ,\\
C_{GG}^{13 \, {\rm TeV}} = & \, 1736.03 \ .
\end{align}
Thus cross sections at $13 \, {\rm TeV}$ are rescaled from the ones at $8 \, {\rm TeV}$ by the following factor
\beq
\mathcal{R}_{GG} \equiv \frac{C_{GG}^{13 \, {\rm TeV}} / (13 \, {\rm TeV})^2}{C_{GG}^{8 \, {\rm TeV}} / (8 \, {\rm TeV})^2} = 4.69 \ .
\label{eq:ratiogluon}
\eeq

Partonic productions through photon fusion have a cross section 
\beq
\sigma_{p p \rightarrow ij}(\sqrt{s}) = r_{\rm inel} \, \frac{8 \pi^2 \, C_{\gamma\gamma}^{\sqrt{s}}}{m_S s} \frac{\Gamma_{S \rightarrow \gamma \gamma} \, \Gamma_{S \rightarrow i j}}{\Gamma_S} \ .
\label{eq:XSgammagamma}
\eeq
The factor $r_{\rm inel}$ accounts for inelastic processes where the proton gets destroyed after radiating a photon. Unfortunately, its precise value suffers from theoretical uncertainties. The recent LO calculation with Madgraph~\cite{Alwall:2014hca} performed in \Ref{Csaki:2016raa} found that the elastic processes are only $4\%$ of the total events, or equivalently $r_{\rm inel} = 25$. This is consistent with the discussion in \Ref{Fichet:2015vvy}, claiming the range $r_{\rm inel} \in [15, 25]$. We normalize our $\sqrt{s} = 13 \, {\rm TeV}$ cross section with the results of \Ref{Csaki:2016raa}. Upon setting $r_{\rm inel} = 25$, we find the contribution from elastic processes
\beq
C_{\gamma\gamma}^{13 \, {\rm TeV}} = 0.04 \ .
\eeq

A key quantity in this regime is the rescaling between cross sections
\beq
\mathcal{R}_{\gamma\gamma} \equiv \frac{C_{\gamma\gamma}^{13 \, {\rm TeV}} / (13 \, {\rm TeV})^2}{C_{\gamma\gamma}^{8 \, {\rm TeV}} / (8 \, {\rm TeV})^2}  \ .
\label{eq:Rgammagamma}
\eeq
The output of the Madgraph calculation in \Ref{Csaki:2016raa} gives $\mathcal R_{\gamma\gamma} \simeq 2$. This quantity, however, is not very well known for instance due to uncertainties regarding the inverse proton radius and the size of $r_{\rm inel}$. Following the discussions in Refs.~\cite{Csaki:2015vek,Fichet:2015vvy,Csaki:2016raa}, we take the range $\mathcal{R}_{\gamma\gamma} \in [2, 5] $, and present our results for the two representative values\footnote{After this work was completed, Ref.~\cite{Harland-Lang:2016qjy} appeared which found $\mathcal{R}_{\gamma\gamma}=2.9$, in the middle of the range we considered.}
\beq
C_{\gamma\gamma}^{8 \, {\rm TeV}} = \left\{ \begin{array}{lcccl}
0.0076 & & & & \mathcal{R}_{\gamma\gamma} = 2 \\ 0.0030 & & & & \mathcal{R}_{\gamma\gamma} = 5 
\end{array} \right. \ . 
\eeq

The expressions in Eqs.~(\ref{eq:XSgg}) and (\ref{eq:XSgammagamma}) are the master equations for this Section. Combined with the decay widths listed in Appendix~\ref{app:DecayAndXS}, they allow us to compute the production cross section for any $ij$-pair. We use them to find the EFT parameters consistent with both the diphoton excess~\cite{ATLAS:2015abc,CMS:2015dxe} and the \LHC\ Run 1 constraints listed in Tab.~\ref{tab:LHC}. For the signal we identify the $1\sigma$ and $2\sigma$ regions consistent with the cross section
\beq
\left.\sigma_{pp \rightarrow \gamma\gamma}\right|_{13 {\rm TeV}} = (10 \pm 3) \, {\rm fb} \ .
\eeq
We do not consider limits from $\bar{t} t$ searches~\cite{Aad:2015fna,Khachatryan:2015sma} since the rate is loop-suppressed in our EFT. In addition, we do not show in our plots $\gamma \gamma$ limits at $\sqrt{s} = 8 \, {\rm TeV}$. They are definitely consistent with the diphoton signal at $\sqrt{s} =13 \, {\rm TeV}$ for the case of gluon fusion, given the rescaling factor in Eq.~(\ref{eq:ratiogluon}). The photon fusion presents tension for $\mathcal{R}_{\gamma\gamma} = 2$, the lowest rescaling factor we consider, but is consistent for the larger value $\mathcal{R}_{\gamma\gamma} = 5$. After this exploration of the parameter space, we will consider constraints from relic density and DM searches.

\begin{table}
\begin{center}
\begin{tabular}{ c || c | c | c | c | c }
 & $\gamma\gamma$~\cite{Aad:2015mna,Khachatryan:2015qba}  & $ZZ$~\cite{Aad:2015kna,Khachatryan:2015cwa} & $WW$~\cite{Aad:2015agg,Khachatryan:2015cwa} & $Z \gamma$~\cite{Aad:2014fha} & $jj$~\cite{Aad:2014aqa,CMS:2015neg} \\ \hline \hline
$\left.\sigma_{pp \rightarrow ij}\right|_{8 {\rm TeV}}$ & 2.4 \, {\rm fb} & 12 \, {\rm fb} & 40 \, {\rm fb} & 4 \, {\rm fb} & 2.5 \, {\rm pb}
\end{tabular}
\end{center}
\caption{$95\%_{\rm CL}$  bounds from \LHC\ Run 1 at $\sqrt{s} = 8 \, {\rm TeV}$ on signals present in our EFT.}
\label{tab:LHC}
\end{table}

\subsection{A SM Dominated Resonance}
\label{sec:SMdom}

\begin{figure}[!t]
\centering
\includegraphics[width=0.495\textwidth]{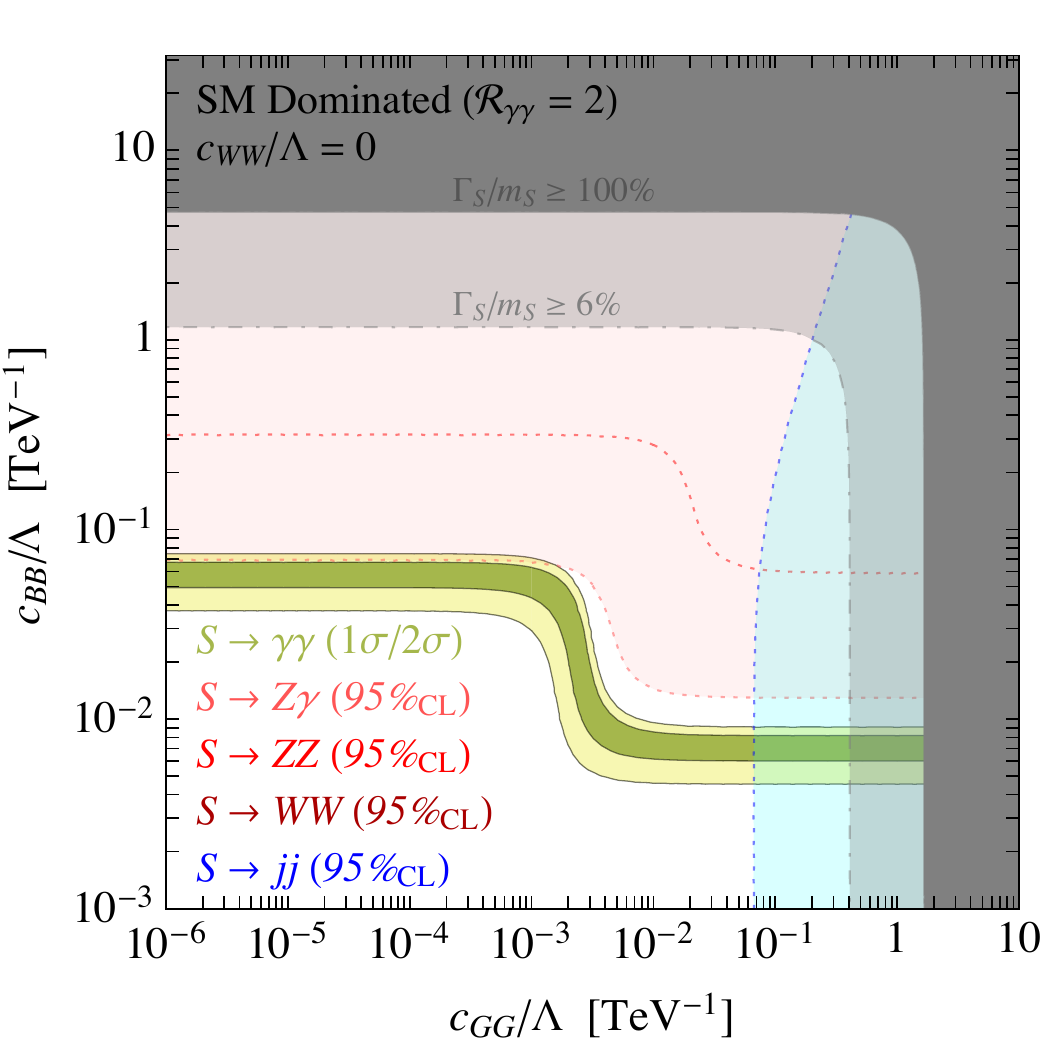} 
\includegraphics[width=0.495\textwidth]{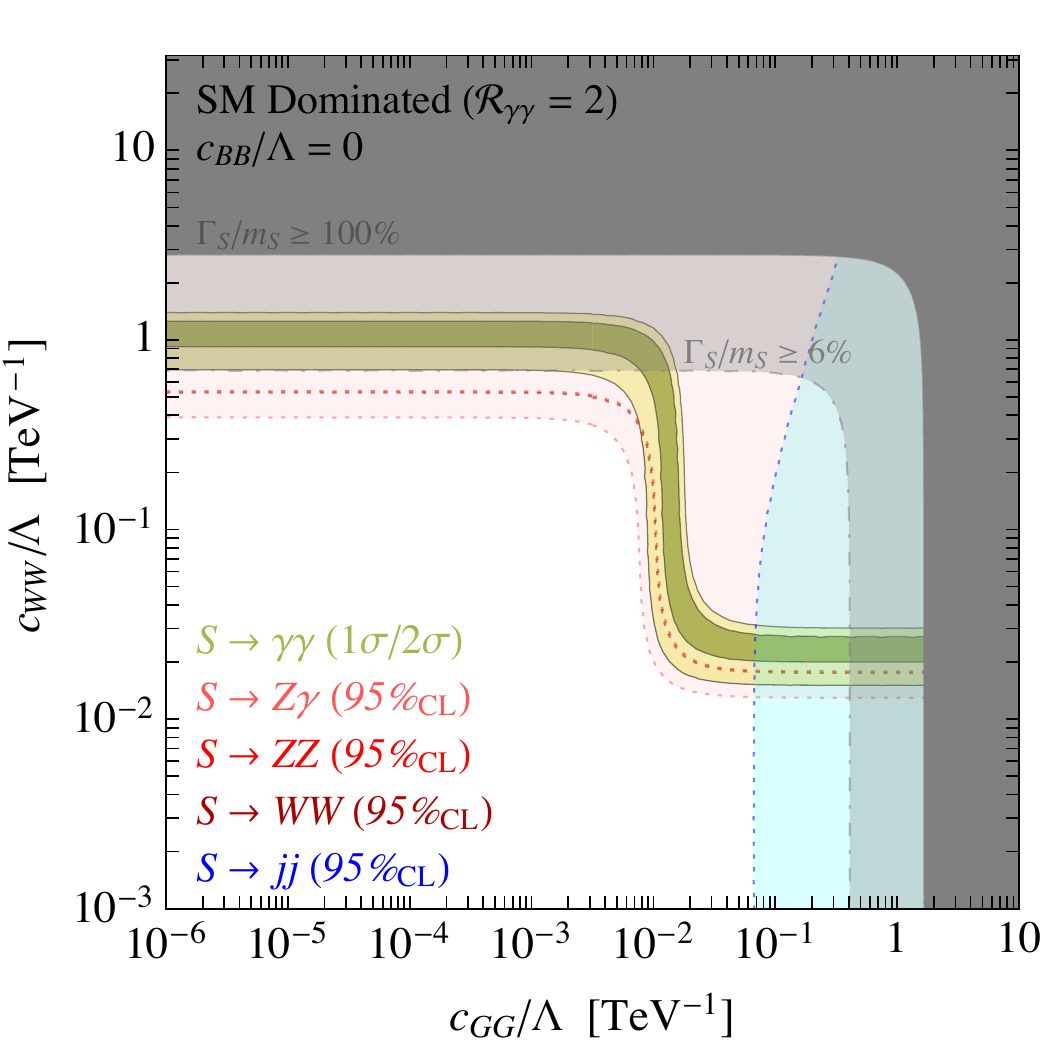} \\
\includegraphics[width=0.495\textwidth]{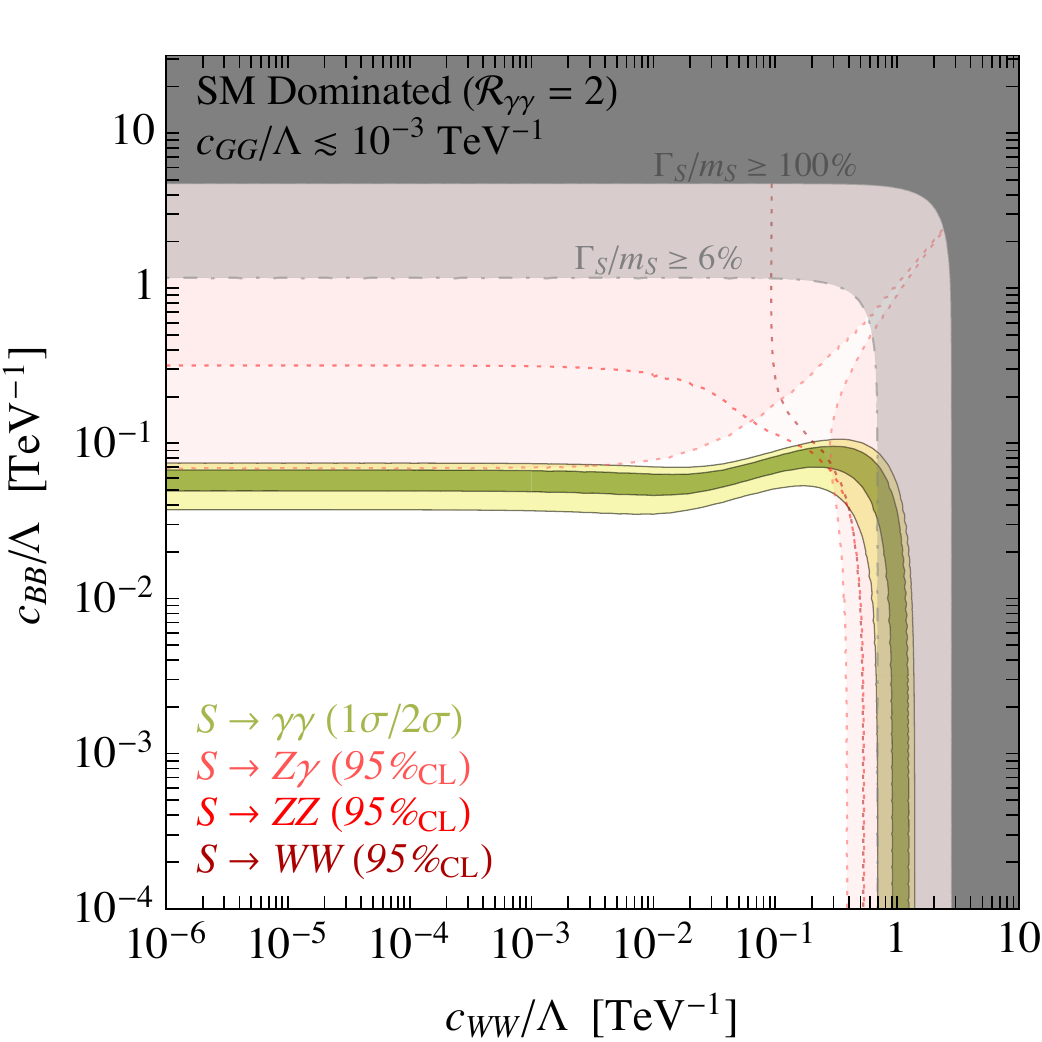}
\includegraphics[width=0.495\textwidth]{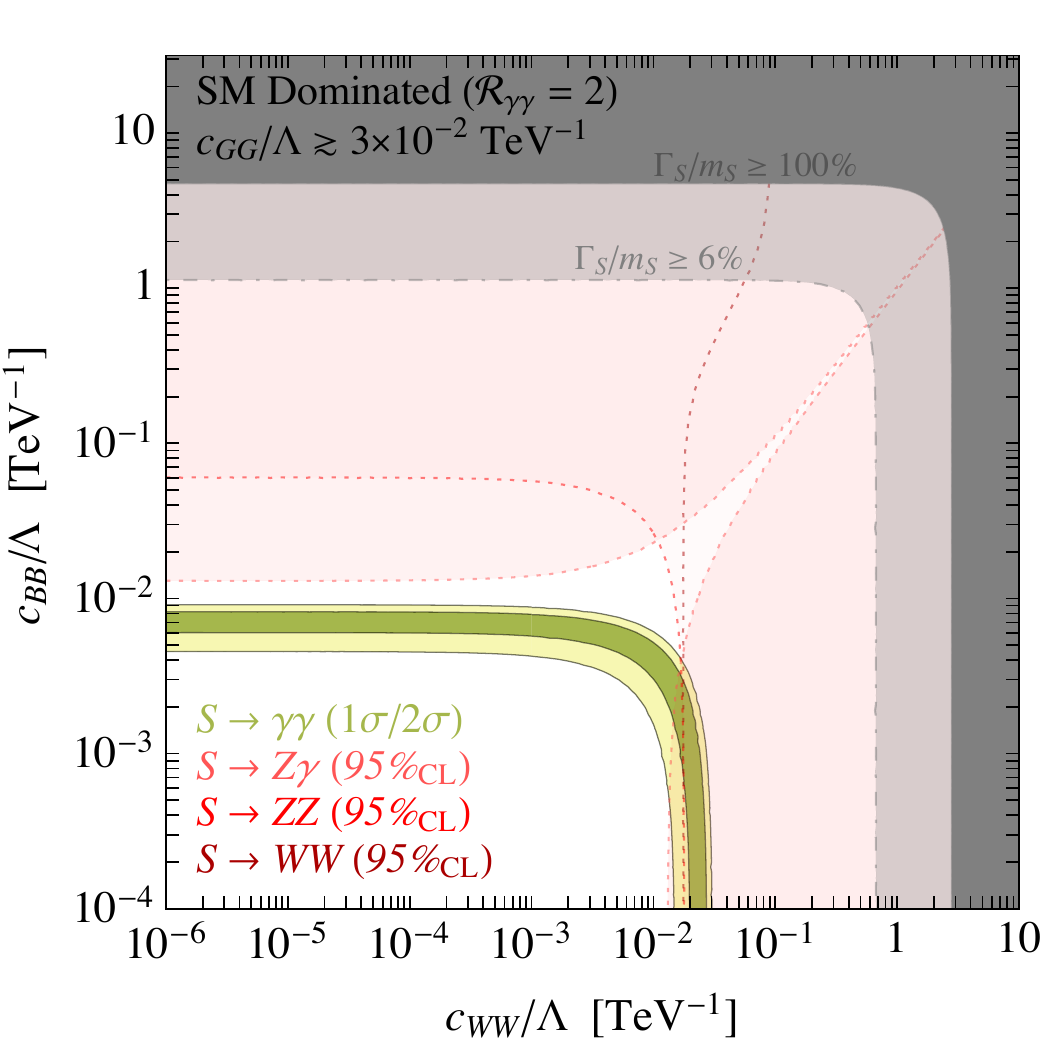} 
\caption{Parameters for diphoton excess (green regions) and excluded by \LHC\ Run 1 searches (red and blue regions). The $\Gamma_S \geq m_S$ region is completely shaded away, whereas the one with $\Gamma_S / m_S \geq 6 \%$ is shaded with light gray. We set the rescaling factor defined in \Eq{eq:Rgammagamma} to $\mathcal{R}_{\gamma\gamma} = 2$. In the upper panels we switch on the coupling to gluons and consider $c_{WW} = 0$ (left) and $c_{BB} = 0$ (right). In the lower panels we assume the production dominated by photon (left) or gluon (right) fusion and visualize the parameter space in the $(c_{WW}, c_{BB})$ plane.}
\label{fig:Narrow_Resonance1}
\end{figure}

\begin{figure}[!t]
\centering
\includegraphics[width=0.495\textwidth]{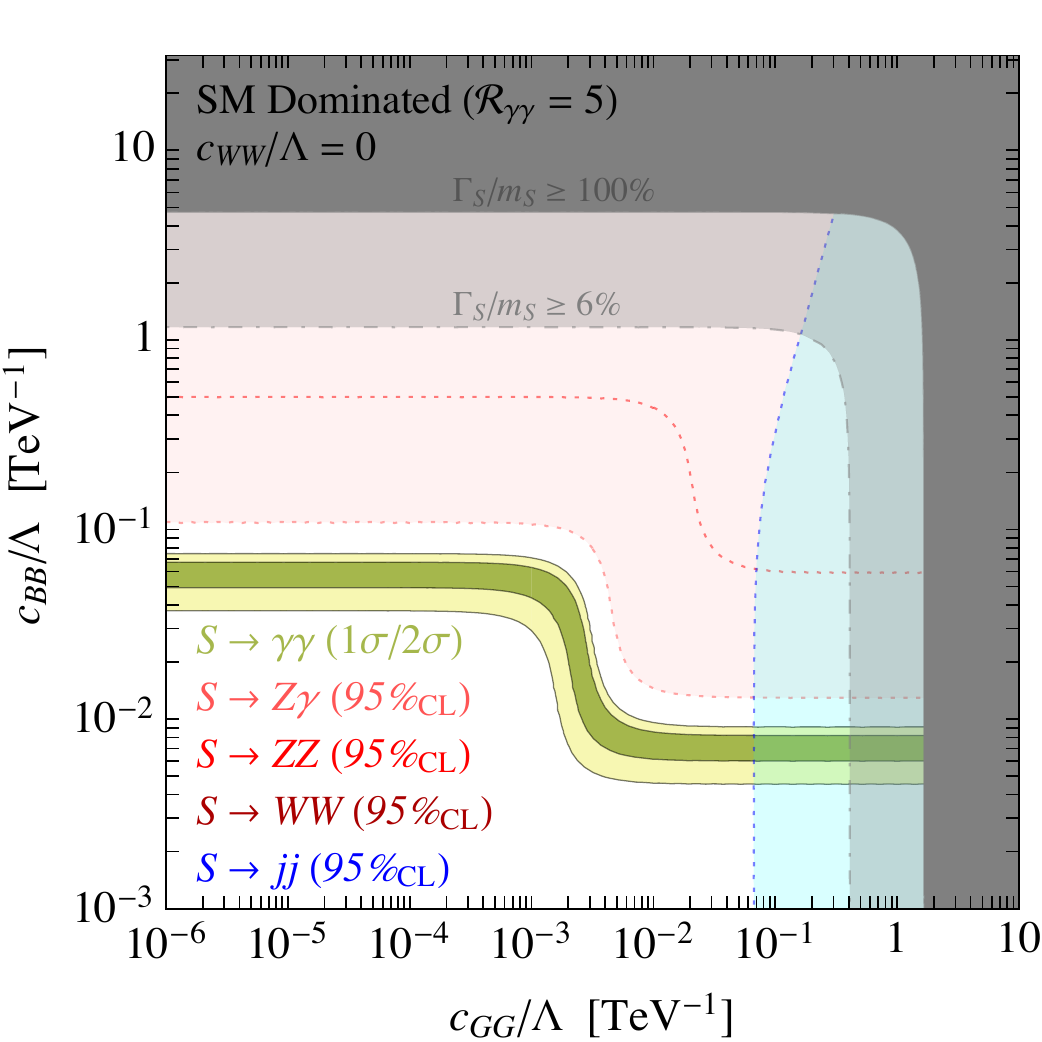}
\includegraphics[width=0.495\textwidth]{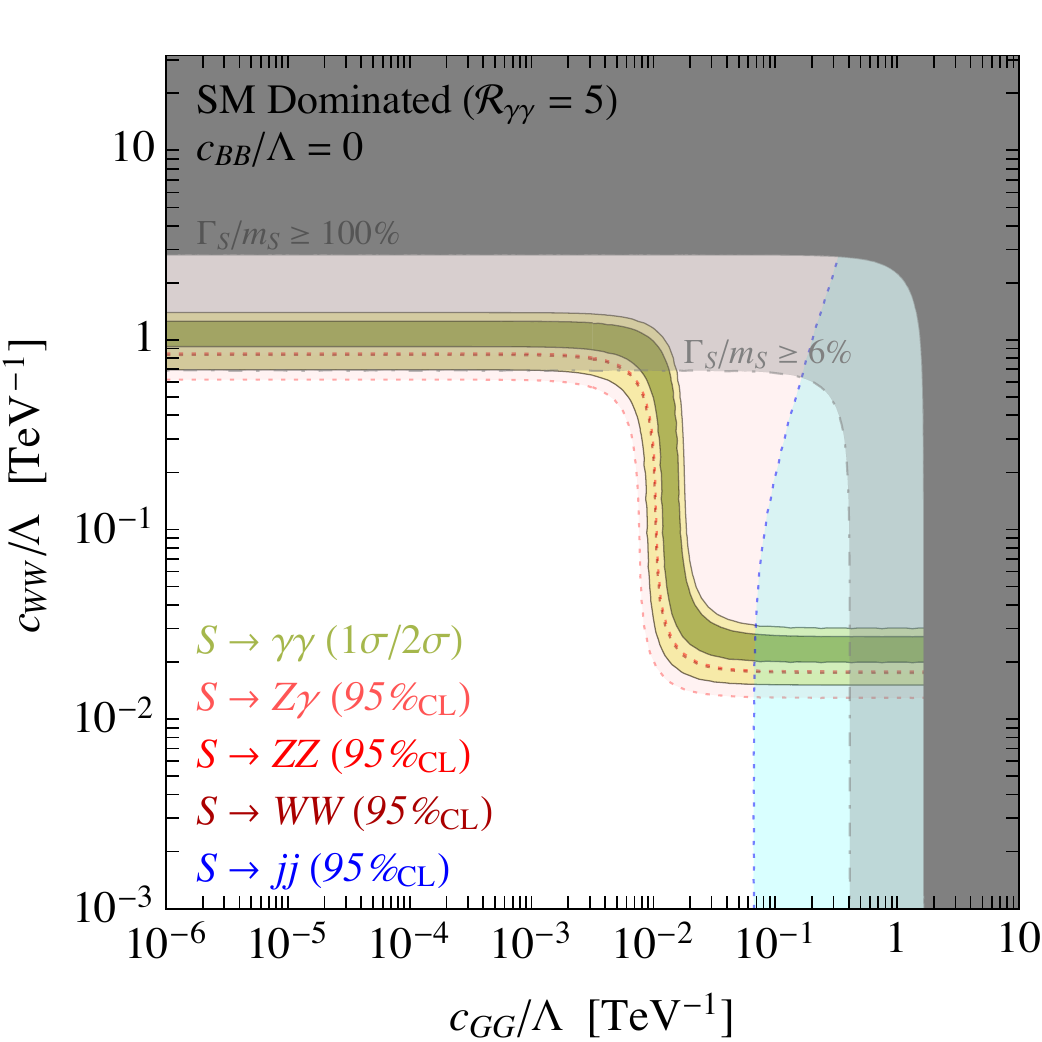} \\
\includegraphics[width=0.495\textwidth]{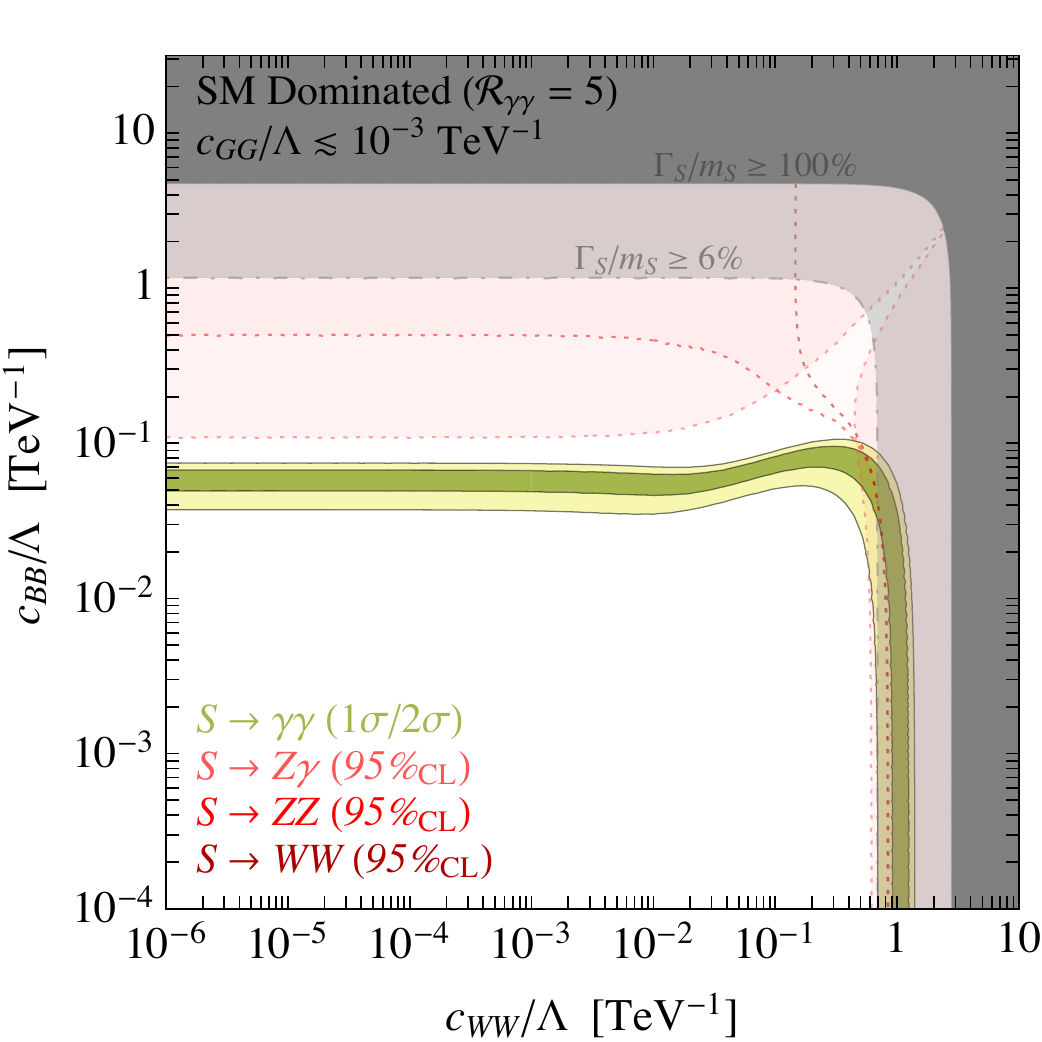}
\includegraphics[width=0.495\textwidth]{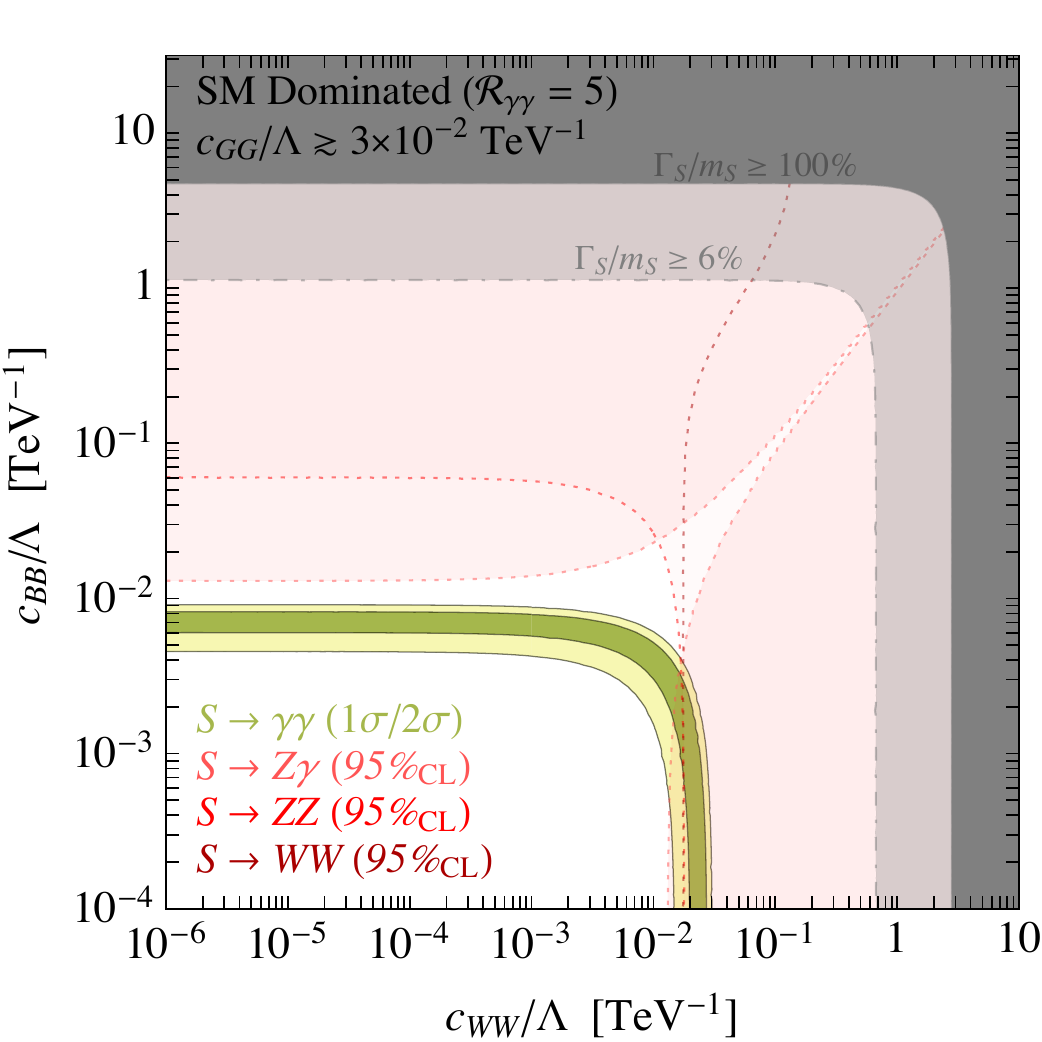}
\caption{Same as \Fig{fig:Narrow_Resonance1} but for $\mathcal{R}_{\gamma\gamma}  = 5$.}
\label{fig:Narrow_Resonance2}
\end{figure}

Our first scenario features DM masses above the kinematical threshold for decays $m_\chi\gtrsim m_S / 2$. The total width of the scalar $S$ results in
\beq
\Gamma_S \simeq \frac{m_S^3}{4 \pi} \left(\frac{8 c_{GG}^2 + 3 c^2_{W W} + c_{BB}^2}{\Lambda^2}\right)  \ ,
\label{eq:GammaSgluons}
\eeq
obtained by using the decay widths given in Appendix~\ref{app:DecayAndXS} in the $m_{W,Z} \ll m_S$ limit. In this scenario, the total width $\Gamma_S$ is quite narrow for a large region of the parameter space because of the $(m_S/\Lambda)^{2}$ suppression, consequence of the non-renormalizable interactions in \Eq{eq:Lintdim5}. 

Gluon and photon fusions are both a potential source for production and which one dominates depends on the relative sizes of the couplings. As discussed in Section~\ref{sec:LHCXS}, the photon fusion cross section suffers from theoretical uncertainties in the inverse proton radius. We summarize our results in \Fig{fig:Narrow_Resonance1} and \Fig{fig:Narrow_Resonance2}, where we choose $\mathcal{R}_{\gamma\gamma} = 2$ and $\mathcal{R}_{\gamma\gamma} = 5$, respectively. These plots show the same quantities with the only difference being the choice for $\mathcal{R}_{\gamma\gamma}$, so we discuss each panel only once and we emphasize whenever the choice for $\mathcal{R}_{\gamma\gamma}$ makes any difference.

In both figures, the green shaded areas corresponds to the $1\sigma$ and $2\sigma$ regions for the diphoton excess. We also shade the areas excluded by \LHC\ Run 1 searches with red ($WW$, $ZZ$, $Z \gamma$) and blue (di-jet) . Finally, we ignore parameters giving $\Gamma_S \geq m_S$ and shade with light gray the regions where $\Gamma_S / m_S$ is above the $6 \%$ value favored by ATLAS.

We start our discussion from the two upper panels, where in the left (right) we show the parameter space for only $c_{GG}$ and $c_{BB}$ ($c_{WW}$) switched on. At very small values of $c_{GG}$, located on the left of the plots, the production is dominated by photon fusion
\beq
\label{eq:FfusionScen1} \sigma_{p p \rightarrow ij}(\sqrt{s}) \simeq  r_{\rm inel} \,  \frac{8 \pi^2 \, C_{\gamma\gamma}^{\sqrt{s}}}{s} \frac{(c_w^2 c_{BB} + s_w^2 c_{WW})^2}{c_{BB}^2 + 3 c_{WW}^2} \, \frac{\Gamma_{S \rightarrow i j}}{m_S} \ ,
\eeq
where $s_w$ ($c_w$) is the sine (cosine) of the weak mixing angle. Not surprisingly, both the region accounting for the diphoton excess and the exclusion limits show up at constant values of $c_{BB}$ (left) or $c_{WW}$ (right). It is always possible to have a consistent explanation of the diphoton excess through photon fusion with only $c_{BB}$, although the $\mathcal{R}_{\gamma\gamma} = 2$ case features some tension with results from $Z\gamma$ searches. However, the case with only $c_{WW}$ is excluded by Run 1 results. As we move toward the right of the two upper panels, eventually we increase $c_{GG}$ enough such that gluon fusion is the dominant production process while still being consistent with dijet searches. In this opposite limit the cross section is approximately
\begin{align}
\label{eq:GfusionScen1} \sigma_{p p \rightarrow ij}(\sqrt{s}) \simeq & \,  \mathcal{K}_{GG} \, \frac{\pi^2 \, C_{GG}^{\sqrt{s}}}{8  \, s} \, \frac{\Gamma_{S \rightarrow i j}}{m_S} \ .
\end{align}
The diphoton excess is again accounted for by a constant value of $c_{BB}$ (left) or $c_{WW}$ (right). Again, having $c_{WW}$ only is excluded by $Z\gamma$ limits, whereas the case with only $c_{BB}$ is allowed regardless of the specific value of $\mathcal{R}_{\gamma\gamma}$ which plays no role for gluon-fusion dominated processes.

A thorough exploration of these two opposite limits is provided in the lower two panels of \Fig{fig:Narrow_Resonance1} and \Fig{fig:Narrow_Resonance2}. More specifically, we consider on the left (right) values of $c_{GG}$ small (big) enough such that the production is dominated by photon (gluon) fusion, and we visualize the allowed regions in the $(c_{WW}, c_{BB})$ plane. First, we notice consistency with our previous findings. Regions with $c_{BB} \gg c_{WW}$ can account for the diphoton events, with again some tension for the case with a rescaling factor $\mathcal{R}_{\gamma\gamma} = 2$. Conversely, the case with mostly $c_{WW}$ is badly excluded by Run 1 searches. An interesting intermediate case, allowed for both gluon and photon fusion, is for couplings to electroweak gauge bosons roughly of the same size $c_{BB} \simeq c_{WW}$. In particular, in the photon fusion case $\mathcal R_{\gamma\gamma}=5$, right at the edge of the $Z\gamma$ limit the couplings $c_{BB}$ and $c_{WW}$ can be large enough to give a relatively broad resonance of $\Gamma_S / m_S \simeq (2\rm-3) \%$. However, it is not possible to reproduce the ATLAS preferred value $\Gamma_S / m_S \simeq 6 \%$ as the $Z\gamma$ limits are too stringent. Nevertheless, this seems to be the only point in parameter space where a sizable width is still possible without having decays into invisible states (as discussed below).

%This configuration is at the edge of $Z\gamma$ limits and it yields a slightly broader resonance, although it is never possible in this scenario to reproduce the ATLAS preferred value $\Gamma_S / m_S \simeq 6 \%$.
 
\subsection{A DM Dominated Resonance}
\label{sec:DMdom}

\begin{figure}[!t]
\centering
\includegraphics[width=0.495\textwidth]{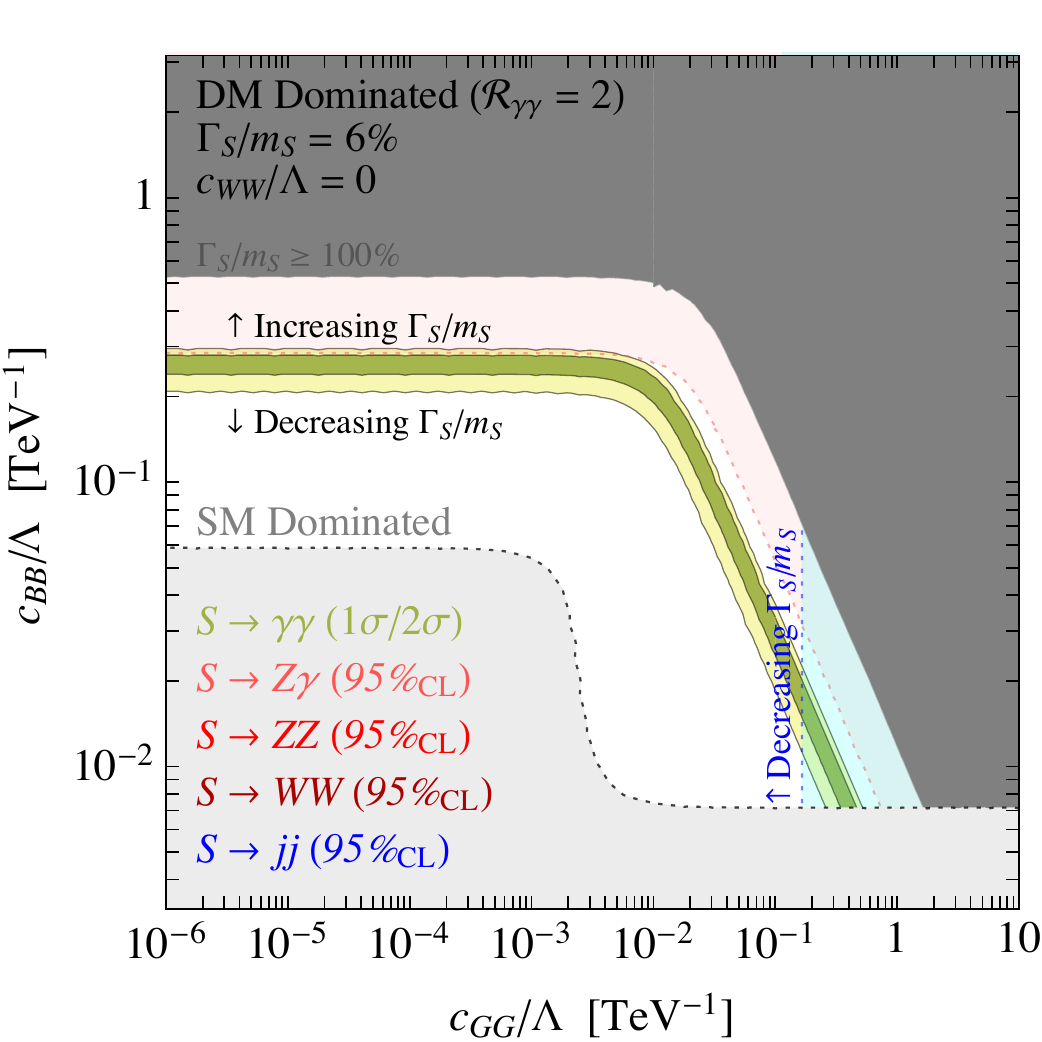}
\includegraphics[width=0.495\textwidth]{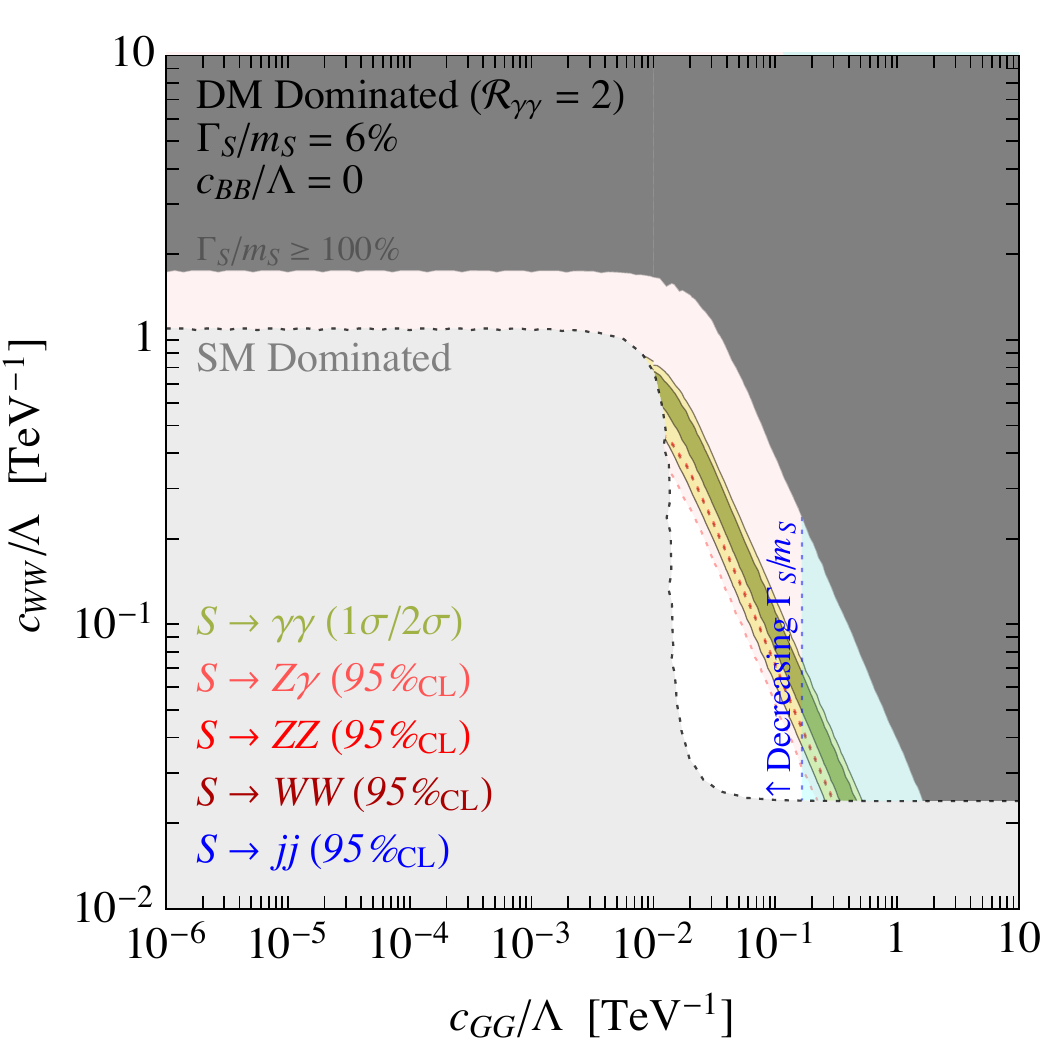} \\
\includegraphics[width=0.495\textwidth]{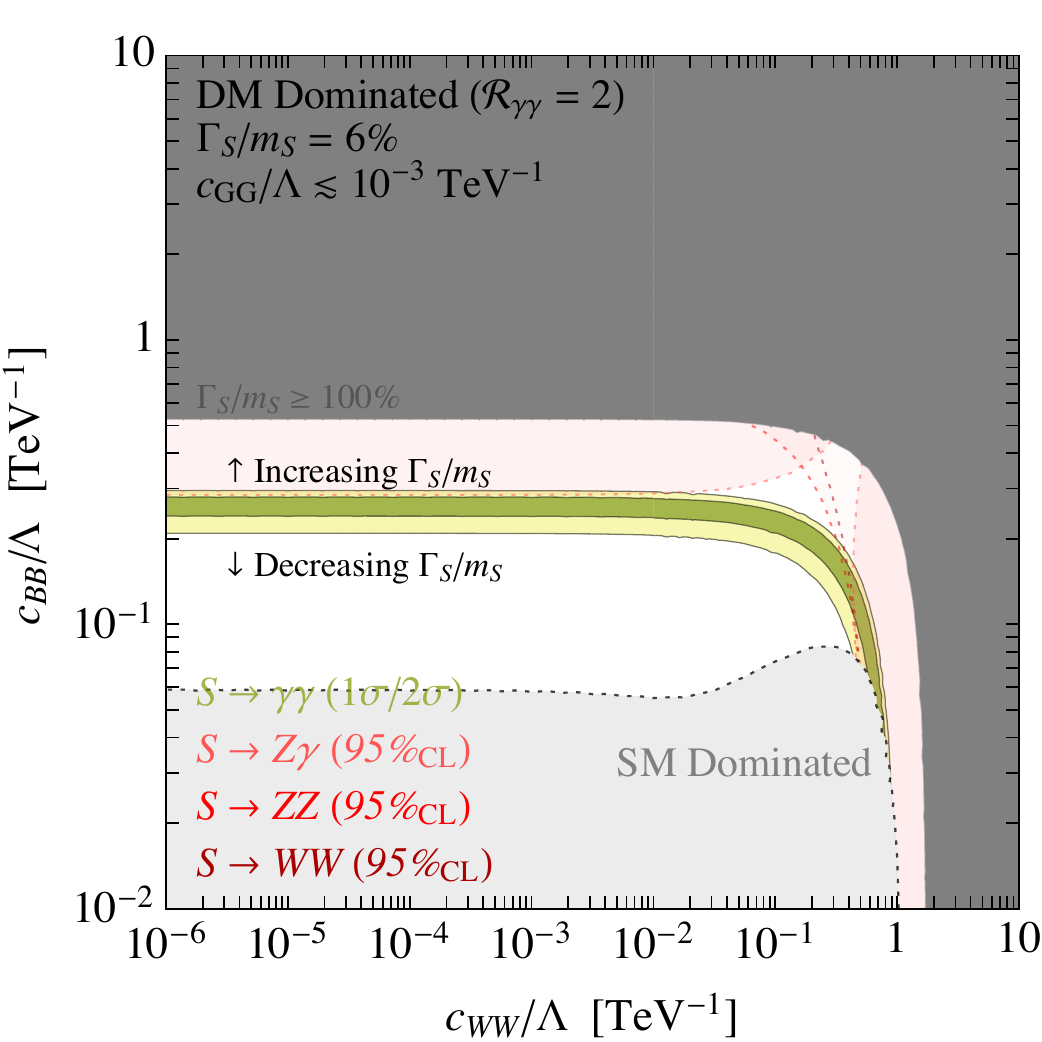}
\includegraphics[width=0.495\textwidth]{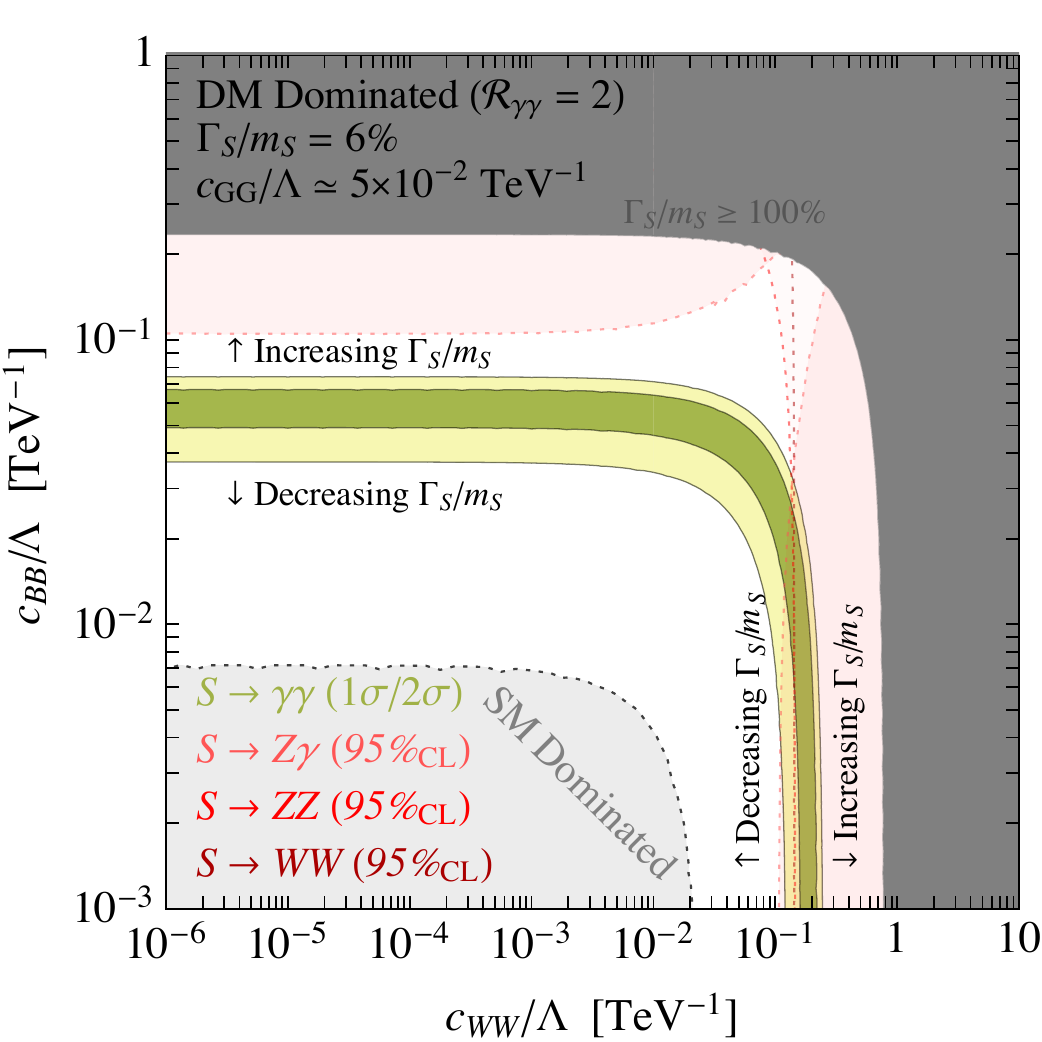}
\caption{Parameter space for the DM dominated scenario for the rescaling factor defined in \Eq{eq:Rgammagamma} equal to $\mathcal{R}_{\gamma\gamma} = 2$. We identify the regions preferred by the diphoton excess (green) and excluded by \LHC\ Run 1 (red and blue). We always set $\Gamma / m_S \simeq 6 \%$, and the little arrows show how our bands moves as we change this value. We shade with dark gray the  $\Gamma_S \geq m_S$ region, and with light gray the region below the boundary where the decay width to SM states alone accounts for the signal. In the upper panels we consider couplings to gluons and $c_{WW} = 0$ (left) or $c_{BB} = 0$ (right).  In the lower panels we assume the production dominated by photon (left) or gluon (right) fusion and visualize the parameter space in the $(c_{WW}, c_{BB})$ plane.}
\label{fig:DMdominated1}
\end{figure}

\begin{figure}[!t]
\centering
\includegraphics[width=0.495\textwidth]{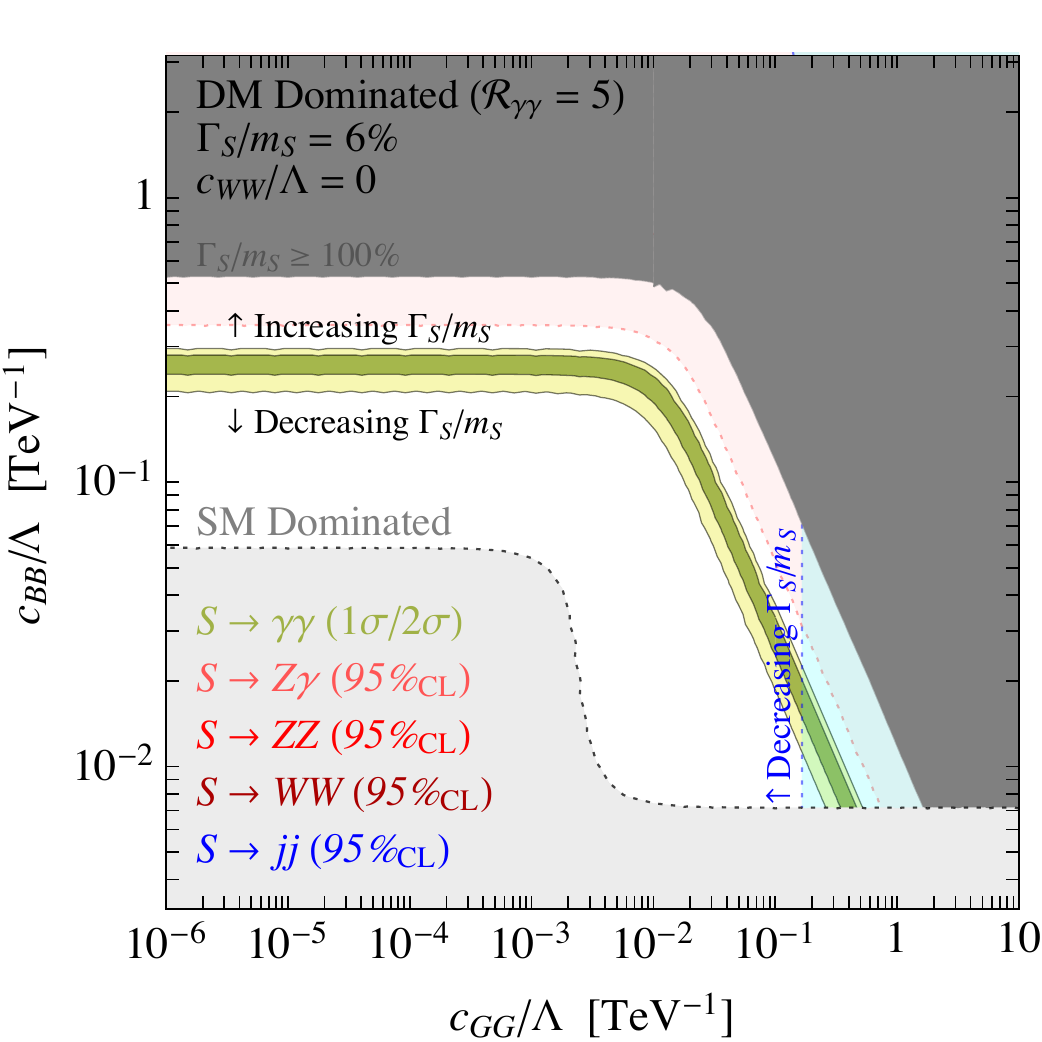}
\includegraphics[width=0.495\textwidth]{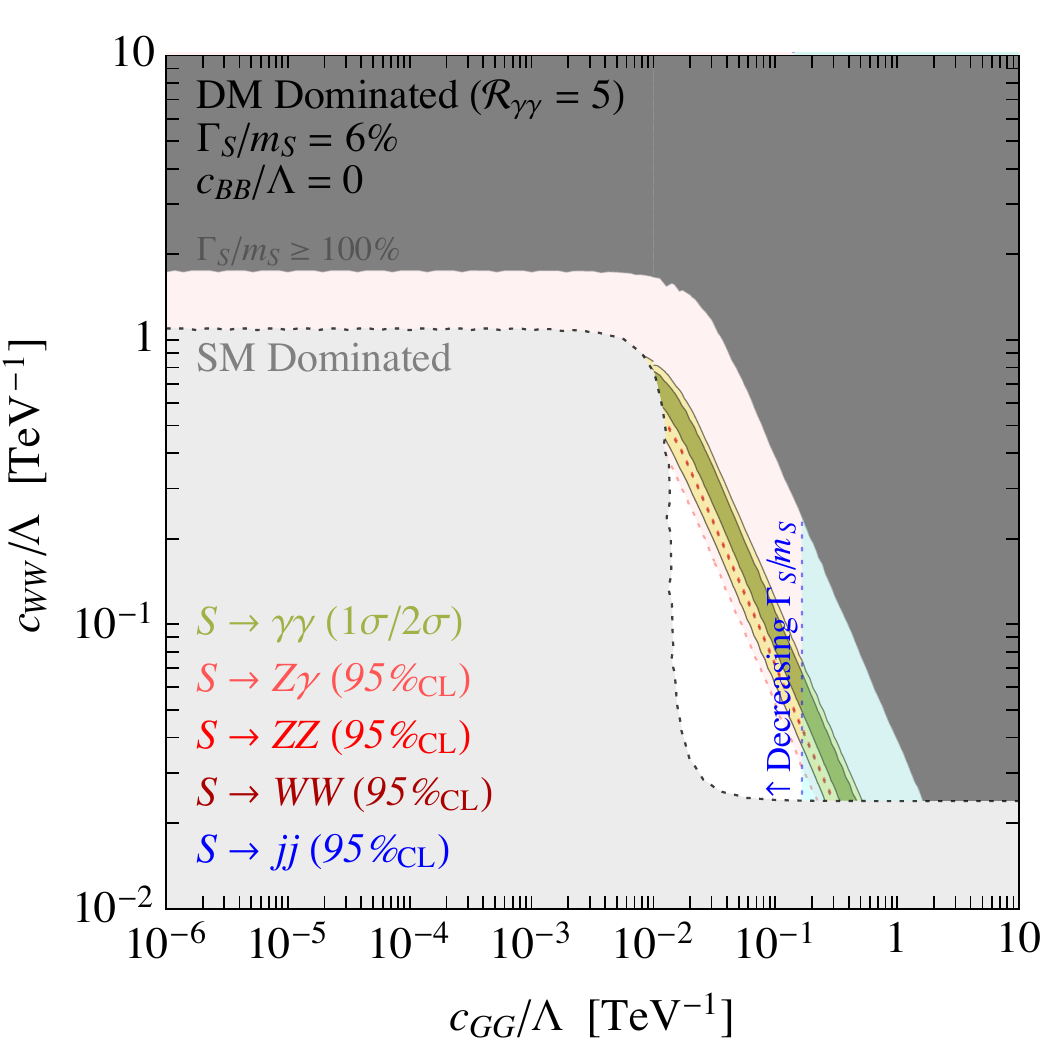} \\
\includegraphics[width=0.495\textwidth]{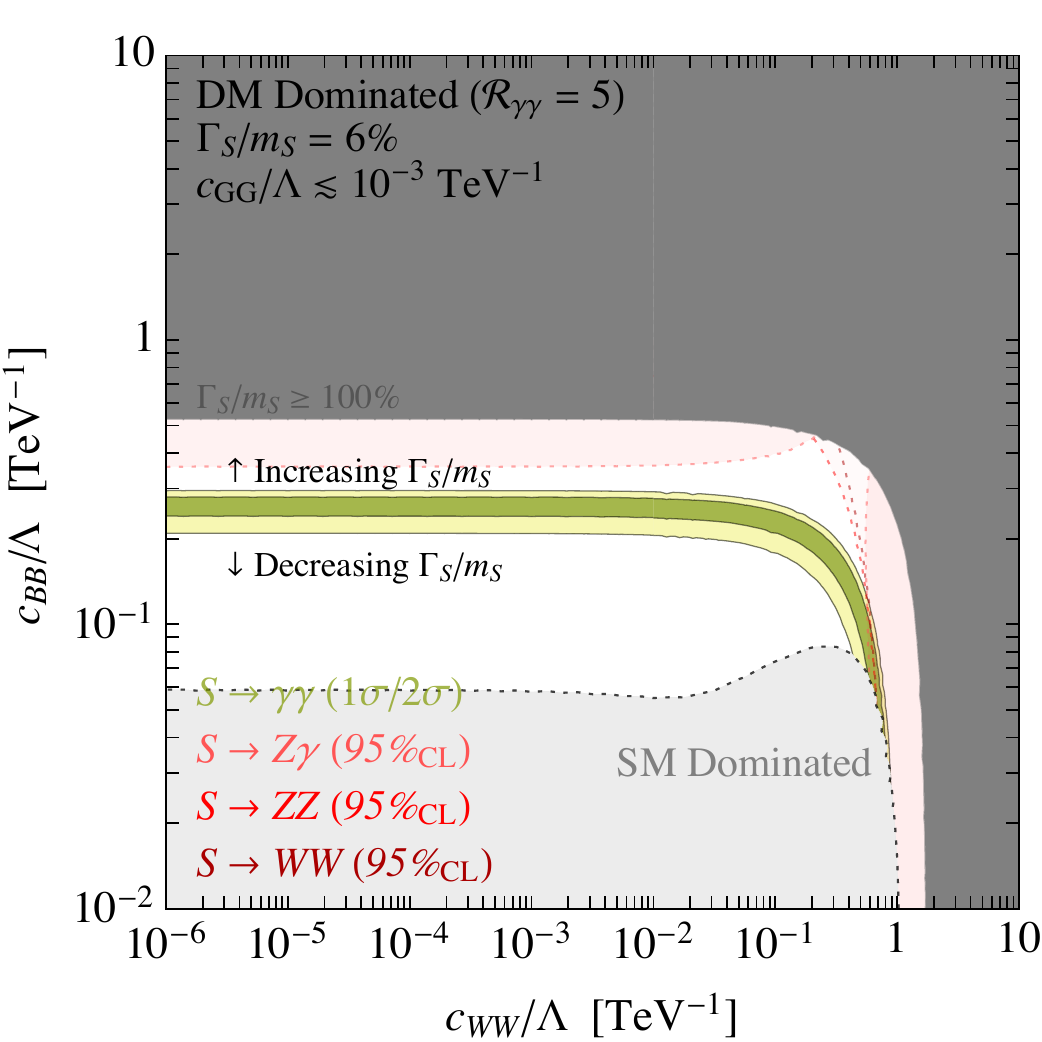}
\includegraphics[width=0.495\textwidth]{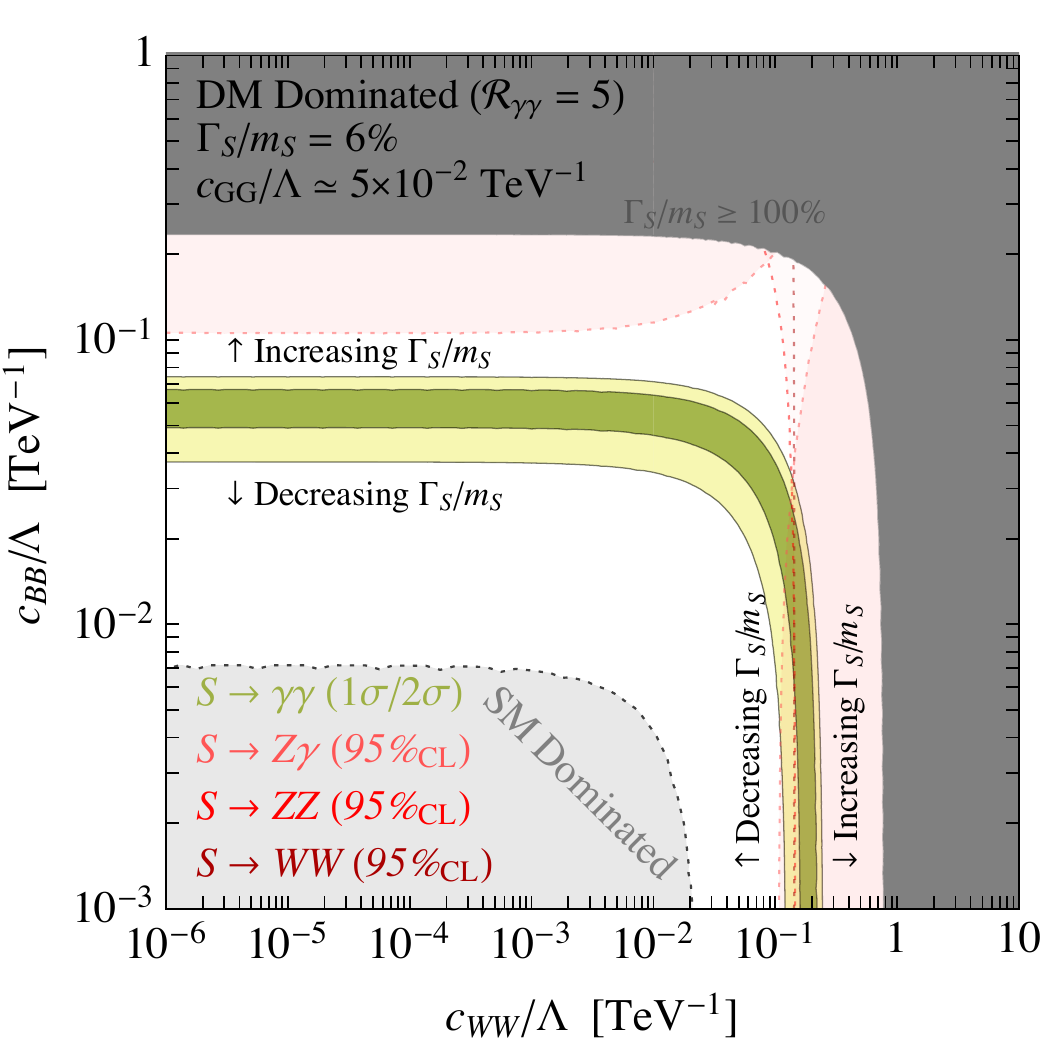}
\caption{Same as \Fig{fig:DMdominated1} but for $\mathcal{R}_{\gamma\gamma}  = 5$.}
\label{fig:DMdominated2}
\end{figure}

For DM mass values below $m_S / 2$ the \LHC\ phenomenology is drastically different. The decay channel to DM is now open and completely dominates the total width
\beq
\Gamma_S \simeq  \frac{c_{\chi S}^2 \, m_S}{8 \pi} \simeq \left(\frac{c_{\chi S}}{1.23}\right)^2  \; 45 \, {\rm GeV} \ .
\label{eq:GammaSDMdom}
\eeq
The above equation is obtained by using the results of Appendix~\ref{app:DecayAndXS} and ignoring the phase space suppression, only relevant for DM masses very close to $m_S / 2 \simeq 375 \, {\rm GeV}$. The ATLAS best fit value for the witdh $\Gamma_S \simeq 45 \, {\rm GeV}$ is easily obtained for couplings to DM of order one.

The analysis proceeds similarly to the SM dominated resonance scenario in Section~\ref{sec:SMdom}, with the important exception that we have also the coupling $c_{\chi S}$ in the game. As a consequence, we can always fix the total width $\Gamma_S$ to any value. Our figures in this Section follow the same conventions as the ones adopted in Section~\ref{sec:SMdom}, with two important differences. First, we present our results in this DM dominated resonance scenario for the ATLAS best fit value $\Gamma_S / m_S \simeq 6 \%$, since we have the freedom to independently choose $\Gamma_S$. We use arrows in our plots to show how our results change if one choses a different value (note that the limits from $Z\gamma$, $ZZ$, and $WW$ searches also follow the arrows). Second, we shade with light gray in each plot the region below the center of green bands in Figs.~\ref{fig:Narrow_Resonance1} and \ref{fig:Narrow_Resonance2}. At the boundary of this ``SM dominated" portion of the parameter space, the SM contribution alone accounts for the signal, therefore we cannot go below it.

The results are shows in Figs.~\ref{fig:DMdominated1} and \ref{fig:DMdominated2}, where the only difference between the two figures is still the choice of $\mathcal{R}_{\gamma\gamma}$. As usual, we start from the class of models where we only have the scalar coupled to gluons and the hypercharge (weak-isospin) gauge boson, with results shown on the top-left(-right) panels. For very small couplings to gluons, on the left of the plots, photon fusion dominates and the total cross section is approximately 
\beq
\sigma_{p p \rightarrow ij}(\sqrt{s}) \simeq  r_{\rm inel} \, \frac{16 \pi^2 \, C_{\gamma\gamma}^{\sqrt{s}}}{s} \frac{(c_w^2 c_{BB} + s_w^2 c_{WW})^2}{c_{\chi S}^2} \, \frac{m_S^2}{\Lambda^2} \, \frac{\Gamma_{S \rightarrow i j}}{m_S}\ .
\label{eq:FfusionScen2} 
\eeq
The case of only couplings to $c_{BB}$ works in this regime, with again some tension with Run 1 bounds for $\mathcal{R}_{\gamma\gamma} = 2$. The case with $c_{WW}$ only, other than being excluded (in agreement with Ref.~\cite{Fichet:2015vvy}), also falls well inside the SM dominated region and therefore we neglect it. On the opposite end of the plots, gluon fusion dominates all productions with cross section 
\beq
\sigma_{p p \rightarrow ij}(\sqrt{s}) \simeq   \mathcal{K}_{GG} \, \frac{2\, \pi^2 \, C_{GG}^{\sqrt{s}}}{s} \, 
\frac{c_{GG}^2}{c_{\chi S}^2} \, \frac{m_S^2}{\Lambda^2}  \, \frac{\Gamma_{S \rightarrow i j}}{m_S} \ .
\label{eq:GfusionScen2} 
\eeq
Unlike the previous scenario, this gluon fusion regime does not pinpoint a specific value of $c_{BB}$ or $c_{WW}$, but the green bands roughly corresponds to $c_{BB} \, c_{GG} \simeq {\rm const}$, with this behavior persisting for values of $\Gamma_S / m_S$ not too close to the SM dominated gray region. This is of course due to the fact that the total width is dominated by invisible decays, and not by decays to gluons as in the SM dominated case. The gluon fusion regime is again consistent with data if we only have the coupling $c_{BB}$, whereas the case with only $c_{WW}$ is excluded by \LHC\ bounds.

As done before, in the bottom panels of Figs.~\ref{fig:DMdominated1} and \ref{fig:DMdominated2} we further explore the $(c_{WW}, c_{BB})$ plane for the opposite photon (left) and gluon (right) fusion regimes. Not surprisingly, the allowed values are $c_{WW} \ll c_{BB}$ up to $c_{WW} \simeq c_{BB}$.

%%%%%%%%%%%%%%%%%%%%%%%%%%%%%%%%%%%%%%%%%%%%%%%%
%%%%%%%%%%%%%%%%%%%% DM %%%%%%%%%%%%%%%%%%%%%%%%%%
%%%%%%%%%%%%%%%%%%%%%%%%%%%%%%%%%%%%%%%%%%%%%%%%

\section{Dark Matter with a (Pseudo-)Scalar Portal}
\label{sec:DM}

With the \LHC\ analysis performed in Section~\ref{sec:LHCscenarios}, we are ready to include the DM in our study. Recent and related DM works on the possibility of a 750 GeV (pseudo-)scalar mediator can be found in Refs.~\cite{Mambrini:2015wyu,Backovic:2015fnp,Bi:2015uqd,Bauer:2015boy,Dev:2015isx,Davoudiasl:2015cuo,Han:2015yjk,Park:2015ysf,Huang:2015svl,Ghorbani:2016jdq}.\footnote{Although a spin-1 mediator cannot directly decay to two photons~\cite{Landau:1948kw,Yang:1950rg}, vector-portal DM models can still be consistent with data if one considers a different decay topology~\cite{Chala:2015cev} or decays of dark Higgs fields~\cite{deBlas:2015hlv,Das:2015enc}. Spin-2 mediators in the context of theories with extra dimensions have been studied in \Ref{Han:2015cty}.} We extend their DM analysis by using the output of our \LHC\ study, where we have considered the full parameter space with both gluon and photon fusion active. We present our results for both cases of scalar and pseudo-scalar mediator. It is useful again to divide our discussion between the two scenarios of SM and DM dominated resonance, presented in Sections~\ref{sec:SMdom} and \ref{sec:DMdom}, respectively. For each case we compute the relic density of the DM as a function of its mass by following the procedure outlined in Appendix~\ref{app:relic} and 
we demand that it makes all of the measured DM in the Universe ($\Omega_\chi h^2 = 0.1188\pm 0.0010$ as inferred by the latest results of \Planck~\cite{Ade:2015xua}).
Furthermore we impose constraints from the following DM searches:
\begin{itemize}
\item Collider searches for events with a singlet jet and missing energy (j+MET)~\cite{Aad:2015zva,Khachatryan:2014rra}, which are suitable only in the gluon fusion regime. We implement our EFT in FeynRules~\cite{Alloul:2013bka} and we generate the associated UFO model file~\cite{Degrande:2011ua}. The signal is then obtained by using MadGraph~\cite{Alwall:2011uj}. We impose the bound in the MET $>$ 500 GeV bin, where the signal must satisfy the bound $\sigma({\rm j+MET}) \lesssim 6 \, {\rm fb}$. The DM analysis is performed by using the full results of our simulations. Here, we give the j+MET production cross section for two opposite limits in order to understand the qualitative behavior of this constraint. At DM masses well below the resonant value $m_{S,P} / 2$, the signal cross section depends only on the coupling to gluons. For CM energy $\sqrt{s} = 8\, {\rm TeV}$ it scales as\footnote{The result for the pseudo-scalar case is identical up to renaming the couplings. Same for \Eq{eq:jMET2}.}
\beq
\left. \sigma^{{\rm MET} \; > \; 500 \, {\rm GeV}}_{pp \rightarrow {\rm j+MET}}\right|_{\rm DM \, dominated} \simeq 6 \, {\rm fb}  \, \left(\frac{c_{GG} / \Lambda}{0.032 \, {\rm TeV}^{-1}} \right)^2  \ .
\label{eq:jMET1}
\eeq
The mediator is produced on-shell in this regime and then it decays to DM pairs with $100\%$ branching ratio. This explain the absence of $c_{\chi S}$ in \Eq{eq:jMET1}, which holds as long as the DM coupling $c_{\chi S}$ is such that the scalar decay width is not too large (see \Eq{eq:GammaSDMdom}). We checked that \Eq{eq:jMET1} correctly describes the parameter space region we are interested in. Conversely, the mediator is way off-shell for DM masses above the resonant value $m_{S,P} / 2$. The process in this case can be approximately described by a contact interaction between gluons and DM particles~\cite{Goodman:2010ku}. For such a heavy DM particle we have 
\beq
\left. \sigma^{{\rm MET} \; > \; 500 \, {\rm GeV}}_{pp \rightarrow {\rm j+MET}}\right|_{\rm SM \, dominated}  \simeq 5.9 \times 10^{-3} \, {\rm fb}  \; c^2_{\chi S}  \left(\frac{c_{GG} / \Lambda}{0.032 \, {\rm TeV}^{-1}} \right)^2 \left( \frac{600 \, {\rm GeV}}{m_\chi} \right)^4 \ .
\label{eq:jMET2}
\eeq
As a consequence, collider limits do not play any role for the SM dominated scenario. The reader interested in further details can find a specific mono-jet analysis for the 750 GeV portal in \Ref{Barducci:2015gtd}.

\item Direct searches, where we impose the most recent \LUX\ bounds~\cite{Akerib:2015rjg} and show \LZ\ projections as extracted from~\Ref{LZTalk}. These limits are only relevant for scalar mediators. Here, we present the spin-independent cross section for a DM Dirac fermion scattering elastically off a nucleus derived from the interactions in Eq.~\eqref{eq:EFTDD}. These low-energy couplings  are connected to those at the \LHC\ scale as explained in Section~\ref{sec:RGE}. For the DM-quark and DM-gluon operators we follow the steps in Ref.~\cite{DelNobile:2013sia}, while for the DM-photon interactions we follow Ref.~\cite{Ovanesyan:2014fha}, and write
\begin{eqnarray}\label{Totalsigma}
\sigma^A_{\rm SI} &=& \frac{1}{\pi} \, \left(\frac{m_\chi \, m_A}{m_\chi + m_A}\right)^2 \bigg\{ (f_\chi^q + f_\chi^G)^2 \langle F_\mathrm{H}^2 \rangle + (f_\chi^q + f_\chi^G)f_\chi^F\langle F_\mathrm{H}F_{\mathrm{R}} \rangle + (f_\chi^F)^2 \langle F_\mathrm{R}^2\rangle\bigg\} \, 
\end{eqnarray}
where $m_A$ is the target nuclear mass. In the rest of the text we denote the scattering cross section of a single nucleon by $\sigma_{\rm SI}$ dividing Eq.~\eqref{Totalsigma} for $(A^2 \langle F_\mathrm{H}^2\rangle)$ and replacing the DM-nucleus reduced mass with the DM-nucleon one. 
Here we have defined
\begin{eqnarray}
f_\chi^q &=& \sum_{u,d,s}\mathcal C_q(\mu_N) \left[Z m_p f^p_q + (A-Z) m_n f^n_q\right] \ ,\\
f_\chi^G&=& -\mathcal C_G(\mu_N)\frac{8\pi}{9 \alpha_s(\mu_N)} \left[Z m_p + (A-Z)m_n\right]f^N_G \ ,\\
f_\chi^F&=& \mathcal C_F(\mu_N)\frac{Z^2\alpha_{\mathrm{em}}}{\pi}\frac{\sqrt{8\pi}}{b(A)} \ ,
\end{eqnarray}
denoting, respectively, the contributions from $\mathcal C_q$, $\mathcal C_G$, and $\mathcal C_F$ to the scattering amplitude. For the up and down scalar couplings we use the recent determination in Refs.~\cite{Crivellin:2013ipa,Hoferichter:2015dsa} based on chiral perturbation theory and a Roy-Steiner analysis
\begin{eqnarray}
f^p_u = (20.8\pm 1.5)\cdot 10^{-3}\ , &\qquad& f^n_u = (18.9\pm 1.4)\cdot 10^{-3}\ ,\nn\\
f^p_d = (41.1\pm 2.8)\cdot 10^{-3}\ , &\qquad& f^n_d = (45.1\pm 2.7)\cdot 10^{-3}\ ,
\end{eqnarray}
in good agreement with Ref.~\cite{Alarcon:2011zs}. 
For the strange scalar coupling we use the lattice QCD calculation $f^p_s = f^n_s = 0.043\pm0.011$ \cite{Junnarkar:2013ac}. In the analysis below we use the central values for these matrix elements. These values then give  the gluon coupling $f^N_G=0.894$. $b(A)$ is the harmonic oscillator parameter defined in Ref.~\cite{Fitzpatrick:2012ix}
\begin{equation}
b(A)=\sqrt{\frac{41.467}{45 A^{-1/3}-25A^{-2/3}}}\ \mathrm{fm} \ .
\end{equation}
Finally, with $\langle F_i F_j \rangle$ we denote the nuclear form factor averaged over the velocity integral and the detector efficiency \cite{Ovanesyan:2014fha}. We follow the analysis of the \LUX\ experiment and use a standard Maxwellian velocity distribution with $v_0 = 220$ km/s. The Helm ($F_H$) and Raleigh ($F_R$) form factors we take, respectively, from Refs.~\cite{Fitzpatrick:2012ix} and \cite{Ovanesyan:2014fha} where, for the latter, we set the two-body parameter $c_2=0$. 

\item Indirect searches, which on the contrary are only relevant for pseudo-scalar mediators since DM annihilations mediated by a scalar field are p-wave suppressed. We impose limits on the annihilation cross section from $\gamma$-ray line searches from both \FERMI~\cite{Ackermann:2015lka}  and \HESS~\cite{Abramowski:2013ax}  considering  peaked and cored DM density profiles, as well as limits on  dwarf spheroidal galaxies (dSphs) observations by \FERMI~\cite{Ackermann:2015zua}. Notice that the \HESS\ collaboration imposes limits~\cite{Abramowski:2013ax} only for an Einasto profile. Since the bound from \HESS\ are very sensitive to the choice of the profile (the region of interest is a small circle of $1^\circ$ centered in the galactic center), we also consider a cored profile (Burkert) in our analysis by rescaling their limits with the $J$-factor given  in Ref.~\cite{Cirelli:2010xx}. We compare the experimental bounds on the annihilation in lines (Fig.~8 of~\cite{Ackermann:2015lka} and Fig.~4 of~\cite{Abramowski:2013ax}) with the predicted annihilation cross section into $\gamma\gamma + \gamma Z /2$. Furthermore, when imposing continuum limits from dSphs we take advantage of the two following facts~\cite{Cirelli:2010xx}: photon spectra from electroweak gauge boson radiation is almost universal (in this case we compare the experimental bounds on $W^+W^-$ (bottom right-panel of Fig.~8 of~\cite{Ackermann:2015zua}) with the predicted annihilation cross section into $W^+W^- + Z Z + \gamma Z /2$), as it is the case for the ones initiated by gluons and light quarks (in this case we compare the limits on $\bar u u$ (top right-panel of Fig.~8 of~\cite{Ackermann:2015zua}) with the predicted annihilation cross section into gluons).    We rescale all indirect limits by a factor of $2$ to account for our choice of Dirac DM.
\end{itemize}

\subsection{SM Dominated Resonance}

We start by considering DM masses above the critical value $m_S / 2$, therefore the (pseudo-)scalar mediator can only decay to SM final states. A thorough exploration of the parameter space in this scenario was performed in Section~\ref{sec:SMdom}, with regions consistent with \LHC\ results shown in Figs.~\ref{fig:Narrow_Resonance1} and \ref{fig:Narrow_Resonance2}. We study DM phenomenology for three representative classes of models.

\begin{figure}[!t]
\centering
\includegraphics[width=0.495\textwidth]{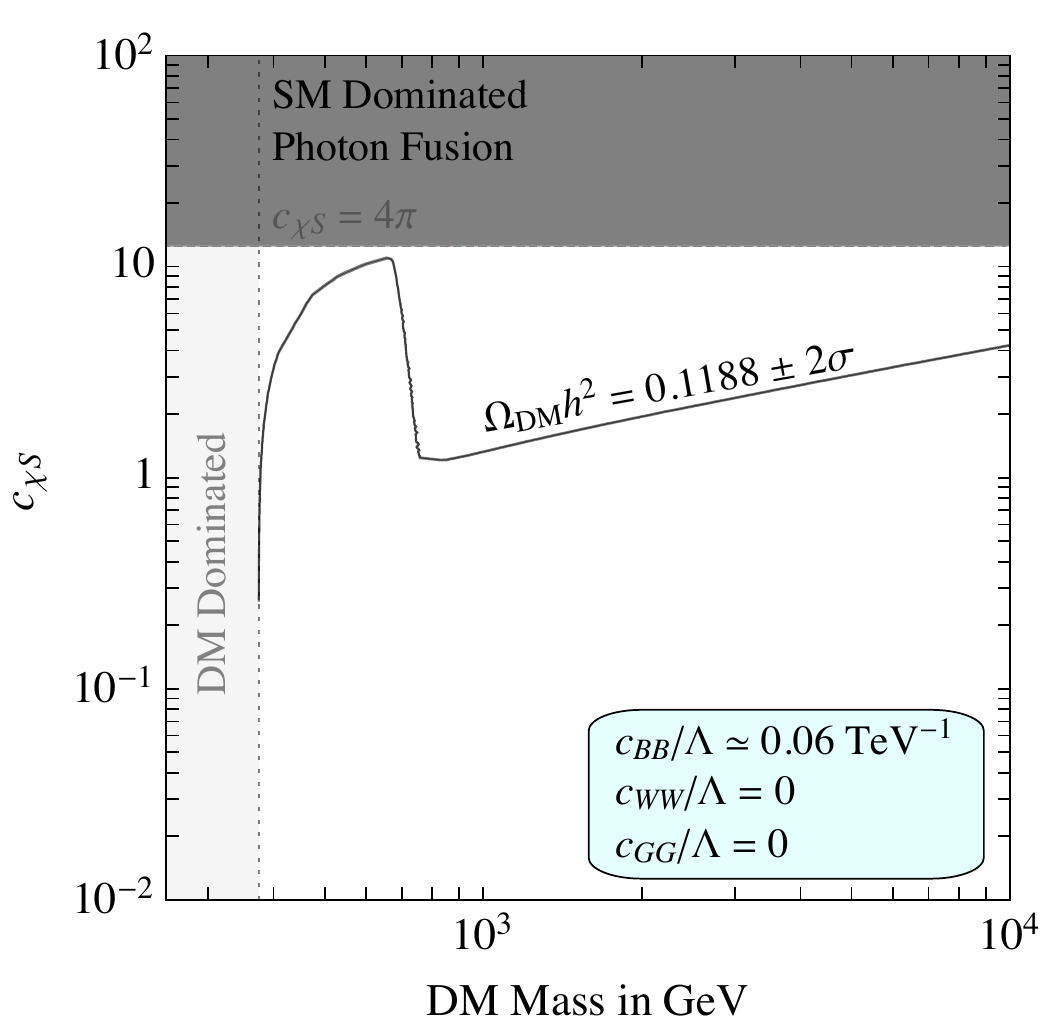}
\includegraphics[width=0.495\textwidth]{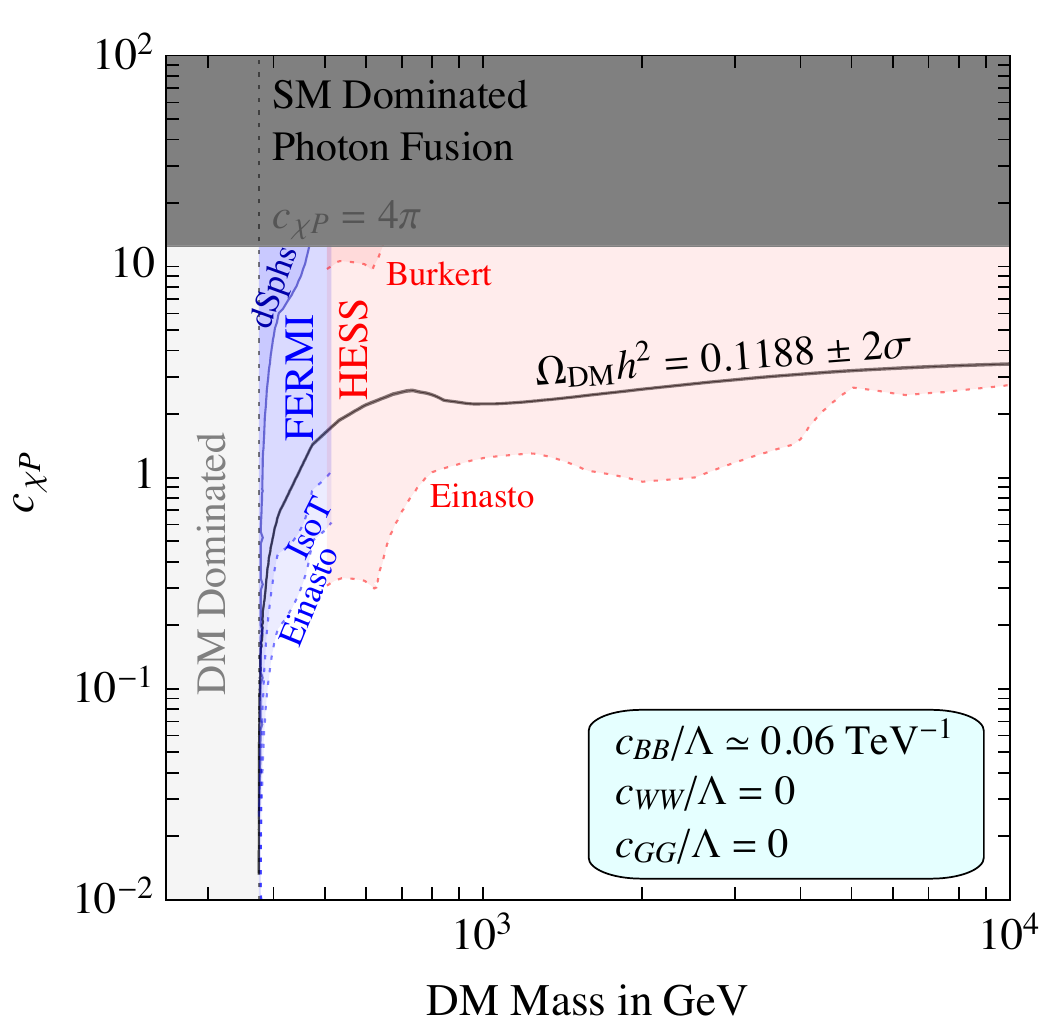}
\caption{DM analysis for scalar (left) and pseudo-scalar (right) mediators in the SM dominated resonance scenario studied in Section~\ref{sec:SMdom}. Here, we consider the photon fusion regime with only the coupling $c_{BB}$ switched on. We visualize the relic density line in the $(m_\chi, c_{\chi S})$ plane, where the top dark region corresponds to non-perturbative values $g_{\chi S}, g_{\chi P} \gtrsim 4 \pi$ of the coupling to DM and the light gray region on the left is the DM dominated scenario not considered here. Shaded regions are excluded by the DM searches relevant for each case. For \FERMI\ limits, solid and dotted lines are for bounds coming from diffuse photons and lines, respectively.}
\label{fig:SMdominatedDM1}
\end{figure}

We start by examining UV completions yielding only the coupling $c_{BB}$ to the hypercharge gauge boson. Our results are shown in \Fig{fig:SMdominatedDM1}. In the scalar mediator case (left), current and projected direct searches are not capable of probing the thermal relic region. In fact, they cannot probe any point of the region where the coupling $c_{\chi S}$ is perturbative, as the radiatively induced couplings to quarks and gluons given in Eqs.~(\ref{eq:RGEWFINALa})-(\ref{eq:RGEWFINALd}) are too small. DM production through thermal freeze-out can be analyzed in two different regimes. DM particles with mass in the range $m_S/2 < m_\chi < m_S$ can only annihilate into SM fields. For DM masses away from the resonance this requires rather large couplings to the scalar portal, almost up to $c_{\chi S} = 4\pi$ for $m_\chi \simeq m_S$, as a consequence of the p-wave suppression. Annihilations to mediators through the process $\chi\chi\rightarrow SS$ open up for DM mass values above $m_S$. The required value for $c_{\chi S}$ suddenly drops above this threshold, and it increases again for larger DM masses. We mention the tantalizing possibility of probing the scalar portal in the photon fusion regime through ID via the process $\chi \chi \rightarrow S S \rightarrow 4 \gamma$. In this case, the photons are distributed in a box centered around $m_S$, and for DM masses not too much larger than $m_S$ they exhibit spectral features similar to the case of a line (see e.g.~\cite{Ibarra:2015tya} for a dedicated study of $\gamma$-ray boxes with the forthcoming Cherenkov Telescope Array). We leave this direction for future work.

Results for the pseudo-scalar case are shown in the right panel of \Fig{fig:SMdominatedDM1}. We observe a similar feature in the relic density line, although far less pronunced. The drop in $c_{\chi P}$ is much smaller because annihilation into SM fields is an s-wave process. This also implies that ID limits are very stringent. In the lower DM region ($m_\chi\lesssim 500 \, {\rm GeV}$) \FERMI\ rules out thermal relics, whereas the \HESS\ line limits are excluding the thermal region only for the choice of an Einasto density profile. 

A potentially interesting intermediate case is photon fusion at the \LHC\ but with both couplings $c_{BB}$ and $c_{WW}$ present. As shown in Figs.~\ref{fig:Narrow_Resonance1} and \ref{fig:Narrow_Resonance2}, the most we can push is for $c_{WW} \simeq c_{BB}$ with a small region where $c_{WW}$ can be a few times larger than $c_{BB}$ such that the scalar width is relatively broad. The conclusions for dark matter phenomenology are pretty similar to the case in \Fig{fig:SMdominatedDM1}. The only differences are that the relic line will move towards lower values of $c_{\chi S}$ and $c_{\chi P}$. For the pseudo-scalar case, limits from lines can be softened as a consequence of the continuum $\gamma$-ray contamination from $c_{WW}$. These cases only form a small part of the allowed parameter space (see Figs.~\ref{fig:Narrow_Resonance1} and \ref{fig:Narrow_Resonance2}), and since we cannot push $c_{WW}$ much above $c_{BB}$ without running into conflicts with $Z\gamma$ limits, we do not further discuss this case.

\begin{figure}[!t]
\centering
\includegraphics[width=0.495\textwidth]{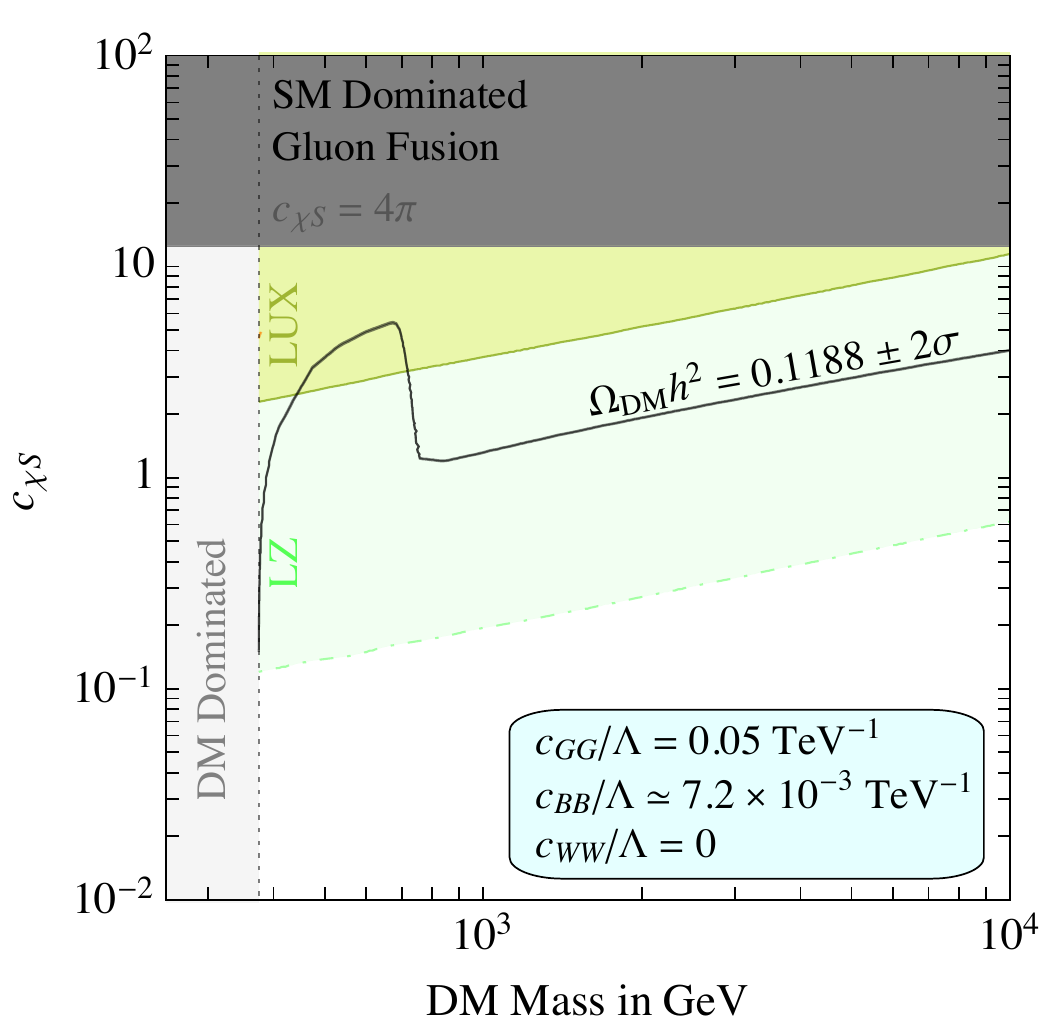}
\includegraphics[width=0.495\textwidth]{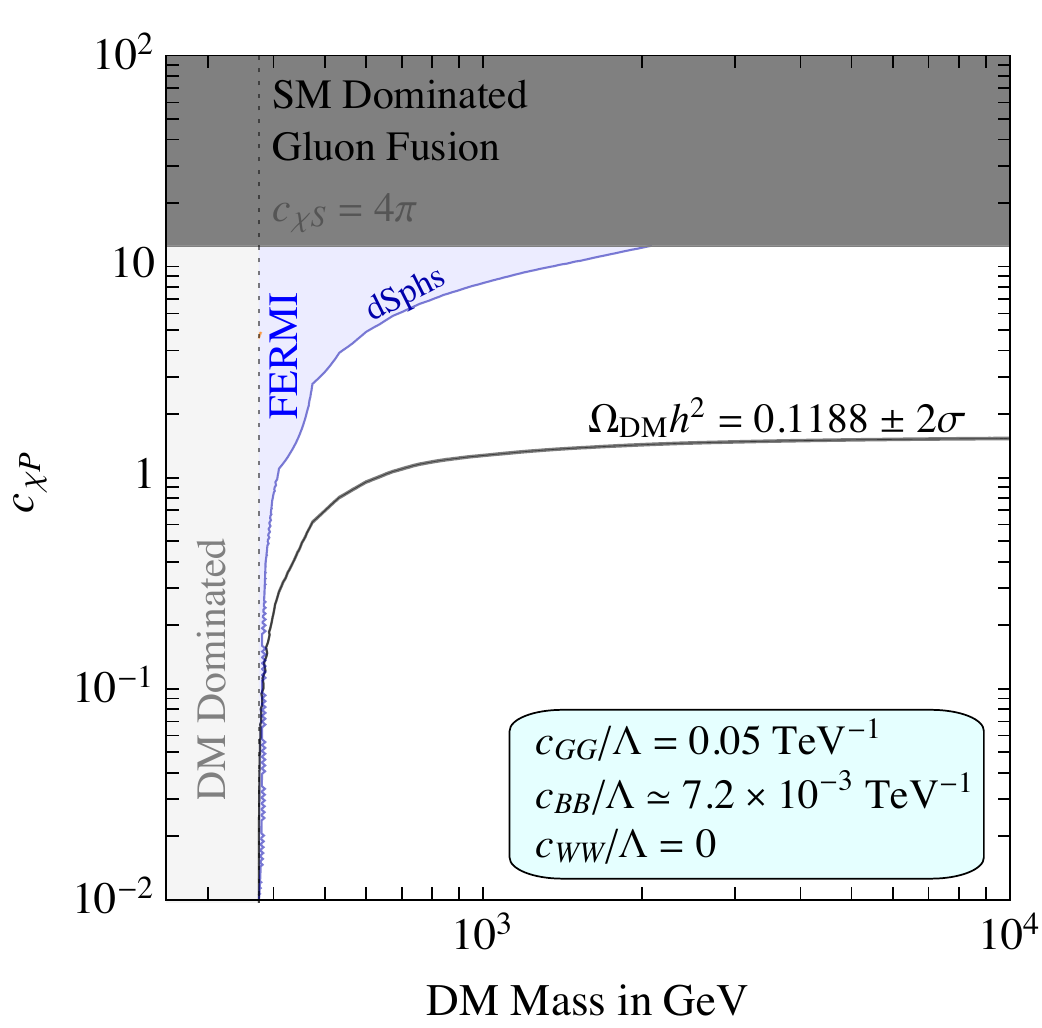}
\caption{DM analysis in the gluon fusion regime. The notation is the same as in \Fig{fig:SMdominatedDM1}.}
\label{fig:SMdominatedDM2}
\end{figure}

We now consider models where the (pseudo-)scalar couples to gluons and \LHC\ productions are dominated by gluon fusion. Unlike the previous case, the couplings are not univocally determined, as gluon fusion dominates and di-jet constraints are satisfied in the whole range $0.03\,\mbox{TeV}^{-1} \lesssim c_{GG}/\Lambda \lesssim 0.07\,\mbox{TeV}^{-1}$ . We choose the value at the center of this allowed range and we show results for this scenario in Fig.~\ref{fig:SMdominatedDM2}. The couplings to gluons for a scalar mediator is responsible for quite large direct detection rates. As an example, the RG analysis in Section~(\ref{sec:DD1}) yields a direct detection cross section $\sigma_{\rm SI}\simeq c^2_{\chi S}\times2.2 \cdot 10^{-46}\ \rm cm^2 $ for $m_\chi = 1 \, {\rm TeV}$ and $c_{GG}/ \Lambda \simeq 0.03 \, {\rm TeV}^{-1} $. Limits from mono-jet events are not relevant in this DM mass range (see \Eq{eq:jMET2}). The thermal relic line for $m_\chi < m_S$ is almost completely excluded by \LUX, except for DM masses extremely close to $m_S/2$. Similar to the photon fusion case, for $m_\chi > m_S$ the required value of $c_{\chi S}$ suddenly drops and a thermal relic is consistent with \LUX\ bounds. However, the entire parameter space will be deeply probed by \LZ. Although the results in \Fig{fig:SMdominatedDM2} are presented for a fixed value of $c_{GG}/\Lambda$, it is straightforward to rescale the results. DD bounds scale linearly with $c_{GG}/\Lambda$. This is true also for the the relic line but only for $m_\chi < m_S$, since for larger DM masses annihilation to mediators dominate and the line is effectively independent on $c_{GG}/\Lambda$.

Unlike the photon fusion case discussed above, a thermal relic with pseudo-scalar mediator (right panel of \Fig{fig:SMdominatedDM2}) is less constrained in the gluon fusion regime. Limits from $\gamma$-ray lines searches are not applicable in this case, since the annihilation cross section in gluons (i.e.~the one responsible for a continuum spectrum of photons) is  up to $200$ times bigger than the one in lines. \FERMI\ limits from $\gamma$-ray continuum  are of course still valid, but they exclude regions way above the thermal relic line. As for the scalar case, mono-jet searches do not put bounds in this DM mass range.  In this case, the relic line is very smooth and the drop around $m_\chi = m_P$ is not visible.

\subsection{DM Dominated Resonance}

The other half of the parameter space corresponds to DM masses below $m_S / 2$, yielding the DM dominated resonance scenario discussed in Section~\ref{sec:DMdom}. We use again the output of our \LHC\ study to identify interesting classes of DM models. 

\begin{figure}[!t]
\centering
\includegraphics[width=0.495\textwidth]{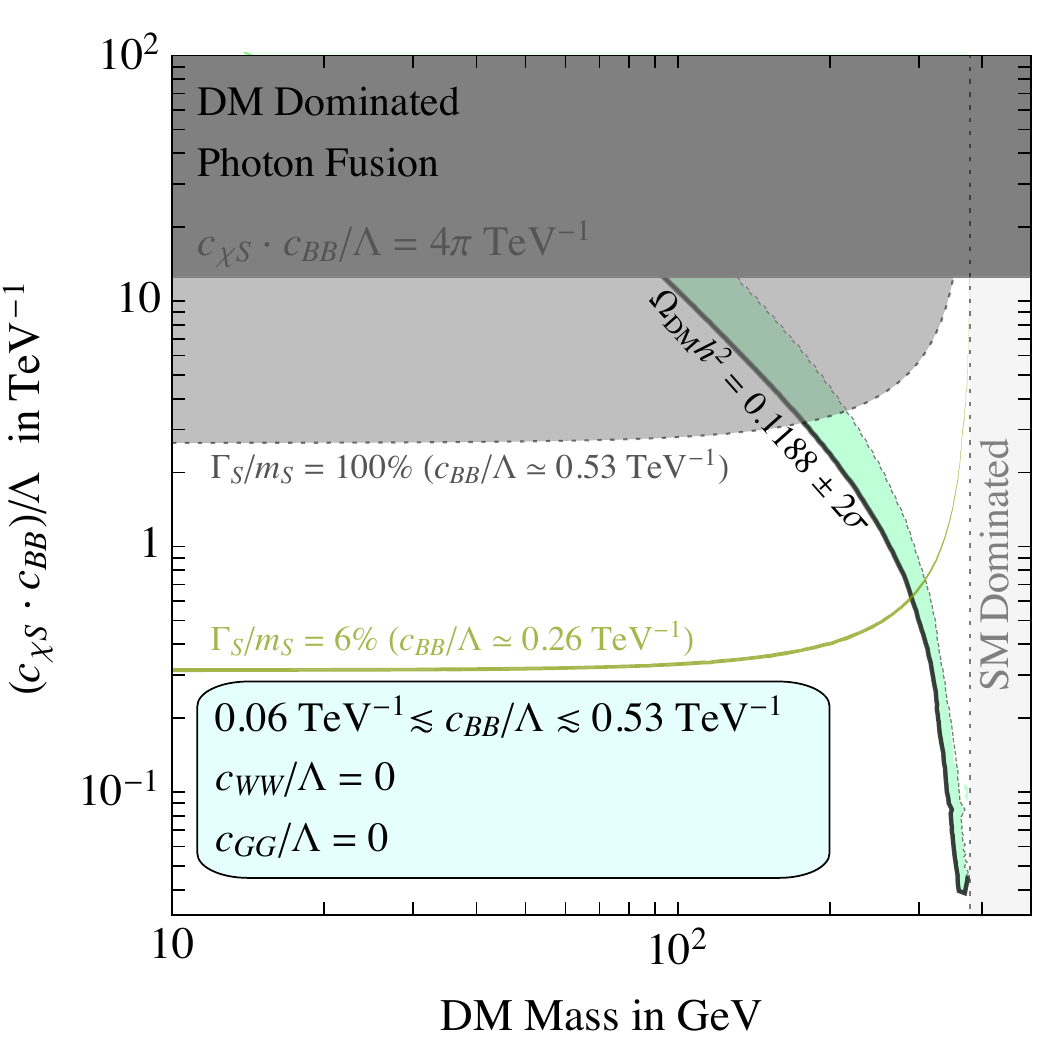}
\includegraphics[width=0.495\textwidth]{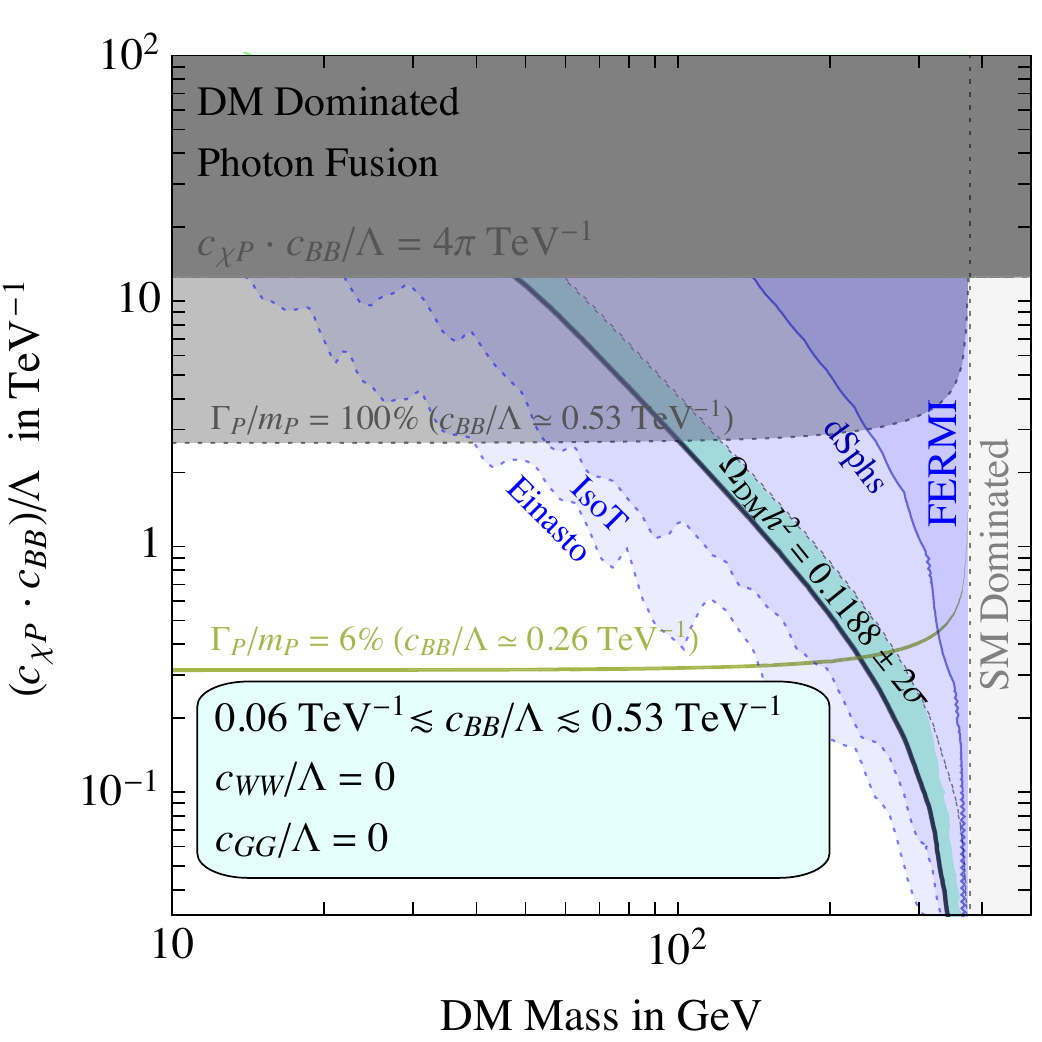}
\caption{DM analysis in the DM dominated resonance scenario at the \LHC\ discussed in Section~\ref{sec:DMdom} for scalar (left) and pseudo-scalar (right) mediators. In this figure we consider the photon fusion regime, with Wilson coefficients $c_{BB}$ in the range shown. We present results in the $(m_\chi, c_{\chi \{S,P\}}  c_{BB}/ \Lambda)$ plane, since rates at DM searches only depend on these two quantities. This is not exact for the thermal line due to resonance effects, and the green band around the thermal relic line quantifies how much this rescaling is violated. We shade away the region where the combination on the vertical axis is above $4 \pi \ {\rm TeV}^{-1}$, and shade with gray where $\Gamma_S \gtrsim m_S$. The light gray region on the right of each figure corresponds to the SM dominated scenario already discussed. We show where thermal freeze-out can reproduce the observed DM density and shade regions excluded by DM searches. For \FERMI\ limits, solid and dotted lines are for bounds coming from diffuse photons and lines, respectively. We identify the line reproducing the ATLAS preferred value for the total width $\Gamma_S \simeq 45 \, {\rm GeV}$. See text for further discussion.}
\label{fig:DMdominatedDM1}
\end{figure}

We now examine the DM phenomenology in the photon fusion regime. Results are shown in \Fig{fig:DMdominatedDM1} for the case where only the coupling $c_{BB}$ is switched on. We present our results in a slightly different way here, putting the combination $c_{\chi \{S,P\}} c_{BB} / \Lambda$ on the vertical axis. Any rate for current and future experiments only depends on this combination, and therefore the same holds for the exclusion regions in the figures. However, it is less obvious that resonance effects on the relic density calculation~\cite{Griest:1990kh} have a small impact, since the resonance is quite broad with a total width $\Gamma_S / m_S \simeq c_{\chi S}^2 / (8 \pi)$, as follows from \Eq{eq:GammaSDMdom}. We address this issue in each case and show that this effect is very small, hence not a concern for our final results.

We start the discussion from the left panel of \Fig{fig:DMdominatedDM1}, corresponding to a scalar mediator in the photon fusion regime. DM searches are powerless for this case\footnote{To give an idea, for $30 \rm \ GeV$ DM and $c_{BB} / \Lambda \simeq 0.26 \, {\rm TeV}^{-1}$ we find a DM-nucleon cross section of $\sigma_{\rm SI}=  6.8 \, c^2_{\chi S} \times 10^{-53}\ \rm cm^2 $, well below the expected LZ sensitivity for $c_{\chi S}<4\pi$. As can be seen from Fig.~\ref{fig:DMdominated2}, in principle $c_{WW}=c_{BB} / \Lambda \simeq 0.26 \, {\rm TeV}^{-1}$ is not excluded and in this case LZ could probe $c_{\chi S}$ values of $\mathcal O(10)$. Clearly, the chances of DD in the photon fusion regime are extremely slim.}.
The solid relic density  line is explicitly obtained for $\Gamma_S \simeq 45 \, {\rm GeV}$. As shown in the top-left panels of Figs.~\ref{fig:DMdominated1} and \ref{fig:DMdominated2}, this corresponds to $c_{BB} / \Lambda \simeq 0.26 \, {\rm TeV}$, and this allows us to identify the iso-contour $\Gamma_S \simeq 45 \, {\rm GeV}$ in the $(m_\chi, c_{\chi S}  c_{BB}/ \Lambda)$ plane of \Fig{fig:DMdominatedDM1}. A thermal relic consistent with $\Gamma_S \simeq 45 \, {\rm GeV}$ then requires $c_{\chi S} \simeq  2.42$. 
If resonance effects are negligible, the relic density line is a universal function of $c_{\chi S} c_{BB} / \Lambda$ and we have explicitly checked that this rescaling invariance works perfectly for lower values of $c_{\chi S}$. We expect it to break down for large enough $c_{\chi S}$, and we estimate the error we could make with this rescaling by computing self-consistently the relic density line for a thermal relic with $\Gamma_S \simeq m_S$. As can be seen from Figs.~\ref{fig:DMdominated1} and \ref{fig:DMdominated2}, this corresponds to a larger coupling $c_{BB} / \Lambda \simeq 0.53 \, {\rm TeV}$, and a thermal relic would then require $c_{\chi S} \simeq 6.77$. The net result on the relic density is a combination of two effects: a large overall coupling in the cross section and a broader width of the mediator. The green bands in the figure show that these combined effects are rather mild. Given the lack of constraints from DM searches, a scalar portal in the photon fusion regime leads again to a viable DM candidate. To summarize this discussion we present here the values of the parameters consistent with a thermal relic and a scalar width of $45$ GeV: 
\begin{equation}\label{DMphotonScalar}
m_\chi \simeq 289\,\mbox{GeV}\ ,\qquad c_{\chi S}\simeq 2.42\ ,\qquad c_{BB}/\Lambda \simeq 0.26\, \mbox{TeV}^{-1}\ ,\qquad c_{WW} = c_{GG}=0\ .
\end{equation}
The \LHC\ analysis allows for values of $c_{WW} \lesssim c_{BB}$, but turning on this coupling does not greatly impact the obtained DM parameters $m_\chi$ and $c_{\chi S}$. 

The pseudo-scalar case is shown in the right panel of \Fig{fig:DMdominatedDM1} with identical conventions. We again give the values of the parameters for a thermal relic and a $45$ GeV width:
\begin{equation}\label{DMphotonPScalar}
m_\chi \simeq 227\,\mbox{GeV}\ ,\qquad c_{\chi P}\simeq1.38\ ,\qquad c_{BB}/\Lambda \simeq 0.26 \, \mbox{TeV}^{-1}\ ,\qquad c_{WW} = c_{GG}=0\ .
\end{equation}
However, in contrast to the scalar case, ID limits are now quite severe, and the $\gamma$-ray line bounds completely rule out a thermal relic even for a cored DM profile like the isothermal one. 

\begin{figure}[!t]
\centering
\includegraphics[width=0.495\textwidth]{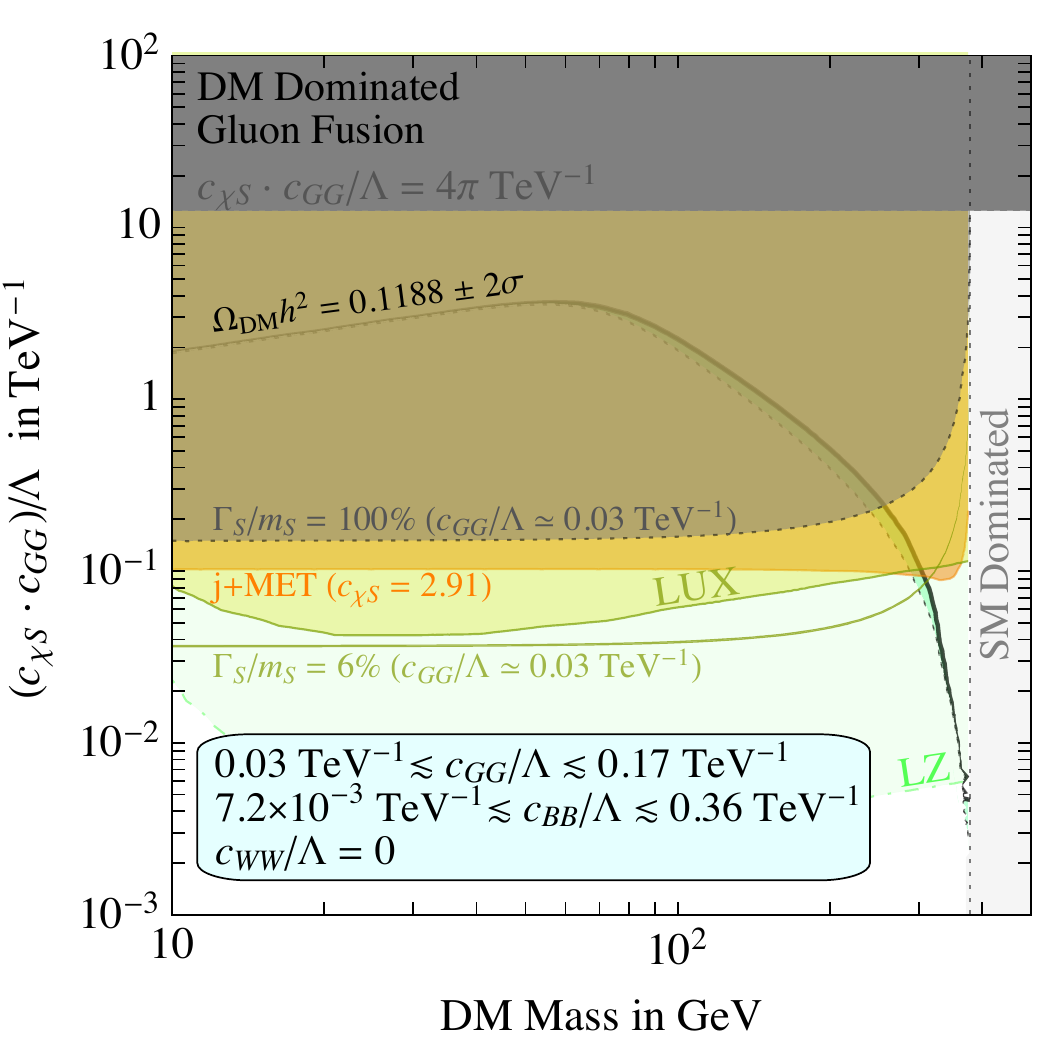}
\includegraphics[width=0.495\textwidth]{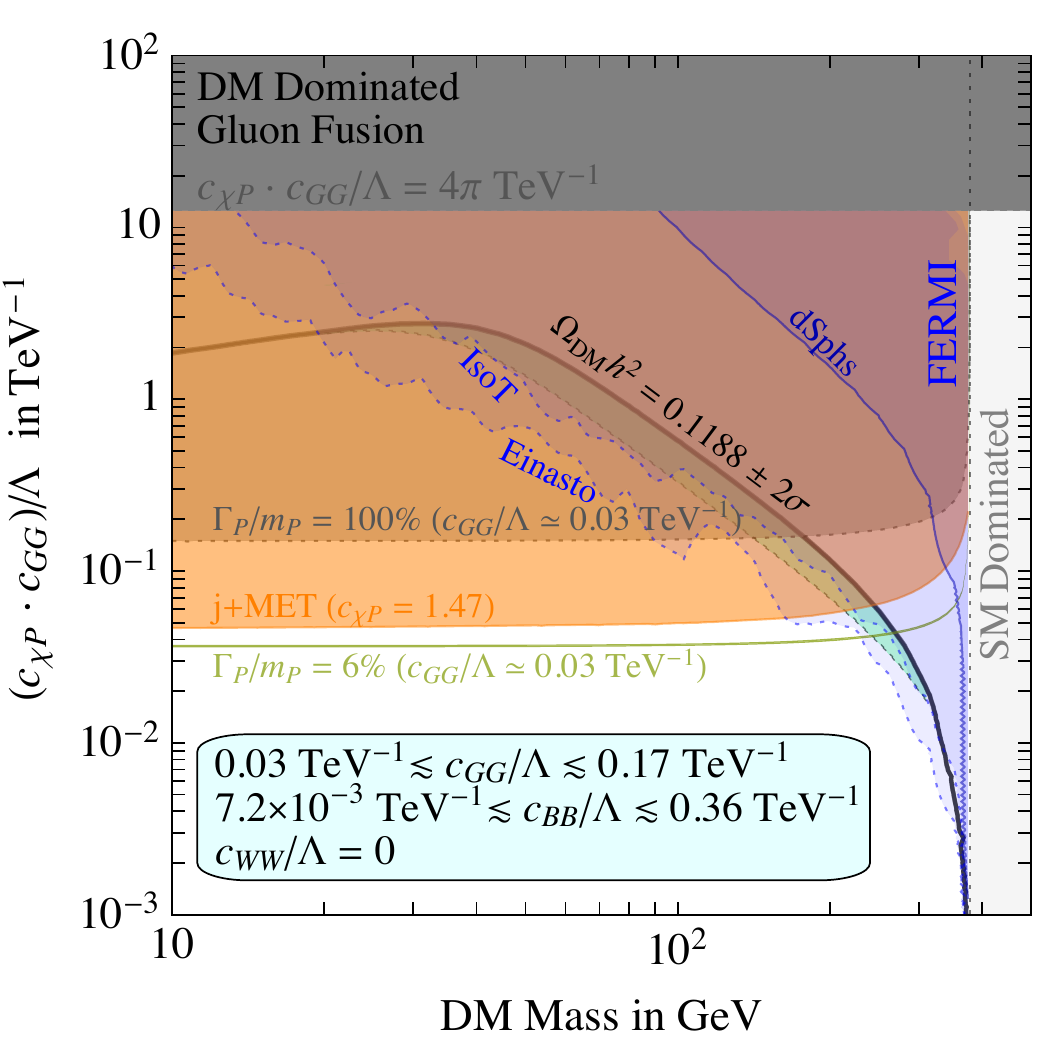}
\caption{Same as \Fig{fig:DMdominatedDM1} but for the gluon fusion case. The indirect detection constraints in the right panel are given for $c_{BB}/\Lambda\simeq 0.01\,\rm TeV^{-1}$.}
\label{fig:DMdominatedDM2}
\end{figure}

Finally, in \Fig{fig:DMdominatedDM2} we consider the gluon fusion regime for a DM dominated resonance. In both panels, $c_{BB} / \Lambda$ and $c_{GG} / \Lambda$ are understood to be within the range written in the label, consistently with what we found in our \LHC\ study for the gluon fusion regime (see Figs.~\ref{fig:DMdominated1} and \ref{fig:DMdominated2}). We present our results in terms of the combination of couplings $c_{\chi S} c_{GG} / \Lambda$. DD cross sections only depend on this combination. The same is true for the relic density line, which we wish to draw again for $\Gamma_S \simeq 45 \ {\rm GeV}$. However, this choice does not identify the value of couplings to SM gauge bosons because the \LHC\ analysis only fixes the product $c_{GG} c_{BB}$, as can be seen from the top-left plot of Figs.~\ref{fig:DMdominated1} and \ref{fig:DMdominated2} or directly in Eq.~\eqref{eq:GfusionScen2}. We therefore take the smallest value of the gluon coupling which is still inside the gluon-dominated regime, $c_{GG} / \Lambda \simeq 0.03 \, {\rm TeV}^{-1}$, which then yields $c_{BB} / \Lambda \simeq0.01 \, {\rm TeV}^{-1}$. We see that \LUX\ constraints are already very close for this gluon coupling and larger values (the \LHC\ di-jet limit is $c_{GG} / \Lambda < 0.17 \, {\rm TeV}^{-1}$) are in conflict with the bounds. Alternatively, keeping $c_{GG} / \Lambda \simeq 0.03 \, {\rm TeV}^{-1}$ fixed, we see that $\Gamma_S/m_S$ cannot be much larger than $6\%$. Lastly, mono-jet searches only put constraints on the Wilson coefficient $c_{GG} / \Lambda$ in this regime (see \Eq{eq:jMET1}). We show collider limits in this plane by choosing the value $c_{\chi S} = 2.91$, giving a $\Gamma_S = 45 \, {\rm GeV}$ width for a thermal relic and $c_{GG} / \Lambda \simeq 0.03 \, {\rm TeV}^{-1}$. Collider bounds are superseded by \LUX\ at small masses, and they become relevant near the resonance. We conclude by giving explicit parameters for a benchmark point not excluded by \LUX\ and \LHC, consistent with a thermal relic and yielding the ATLAS preferred width:
\begin{equation}\label{DMgluonScalar}
m_\chi = 310\,\mbox{GeV}\ ,\,\, c_{\chi S}=2.91\ ,\,\, c_{GG}/\Lambda = 0.03 \,\mbox{TeV}^{-1}\ ,\,\,  c_{BB}/\Lambda = 0.01 \, \mbox{TeV}^{-1}\ ,\,\, c_{WW} = 0\ .
\end{equation}
These values are in excellent agreement with those found in Ref.~\cite{Backovic:2015fnp} and give a spin-independent DM-nucleon cross section $\sigma_{\rm SI}\simeq 1.88 \cdot 10^{-45}\ \rm cm^2$ that can be probed in the near future by the \LUX\ experiment (the current bound is  $2.6 \cdot 10^{-45}\ \rm cm^2$ for $m_\chi\simeq 310$ GeV~\cite{Akerib:2015rjg}). They could also yield a mono-jet signal at the \LHC\ Run 2. As before, nonzero values of $c_{WW}\lesssim c_{BB}$ are not excluded but do not greatly impact the DM phenomenology. 

The right panel shows the pseudo-scalar case. In this scenario, the ID limits from $\gamma$-ray lines depend on the specific value of $c_{BB}$ which is not fixed unlike the analysis in \Fig{fig:SMdominatedDM2}. We present our limits for $c_{BB}/\Lambda = 0.01 \, \mbox{TeV}^{-1}$, which reproduces the ATLAS preferred width if one chooses the smallest allowed coupling to gluons in the gluon-fusion regime. The values of the parameters for a thermal relic and a $45$ GeV width read:
\begin{equation}\label{DMgluonPScalar}
m_\chi \simeq 268\,\mbox{GeV}\ ,\,\, c_{\chi P}\simeq1.47\ ,\,\, c_{GG} / \Lambda \simeq 0.03 \, {\rm TeV}^{-1}\ , \,\, c_{BB}/\Lambda \simeq 0.01 \, \mbox{TeV}^{-1}\ , \,\, c_{WW} = 0\ .
\end{equation}
As one can see from \Fig{fig:DMdominatedDM2}, ID limits are very stringent and the $\gamma$-ray line bounds rule out a thermal relic. However, the \LHC\ analysis only fixes the value of $c_{GG} c_{BB}$ in this gluon fusion regime, unlike the photon-fusion case with parameters for DM given in Eq.~\eqref{DMphotonPScalar}. Thus we can choose a larger $c_{GG}$ and a smaller $c_{BB}$, which of course makes the \FERMI\ $\gamma$-ray line bounds less stringent. On the other hand, a larger value of $c_{GG}$ will also move up the $45$ GeV width line in the $(m_\chi, c_{\chi P} c_{GG}/\Lambda)$ plane by the same factor as the $\gamma$-ray line constraints, crossing the relic density line for smaller DM mass. We explicitly checked that a $45$ GeV width can be obtained while evading the \FERMI\ bounds for a DM mass of roughly 220 GeV and coupling to gluons $c_{GG}/\Lambda\simeq 0.05\,\mbox{TeV}^{-1}$. Here is where mono-jet bounds come into play. In analogy to what we have done for the scalar case, we present j+MET limits for $c_{\chi P} \simeq 1.47$ as in the benchmark point of \Eq{DMgluonPScalar}. We observe a different shape of the mono-jet bound as compared to the scalar case, consequence of the fact that the width of the resonance has a different dependence on the DM mass for scalar and pseudo-scalar. In particular, the j+MET limits are less stringent around the resonance for the pseudo-scalar case, because for fixed DM mass $m_\chi \simeq 375 \, {\rm GeV}$ and same couplings the decay width of the pseudo-scalar is typically larger. Nevertheless, this limits how much we can increase the coupling to gluons and therefore it makes a DM candidate with pseudo-scalar portal in the gluon fusion regime quite unlikely.

%%%%%%%%%%%%%%%%%%%%%%%%%%%%%%%%%%%%%%%%%%%%%%%%
%%%%%%%%%%%%%%%%%% OUTLOOK %%%%%%%%%%%%%%%%%%%%%%%%
%%%%%%%%%%%%%%%%%%%%%%%%%%%%%%%%%%%%%%%%%%%%%%%%

\section{Outlook}
\label{sec:con}

In this paper we have studied the minimal EFT for the diphoton events recently observed at the \LHC\ and DM. The field content is the same as the SM one with the addition of two gauge singlets, a (pseudo-)scalar and a Dirac fermion. We coupled the two singlets via a portal Yukawa interaction, and we also coupled the (pseudo-)scalar to SM gauge bosons via dimension 5 contact interactions. Due to observed decays to two photons, a coupling of the (pseudo-)scalar to electroweak gauge bosons is mandatory. On the contrary, the coupling to gluons is optional, as the new scalar can be produced through photon fusion in proton-proton collisions. 

The \LHC\ phenomenology turns out to be identical for scalar and pseudo-scalar, and we presented a study valid in both cases in Section~\ref{sec:LHCscenarios}. The knowledge of the resonance mass splits in two the possible values of the DM mass, according to whether invisible decays are kinematically accessible. We dubbed these two scenarios SM and DM dominated resonance, corresponding to DM masses that make invisible decays forbidden and allowed, respectively. Despite the six free EFT parameters (after fixing the resonance mass to 750 GeV), the parameter space region consistent with both the diphoton excess and bounds from \LHC\ Run 1 are compactly summarized in Figs.~\ref{fig:Narrow_Resonance1}-\ref{fig:DMdominated2}. Remarkably, the Wilson coefficients are quite constrained and either gluon or photon fusion dominates the total production, unless we choose very specific ratios of the couplings. In the SM dominated scenarios we typically find a very small width for the new resonance. We have not attempted to construct explicit UV completions realizing the parameter space configuration identified by our analysis, consistently with the EFT spirit of this work. Explicit models in the gluon fusion regime have been studied in Refs.~\cite{Harigaya:2015ezk,Angelescu:2015uiz,Nakai:2015ptz,Knapen:2015dap,Buttazzo:2015txu,Franceschini:2015kwy,DiChiara:2015vdm,McDermott:2015sck,Ellis:2015oso,Low:2015qep,Bellazzini:2015nxw,Gupta:2015zzs,Patel:2015ulo,Craig:2015lra,Hall:2015xds,Goertz:2015nkp,Jiang:2015oms,Bardhan:2015hcr}, and we think it would be very interesting to find some explicit realization of the photon fusion regime as well. Considering the large coefficients, the photon fusion scenario probably requires a strongly-coupled UV completion, see for instance Refs.~\cite{Fichet:2015vvy,Heckman:2015kqk}. Every sensible UV completion with a cutoff $\Lambda \gtrsim {\rm few \; TeV}$ should return Wilson coefficients at the \LHC\ scale within the bounds identified in Figs.~\ref{fig:Narrow_Resonance1}-\ref{fig:DMdominated2}. Upon specifying a UV complete theory, these bounds can be easily translated into limits on masses and couplings of new particles inducing the dimension~5 operators. It has been shown in specific UV completions (see e.g. Refs~\cite{Franceschini:2015kwy,Falkowski:2015swt}) that (few) new vector-like fermions with TeV scale masses and Yukawa couplings to the resonance of order one can reproduce the signal.

The results presented in Section~\ref{sec:LHCscenarios} have a range of validity beyond DM models. Even in what we call the DM dominated case, our only assumption is the presence of some additional decay channels that does not have to be to neither stable nor cosmically abundant particles. But other than being interesting by itself, it significantly simplified our DM analysis in Section~\ref{sec:DM}. We found it convenient again to split the DM discussion for the two different scenarios of SM and DM dominated resonance. Moreover, the two cases of scalar and pseudo-scalar mediator lead to drastically different DM phenomenology. Our findings can be compactly summarized by the following four classes of DM models:
\begin{itemize}
%%%%%%%%%%%%%%%%%%%%%%%%%%%%%%%%%
\item[$\diamond$] {\bf Scalar with Photon Fusion:} 
DM searches cannot probe this parameter space region. Mono-jet bounds do not apply and ID rates are p-wave suppressed. The only hope would be direct searches, but the RG induced couplings given in Section~\ref{sec:DD2} are below the \LZ\ projected sensitivity. A thermal relic for DM masses above $m_S / 2$ but below $m_S$ can only be attained for DM couplings close to the perturbative limit, while for larger DM masses perturbative values of $c_{\chi S}$ are allowed. On the other hand, a thermal relic is totally viable for DM masses below the resonance, and the ATLAS preferred value $\Gamma_S / m_S \simeq 6 \%$ can be achieved with invisible decays. Results are summarized on the left panels of \Fig{fig:SMdominatedDM1} and \ref{fig:DMdominatedDM1}.
%%%%%%%%%%%%%%%%%%%%%%%%%%%%%%%%%
\item[$\diamond$] {\bf Scalar with Gluon Fusion:} DM annihilations are still p-wave suppressed,  but \LHC\ and DD experiments can put strong constraints. \LUX\ bounds, evaluated through the RG prescription given in Section~\ref{sec:DD1}, are typically stronger than the ones from mono-jet. The only exception is for DM masses right below the resonant value $m_S / 2$, where \LHC\ limits slightly overtakes the ones from DD. In this mass region a thermal relic is consistent with collider and direct searches, and it would give a signal in future experiments. On the other hand, \LUX\ rules out most of the parameter space for masses between $m_S/2$ and $m_S$, while for masses above $m_S$ the parameter space is currently viable. This entire scenario, regardless of the specific value of the DM mass, will be deeply probed by \LZ.  Results are summarized on the left panels of \Fig{fig:SMdominatedDM2} and \ref{fig:DMdominatedDM2}.
%%%%%%%%%%%%%%%%%%%%%%%%%%%%%%%%%
\item[$\diamond$] {\bf Pseudo-scalar with Photon Fusion:} DM annihilations mediated by a pseudo-scalar particles are s-wave processes. ID constraints are the only meaningful ones in this case, since DD rates are very suppressed and mono-jet limits do not apply. For $m_\chi \lesssim 500 \, {\rm GeV}$, \FERMI\ searches for photon lines basically rule out a thermal relic. For larger masses the implications of \HESS\ limits are less obvious as they are quite sensitive to the density profile assumption. Results are summarized on the right panels of \Fig{fig:SMdominatedDM1} and \ref{fig:DMdominatedDM1}.
%%%%%%%%%%%%%%%%%%%%%%%%%%%%%%%%%
\item[$\diamond$] {\bf Pseudo-scalar with Gluon Fusion:} Introducing a pseudo-scalar coupling to gluons has two main effects on DM phenomenology: making mono-jet searches meaningful and contaminating the line signals with consequent weakening of the ID constraints. Neither of these bounds quite gets to freeze-out line for DM masses above the resonance. The situation is rather different for DM masses smaller than $m_S /2$, where the combination of limits from \FERMI\ $\gamma$-ray line searches and \LHC\ mono-jet searches is strong enough to rule out a thermal relic. Results are summarized on the right panels of \Fig{fig:SMdominatedDM2} and \ref{fig:DMdominatedDM2}.
\end{itemize}

If the diphoton excess turns out to be more than just a statistical fluctuation, we may have started a new era of discoveries in particle physics. Among other things, such as being part of a theoretical construct that solves the gauge hierarchy problem, this new particle could be the connector between the SM and the dark sector. Our general EFT analysis identified a broad class of DM models with a 750 GeV (pseudo-)scalar portal consistent with current experimental limits. Although the study of a specific DM theory goes beyond the purpose of this work, our results in Section~\ref{sec:DM} clearly pinpoints preferred models. The most appealing one is presumably the scalar mediator case in the gluon fusion regime, since it could soon give a signal in direct and collider searches. Contrarily, scalar portals in the photon fusion regime are unattainable by all DM experiments. In these cases, as well as pseudo-scalar cases in the gluon fusion regime and for large DM masses where ID limits are not as powerful, the most promising strategy to probe the models is to accumulate more evidence through other \LHC\ channels such as $Z \gamma$ searches. We believe it would be very interesting to further investigate the associated phenomenology of specific UV-complete DM models reproducing our EFT framework at lower energies.

\section*{Acknowledgments} 

We thank Lawrence Hall, Gordan Krnjaic, Marco Nardecchia, Duccio Pappadopulo, Diego Redigolo, Alessandro Strumia, Tim Tait for useful discussions. This work was supported by the U.S. Department of Energy grant number DE-SC0010107 (FD) and the Dutch Organization for Scientific Research (NWO) through a VENI grant (JdV). P.P. acknowledges support of the European Research Council project 267117 hosted by Universit\'e Pierre et Marie Curie-Paris 6, PI J. Silk. 
\appendix

\section{Decay Widths and Cross Sections}
\label{app:DecayAndXS}

In this Appendix we give all the details of the results for decay widths and cross sections for both \LHC\ production and DM annihilation.

\subsection*{Interactions for Mass Eigenstates}

In this first Appendix we express the interactions in \Eq{eq:Lintdim5} in terms of the mass eigenstates, and provide all the squared matrix elements for decays of the scalar $S$ to any possible final state. The results contained here will be the building blocks to easily obtain decay widths and cross sections.

The SM charged EW bosons are obtained by a $\pi/ 4$ rotations among the $SU(2)_L$ gauge bosons
\beq
\left(\begin{array}{c} W_\mu^+ \\ W_\mu^- \end{array} \right) = 
\left( \begin{array}{cc} \frac{1}{\sqrt{2}} & - \frac{i}{\sqrt{2}} \\ 
\frac{1}{\sqrt{2}} & \frac{i}{\sqrt{2}} \end{array} \right)
\left(\begin{array}{c} W_\mu^1 \\ W_\mu^2 \end{array} \right)  \ .
\eeq
The neutral gauge bosons are expressed by a weak mixing angle rotation
\beq
\left(\begin{array}{c} Z_\mu \\ A_\mu \end{array} \right) = 
\left( \begin{array}{cc} c_w & - s_w \\ 
s_w & c_w \end{array} \right)
\left(\begin{array}{c} W_\mu^3 \\ B_\mu \end{array} \right)  \ .
\eeq
The interactions in \Eq{eq:Lintdim5} can be rewritten as a function of the mass eigenstates
\beq
 \mathcal{L}^{\rm d = 5 }_{(\rm int)} =  \frac{S}{\Lambda} \left[ c_{GG} \, G^{A\,\mu\nu} G^A_{\mu\nu} + c_{W^+ W^-} \, 
W^{+\,\mu\nu} W^-_{\mu\nu} + c_{Z Z} \, Z^{\mu\nu} Z_{\mu\nu}  + c_{Z \gamma} \, Z^{\mu\nu} F_{\mu\nu} +  c_{\gamma \gamma} \, F^{\mu\nu} F_{\mu\nu}
\right]  \ ,
\eeq 
with Wilson coefficients connected to the gauge invariant ones as follows
\begin{align}
c_{W^+ W^-} = & \, 2 \, c_{WW}  \ , \\
c_{Z Z} = & \, c_w^2 c_{WW} + s_w^2 c_{BB} \ , \\
c_{Z \gamma} = & \,  2 c_w s_w (c_{WW} - c_{BB})\ , \\
c_{\gamma \gamma} = & \, s_w^2 c_{WW} + c_w^2 c_{BB} \ .
\end{align}
The couplings for the pseudo-scalar case are analogous with just the replacement $c_{XX} \rightarrow \tilde{c}_{XX}$, where $X = \{G, W, B\}$

\begin{figure}
\begin{center}
\includegraphics[scale=0.65]{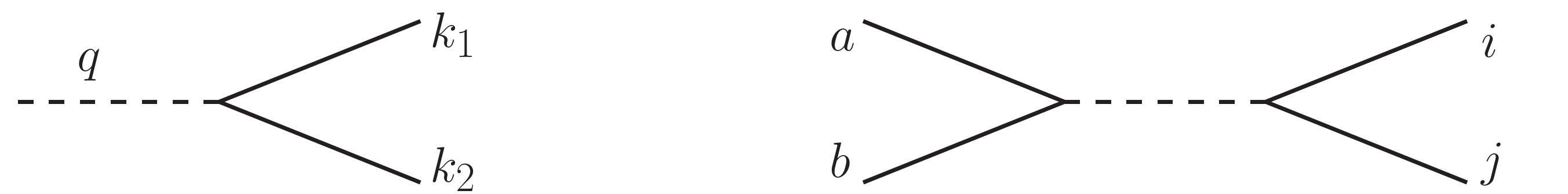} 
\end{center}
\caption{Left: Feynman diagram for the $S$ decay process to a generic two-body final state. In this Appendix we give the squared matrix elements for arbitrary external four momentum $q$ of the scalar, and we define $s = q^2$. The on-shell case corresponds to $s = m_S^2$. Right: Feynman diagrams for arbitrary annihilations $2 \rightarrow 2$.} 
\label{fig:FeynDiag}
\end{figure}

\subsection*{Squared Matrix Elements}

We use the above Lagrangian to compute the squared matrix elements for the decays process $S \rightarrow i j$. The Feynman diagram is shown on the left of \Fig{fig:FeynDiag}. We keep the external scalar state off-shell, and we define its invariant mass to be $q^2 = s$. In the on-shell limit, which we will take for example to compute decay widths, we will have $s = m_S^2$. We also sum over all the possible final polarizations. All the following calculation have been performed by hand and cross-checked with FeynCalc~\cite{Mertig:1990an}. Denoting by $k_1$ and $k_2$ the four momenta of the final state particles, we find the following squared matrix elements for decay processes to SM final states
\begin{align}
\squared{\mathcal{M}_{S \; \rightarrow \; G G}}  =  & \,  256 \, \frac{c^2_{GG} \left( k_1 \cdot k_2 \right)^2}{\Lambda^2}
= 64 \, \frac{c^2_{GG}\, s^2}{\Lambda^2} \ , \\
\squared{\mathcal{M}_{S \; \rightarrow \; W^+ W^-}}  = & \, 
4 \, \frac{c^2_{W^+ W^-}}{\Lambda^2} \left[ 2 \left(k_1 \cdot k_2\right)^2 + m_W^4 \right] =
2 \frac{c^2_{W^+ W^-} \, s^2}{\Lambda^2} \left(1  - \frac{4 m_W^2}{s} + \frac{6 m_W^4}{s^2} \right) \ , \\
\squared{\mathcal{M}_{S \; \rightarrow \; Z Z}}  = & \, 16 \, \frac{c^2_{ZZ}}{\Lambda^2} \left[ 2 \left(k_1 \cdot k_2\right)^2 + m_Z^4 \right] = 8 \frac{c^2_{ZZ} \, s^2}{\Lambda^2} \left(1  - \frac{4 m_Z^2}{s} + \frac{6 m_Z^4}{s^2} \right)  \ , \\
\squared{\mathcal{M}_{S \; \rightarrow \; Z \gamma}}  =  & \, 8 \frac{c^2_{Z \gamma}}{\Lambda^2} \left( k_1 \cdot k_2 \right)^2 = 2 \frac{c^2_{Z \gamma} \, s^2}{\Lambda^2} \left(1 - \frac{m_Z^2}{s} \right)^2 \ , \\
\squared{\mathcal{M}_{S \; \rightarrow \; \gamma\gamma}}  = & \, 32 \, \frac{c^2_{\gamma\gamma} \left( k_1 \cdot k_2 \right)^2}{\Lambda^2}
= 8 \, \frac{c^2_{\gamma\gamma}\, s^2}{\Lambda^2}  \ .
\end{align}
Likewise, we can evaluate the squared matrix element for decay to DM pairs
\beq
\squared{\mathcal{M}_{S \; \rightarrow \; \chi\chi}} = 4 \, c_{\chi S}^2 \, \left(k_1 \cdot k_2 - m_\chi^2 \right) = 
2 \, c_{\chi S}^2 \, s \left(1 - \frac{4 m_\chi^2}{s} \right)   \ .
\label{eq:M2Schichi}
\eeq

We switch to the case of a pseudo-scalar, with matrix elements for decays to SM states
\begin{align}
\squared{\mathcal{M}_{P \; \rightarrow \; G G}} = & \, 256 \frac{\tilde{c}^2_{GG}}{\Lambda^2} \left( k_1 \cdot k_2 \right)^2 =   64 \frac{\tilde{c}^2_{GG} \, s^2}{\Lambda^2}   \ ,  \\
\squared{\mathcal{M}_{P \; \rightarrow \; W^+ W^-}} = & \,  8 \frac{\tilde{c}^2_{W^+ W^-}}{\Lambda^2} \left[ (k_1 \cdot k_2)^2 - m_W^4 \right] = 2 \frac{\tilde{c}^2_{W^+ W^-}\,s^2}{\Lambda^2} \left(1 - \frac{4 m_W^2}{s} \right) \ , \\
\squared{\mathcal{M}_{P \; \rightarrow \; Z Z}} = & \,  64 \frac{\tilde{c}^2_{ZZ}}{\Lambda^2} \left[ (k_1 \cdot k_2)^2 - m_Z^4 \right] = 8 \frac{\tilde{c}^2_{ZZ} \, s^2}{\Lambda^2} \left(1 - \frac{4 m_Z^2}{s} \right) \ , \\
\squared{\mathcal{M}_{P \; \rightarrow \; Z \gamma}}  = & \, 8 \frac{\tilde{c}^2_{Z \gamma}}{\Lambda^2} (k_1 \cdot k_2)^2  = 
2 \frac{\tilde{c}^2_{Z \gamma} \, s^2}{\Lambda^2} \left( 1 - \frac{m_Z^2}{s} \right)^2 \ , \\
\squared{\mathcal{M}_{P \; \rightarrow \; \gamma\gamma}} = & \, 32 \frac{\tilde{c}^2_{\gamma\gamma}}{\Lambda^2} \left( k_1 \cdot k_2 \right)^2 =   8 \frac{\tilde{c}^2_{\gamma\gamma} \, s^2}{\Lambda^2}   \ .
\end{align}
Finally, the squared matrix element for decay to DM results in
\beq
\squared{\mathcal{M}_{P \; \rightarrow \; \chi\chi}} = 4 \, c_{\chi P}^2 \, \left(k_1 \cdot k_2 + m_\chi^2 \right) = 2 \, c_{\chi P}^2 \,  s    \ .
\label{eq:MPchichi} 
\eeq

\subsection*{Decays Rates}

With the squared matrix elements in hands, it is straightforward to compute the partial decay width for a generic channel. We have the general expression for scalar decays
\beq
\Gamma_{S \rightarrow i j} = s_{ij} \,  \frac{\squared{\mathcal{M}_{S \rightarrow i j} }}{16 \pi \,  m_S}   \sqrt{1- 2 \frac{(m_i^2 + m_j^2)}{m_S^2} + \frac{(m_i^2 - m_j^2)^2}{m_S^4} }  \ ,
\label{eq:dGamma}
\eeq
where the statistical factor $s_{ij}$ accounts for identical particles in the final state. We find
\begin{align}
\Gamma_{S \rightarrow G G} = & \, \frac{2 \, c^2_{GG} \, m_S^3}{\pi \, \Lambda^2} \ , \\ 
\Gamma_{S \rightarrow W^+ W^-} = & \, \frac{c^2_{W^+ W^-} \, m_S^3}{ 8\, \pi \, \Lambda^2} \left(1  - \frac{4 m_W^2}{m_S^2} + \frac{6 m_W^4}{m_S^4} \right) \sqrt{1 - \frac{4 m_W^2}{m_S^2}} \ , \\
\Gamma_{S \rightarrow Z Z} = & \,  \frac{c^2_{Z Z}\, m_S^3}{4 \, \pi \, \Lambda^2}\left(1  - \frac{4 m_Z^2}{m_S^2} + \frac{6 m_Z^4}{m_S^4} \right)   \sqrt{1 - \frac{4 m_Z^2}{m_S^2}} \ , \\
\Gamma_{S \rightarrow Z \gamma} = & \, \frac{c^2_{Z \gamma} \, m_S^3}{8 \, \pi \, \Lambda^2} \left(1 - \frac{m_Z^2}{m_S^2}\right)^3 \ , \\
\Gamma_{S \rightarrow \gamma \gamma} = & \, \frac{c_{\gamma \gamma}^2 \, m_S^3}{ 4\, \pi \, \Lambda^2} \ , \\
\Gamma_{S \rightarrow \chi \chi} = & \,  \frac{c_{\chi S}^2 \, m_S}{8 \pi} \left( 1 - \frac{4 m_\chi^2}{m_S^2} \right)^{3/2}  \ .
\end{align}
The expression for pseudo-scalar decays can be obtained identically. 

\subsection*{\LHC\ Cross Sections}

The total cross section for the process $p p \rightarrow i j$ is obtained from the factorization theorem
\beq
\sigma_{p p \rightarrow i j}(\sqrt{s}) = \int_0^1 dx_a \int_0^1 dx_b \; \left[ f_{a/p}(x_a) f_{b/p}(x_b) + a \leftrightarrow b \; ({\rm if} \; a \neq b) \right] \, \sigma_{a b \rightarrow i j}(\sqrt{x_a x_b \, s}) \ ,
\label{eq:sigmaTOTALdef}
\eeq
where $f_{a/p}$ and $f_{b/p}$ are the $a$ and $b$ parton distribution function (pdf) inside the proton. We introduce a new final state variable $x$, defined as the invariant mass of the $ij$-pair in units of $m_S$. By using energy-momentum conservation we can rewrite this variable as follows 
\beq
x \equiv \frac{m_{ij}}{m_S} = \frac{\sqrt{x_a x_b \, s} }{m_S} \ .
\eeq
The total cross section can written in a compact form
\beq
\sigma_{p p \rightarrow ij}(\sqrt{s}) = \frac{2 m_S^2}{s}  \, \int_{0}^{\frac{\sqrt{s} }{m_S}} dx \; x \; \left[ F_{ab}^{\sqrt{s}}(x) + a \leftrightarrow b \; ({\rm if} \; a \neq b) \right] \; \sigma_{a b \rightarrow i j}(m_S \, x)   \ ,
\label{eq:sxLHCgeneral}
\eeq
where we define the flux function at fixed CM energy for the pp collision
\beq
F_{ab}^{\sqrt{s}}(x)  = \int_{\frac{m_S^2 x^2}{s} }^1 \frac{d X}{X} \; f_{a/p}(X) \; f_{b/p} \left(\frac{m_S^2 x^2}{s X}\right)  \ .
\label{eq:Fdef}
\eeq
The partonic cross section for a $2 \rightarrow 2$ collisions takes the general form
\beq
\sigma_{a b \rightarrow i j}(\sqrt{s}) = s_{ij} \,  \frac{\squared{\mathcal{M}_{a b \rightarrow i j}}}{16 \pi \, s} \,  
\frac{g(m_i^2/s, m_j^2/s)}{g(m_a^2/s, m_b^2/s)} \ ,
\eeq
where as usual $s_{ij}$ accounts for identical particles in the final state and we define the function
\beq
g(x, y) = \sqrt{1 - 2 (x+y) + (x-y)^2} \ .
\eeq

The above equations are general. We specialize now to the case of a s-channel resonance shown in \Fig{fig:FeynDiag}, where the partonic matrix elements always take the form
\beq
\squared{\mathcal{M}_{a b \rightarrow i j} } = \frac{\mathcal{N}_{ab}}{4} \frac{\squared{\mathcal{M}^*_{S \rightarrow a b}} \squared{\mathcal{M}_{S \rightarrow  i j}}}{(s - m_S^2)^2 + m_S^2 \Gamma_S^2} \ ,
\label{eq:generalMggfusion}
\eeq
The expression for a s-channel pseudo-scalar resonance $P$ is identical. Here, the factor of $1/4$ average over the initial polarizations, since every possible initial state has always $2$ polarizations. We also account for a possible average over the color number $\mathcal{N}_{ab}$, as we can have gluons in the initial state. Furthermore, we specialize to the case of only gluons and photons in the initial state, and we write the partonic cross section 
\beq
\sigma_{a b \rightarrow i j}(\sqrt{s}) = 8 \pi\, \mathcal{N}_{ab} \, \frac{\Gamma_{S \rightarrow a b}(s) \, \Gamma_{S \rightarrow i j}(s)}{(s - m_S^2)^2 + m_S^2 \Gamma_S^2} \ .
\label{eq:partonicXS}
\eeq
The partial decay widths in the above equation have to be computed as we would for a scalar particle $S$ of mass $\sqrt{s}$.

The invariant mass of the diphoton pairs observed at the \LHC\ is never far from $m_S$, therefore we can further simplify our expression by employing the narrow width approximation
\beq
\frac{1}{(s - m_S^2)^2 + m_S^2 \Gamma_S^2} \quad \rightarrow \quad \frac{\pi}{m_S \Gamma_S} \delta(s - m_S^2) \ .
\eeq
The $dx$ integral in \Eq{eq:sxLHCgeneral} is straighforward
\beq
\sigma_{p p \rightarrow ij}(\sqrt{s}) = \frac{8 \pi^2}{m_S s} \; \mathcal{N}_{ab} \, C_{ab}^{\sqrt{s}}\, \frac{\Gamma_{S \rightarrow a b} \, \Gamma_{S \rightarrow i j}}{\Gamma_S}  \ ,
\label{eq:sxLHCschannelNarrow2}
\eeq
where we define 
\beq
C_{ab}^{\sqrt{s}}  \equiv  F_{ab}^{\sqrt{s}}( x = 1) = \int_{\frac{m_S^2}{s} }^1 \frac{d X}{X} \; f_{a/p}(X) \; f_{b/p} \left(\frac{m_S^2}{s X}\right)  \ .
\label{eq:Cabdef}
\eeq

\subsection*{DM Annihilation Cross Sections I: SM Final States}

We collect the total cross sections for DM annihilation to SM final states. Considering the process $\chi \chi \rightarrow i j$, the cross section formally reads
\beq
\sigma_{\chi \chi \rightarrow i j}(s) = \frac{s_{ij}}{16 \pi} \, \squared{\mathcal{M}_{\chi \chi \rightarrow  i j} } \, \frac{\sqrt{1- 2 \frac{(m_i^2 + m_j^2)}{s} + \frac{(m_i^2 - m_j^2)^2}{s^2} }}{\sqrt{s} \, \sqrt{s - 4 m_\chi^2} } \ .
\eeq
Here, $\sqrt{s}$ is the energy in the CM frame of the collision and the statistical factor $s_{ij} = 1/2$ for identical particles in the final state. 

The squared matrix element for the collision mediated by a scalar exchanged in the s-channel can be expressed as follows (see \Fig{fig:FeynDiag})
\beq
\squared{\mathcal{M}_{\chi \chi \rightarrow i j} } = \frac{1}{4} \frac{\squared{\mathcal{M}^*_{S \rightarrow \chi \chi}} \squared{\mathcal{M}_{S \rightarrow  i j}}}{(s - m_S^2)^2 + m_S^2 \Gamma_S^2} \ ,
\label{eq:generalM}
\eeq
where the factor $1/4$ averages over the initial DM polarizations. Plugging the squared matrix element for $S \rightarrow i j$ as given in \Eq{eq:M2Schichi}, we find the general expression for the DM annihilation cross section
\beq
\sigma_{\chi \chi \rightarrow i j}(s) = s_{ij} \, \frac{c_{\chi S}^2 }{32 \pi} \, \frac{\squared{\mathcal{M}_{S \rightarrow  i j}}}{(s - m_S^2)^2 + m_S^2 \Gamma_S^2} \, \sqrt{1 - \frac{4 m_\chi^2}{s}} \, \sqrt{1- 2 \frac{(m_i^2 + m_j^2)}{s} + \frac{(m_i^2 - m_j^2)^2}{s^2}}  \ .
\eeq
The result for each channel $ij$ can be found by plugging the squared matrix elements given in this Appendix. Likewise, the expression for processes mediated by a pseudo-scalar results in
\beq
\sigma_{\chi \chi \rightarrow i j}(s) = s_{ij} \, \frac{c_{\chi P}^2 }{32 \pi} \, \frac{\squared{\mathcal{M}_{P \rightarrow  i j}}}{(s - m_P^2)^2 + m_P^2 \Gamma_P^2} \, \frac{\sqrt{1- 2 \frac{(m_i^2 + m_j^2)}{s} + \frac{(m_i^2 - m_j^2)^2}{s^2}}}{\sqrt{1 - \frac{4 m_\chi^2}{s}}}  \ .
\eeq
For the last two equations, we see that annihilations mediated by the scalar and the pseudo-scalar are p- and s-wave processes, respectively.

\subsection*{DM Annihilation Cross Sections II: Mediators Final States}

DM annihilations to mediator particles become kinematically accessible for  $m_\chi \gtrsim 750 \, {\rm GeV}$. These processes are mediated by a virtual DM particle exchanged in both t- and u-channels. We computed the full cross section as a function of the CM energy $\sqrt{s}$ and used them for the relic density calculation. The general expressions are quite involved. In this Appendix we only report the non-relativistic limits for annihilation to scalar and pseudo-scalars
\begin{align}
\sigma_{\chi \chi \rightarrow S S} \, v_{\rm rel} \simeq & \,  \frac{3 \, c_{S\chi}^4}{128 \pi \, m_\chi^2}  \; v_{\rm rel}^2 \;
\left( 1 - \frac{8 \epsilon_S}{9} + \frac{2 \epsilon_S^2}{9} \right) (1 - \epsilon_S)^{1/2} \left(1 - \frac{\epsilon_S}{2} \right)^{-4}  \ , \\
\sigma_{\chi \chi \rightarrow P P} \, v_{\rm rel} \simeq & \,  \frac{c_{P\chi}^4}{384 \pi \, m_\chi^2}  \; v_{\rm rel}^2 \;
(1 - \epsilon_P)^{5/2} \left(1 - \frac{\epsilon_P}{2} \right)^{-4}  \ ,
\end{align}
where
\beq
\epsilon_{S,P} \equiv \frac{m_{S,P}^2}{m_\chi^2} \ .
\eeq
The processes are p-wave suppressed in both cases. These approximate results are quite accurate since we are away from the resonant value $m_\chi \simeq 375 \, {\rm GeV}$, but nevertheless we use the full expressions for our numerical analysis.

\section{RGE: Equations and Solutions}
\label{app:RG}

In this Appendix we give the details of our RG analysis. As done in Section~\ref{sec:RGE}, we divide the discussion into two cases according to whether we have a coupling to gluons at the cutoff scale.

\subsection*{RGE with coupling to gluons at the cutoff}

For non zero values of $c_{GG}(\Lambda)$ and in the renormalization scale range $m_S < \mu < \Lambda$ we limit ourselves to the following 
effective Lagrangian 
\begin{equation}
\mathcal{L}_{\rm EFT}^{m_S < \mu < \Lambda}  = \sum_{q = u, d, s,c,b,t} \frac{c_{yq} \, y_q}{\Lambda} S \, \left(\overline{q}_L H q_R + \mathrm{h.c.} \right) + \frac{c^\prime_{GG} \, \alpha_s}{\Lambda} S \, G^{A\,\mu\nu} G^A_{\mu\nu}\ ,
\end{equation}
with $H$ the SM Higgs doublet and $y_q$ the quark Yukawa couplings.

For this RGE analysis we find it convenient to employ a different normalization for the coupling to gluons
\beq
 c^\prime_{GG}(\mu) \alpha_s(\mu) =  c_{GG}(\mu) \ , 
 \label{eq:cprimevsc}
\eeq
as it yields a simple anomalous dimension matrix. Likewise, factorizing out the Yukawa couplings ensures that the evolution of $c_{yq}$ is the same for every quark.

We define the two dimensional array of Wilson coefficients $\vec{c}(\mu)^T = (c_{yq} (\mu) \,\,\, c_{GG}^\prime(\mu))$ and write the RG equation as
\begin{equation} \label{RGE}
\frac{d\vec c(\mu)}{d\ln \mu} = \gamma(\mu) \,  \vec c(\mu)\,\,\,,
\end{equation}
in terms of the anomalous dimension matrix $\gamma$. As anticipated, the normalization chosen in \Eq{eq:cprimevsc} ensures the simple form 
\begin{equation}
\gamma(\mu) = \bma 0 & \gamma_{Gq}(\mu) \\
			0  & 0  
		        \ema,
\end{equation}
where $\gamma_{Gq}$ describes the mixing from $ c^\prime_{GG}$ into $c_{yq}$. We work in a mass independent renormalization scheme (such as $\overline{\rm MS}$) where the anomalous dimension matrix depends on the renormalization scale $\mu$ only through SM couplings. Throughout this work we use leading-order (LO) QCD evolution such that~\cite{Collins:1976yq}
\begin{equation}
\gamma_{Gq}(\mu) = \frac{\alpha_s^2(\mu)}{4\pi} \gamma_0\ ,
\end{equation}
where $\gamma_0 =32$. 

We always take $c_{yq}(\Lambda)=0$ as our boundary condition. The solution of the RG system in the energy range $m_S < \mu < \Lambda$ allows us to obtain the couplings at the scale $\mu = m_S$. We use the output of this operation in Section~\ref{sec:RunningLambdatomS} to justify how such a running has a negligible impact on \LHC\ phenomenology. For this reason, we neglect this running and start our actual RG analysis at the scale $\mu = m_S$, with the coupling $c_{yq}$ still set to vanish. At this scale, we integrate out the scalar resonance and obtain the effective Lagrangian
\begin{equation}
\mathcal{L}_{\rm EFT}^{m_t < \mu < m_S} = \sum_{q = u, d, s,c,b,t} \mathcal{C}_{yq} y_q \, \bar \chi \chi \, \left(\overline{q}_L H q_R + \mathrm{h.c.} \right) +  \mathcal{C^\prime}_{GG}\alpha_s \, \bar \chi \chi\, G^{A\,\mu\nu} G^A_{\mu\nu}\ ,
\end{equation}
with boundary conditions
\begin{align}
\label{LambdaTomS} \mathcal C_{yq} (m_S) \simeq & \, 0 \ , \\
%\mathcal C_{yq} (m_S) = & \,  \frac{c_{\chi S}}{\Lambda m_S^2} c_{yq}(m_S) = \frac{c_{\chi S}}{\Lambda m_S^2} \left[ \frac{\gamma_0}{2\beta_0}\left(\alpha_s(\Lambda)-\alpha_s(m_S)\right)\right] c^\prime_{GG}(\Lambda)\  , \\
\label{LambdaTomS2} \mathcal {C}^\prime_{GG}(m_S)  = & \, \frac{c_{\chi S}}{\Lambda m_S^2}c^\prime_{GG}(m_S)  \ .
%= \frac{c_{\chi S}}{\Lambda m_S^2} c^\prime_{GG}(\Lambda)
\end{align}
With this choice of boundary conditions we start our RG evolution at $\mu = m_S$ and connect the couplings in Eqs~(\ref{LambdaTomS}) and (\ref{LambdaTomS2}) with the ones at the nuclear scale.

The evolution down to the electroweak scale, where we integrate out the top quark, goes along almost the same exact lines. The main difference is that  a threshold correction to $\mathcal {C}^\prime_{GG}$ is induced after the top quark is integrated out \cite{Shifman:1978zn}. At slightly lower energies, we break the electroweak $SU(2)_L \times U_Y(1)\rightarrow U(1)$ via the Higgs vacuum expectation value. For simplicity, we perform these steps at the same scale $m_t$ and we match to the effective Lagrangian
\begin{equation}
\mathcal{L}_{\rm EFT}^{m_b < \mu < m_t}  =\sum_{q = u, d, s,c,b} \mathcal{C}_{q} m_q \, \bar \chi \chi \,\overline{q} q +
 \mathcal{C^\prime}_{GG}\alpha_s \, \bar \chi \chi\, G^{A\,\mu\nu} G^A_{\mu\nu}\ ,
\end{equation}
where, at a scale right below $\mu = m_t$, 
\begin{align}
\mathcal C_q (m_t) = & \, \mathcal C_{yq} (m_t) = \left[ \frac{\gamma_0}{2\beta_0}\left(\alpha_s(m_S)-\alpha_s(m_t)\right)\right] \mathcal C^\prime_{GG}(m_S)\ , \\
\mathcal {C}^\prime_{GG}(m_t) = &  \, \mathcal {C}^\prime_{GG}(m_S) -\frac{1}{12\pi}\mathcal C_{yt}(m_t)\ .
\end{align}
Here, $\beta_0 = 11 -2 n_f/3$ and $n_f$ is the number of active flavors ($n_f =6$ for $\mu > m_t$). 

The evolution to $\mu_N$ is now straightforward, with the main difference being that the number of active quark flavors is reduced by one after each quark threshold. We give here the numerical result for the evolution from $m_S$ to $\mu_N$ which is independent of the cut-off scale $\Lambda$. For $q=\{u,d,s\}$, we obtain
\begin{align}
\mathcal C_q (\mu_N)  = & \, -0.54 \, \mathcal C^\prime_{GG}(m_S)\ , \\
\label{CGprime}
\mathcal {C}^\prime_{GG}(\mu_N) =  & \, 1.02\,\mathcal C^\prime_{GG}(m_S) \ . 
\end{align}
We note that the $2\%$ correction in Eq.~\eqref{CGprime} is actually a two-loop effect as it arises from $\mathcal C^\prime_{GG}$ mixing into $\mathcal C_q$ and a subsequent threshold correction. Its small size indicates that the LO analysis is sufficient for our purposes.
Using the values $\alpha_s(m_S) = 0.092$ and $\alpha_s(\mu_N) = 0.362$, we can derive the low-energy couplings in the language of the basis of Eq.~\eqref{eq:EFTDD}, as given in Eqs.(\ref{eq:RGgluonFINALa}) and (\ref{eq:RGgluonFINALb}) of Section~\ref{sec:RGE}.

\subsection*{RGE without coupling to gluons at the cutoff}

In the second scenario we assume that $S$ does not couple to gluons at the cutoff scale $\Lambda$. That is, the effective Lagrangian at the scale $\mu=\Lambda$ has only the interactions
\beq
\mathcal{L}_{\rm EFT} = \frac{S}{\Lambda} \left[ c_{WW} \, W^{I\,\mu\nu} W^I_{\mu\nu} + c_{BB} \, B^{\mu\nu} B_{\mu\nu} \right]  \ .
\eeq
The RGE to lower energies requires the inclusion of two additional operators~\cite{Crivellin:2014gpa}
\beq
\Delta \mathcal{L}_{\rm EFT} = \frac{S}{\Lambda} \left[\sum_{q = u, d, s,c,b,t} c_{yq} \, y_q \,\left(\overline{q}_L H q_R + \mathrm{h.c.} \right)  + c_H \,(H^\dagger H)^2 \right] \ .
 \eeq
Also in this case we take $c_{yq}(\Lambda)=c_H(\Lambda)=0$ as our boundary conditions. However, they are radiatively induced at lower scales and must be kept in order to have a consistent RG analysis. 

Analogous to the previous scenario, we define $\vec {c}(\mu)^T = (c_{yq}(\mu) \,\,\,c_H(\mu) \,\,\,c_{BB}(\mu) \,\,\,c_{WW}(\mu))$ and write the RG equation as
\begin{equation} \label{RGE}
\frac{d\vec c(\mu)}{d\ln \mu} = \gamma(\mu) \, \vec c(\mu)\,\,\,,
\end{equation}
in terms of the anomalous dimension matrix $\gamma(\mu)$. As we are now considering electroweak corrections, we only consider the mixing of $c_{BB}$ and $c_{WW}$ into $c_{yq}$ and $c_H$. We neglect the evolution of $g$, $g^\prime$, $c_{BB}$ and $c_{WW}$ themselves as this would correspond to $\alpha^2_{\mathrm{em}}$ corrections to direct detection cross sections. That is, we approximate 
\begin{equation}
\gamma(\mu) \simeq \bma 0 &0 & \gamma_{Bq}(\mu) & \gamma_{Wq}(\mu) \\
			0  & 0 & \gamma_{BH}(\mu) & \gamma_{WH}(\mu) \\
		        0 & 0 & 0 & 0 \\
		        0 & 0 & 0 & 0 \\
		        \ema\,.
\end{equation}
The anomalous dimensions can be straightforwardly calculated and we find\footnote{Our results of $\gamma_{Bq}$ and $\gamma_{Wq}$ agree with Ref.~\cite{Crivellin:2014gpa}. However, all the other terms with $g^\prime$ were not reported in that reference. We include them, and we also find an extra factor of $2$ for the piece proportional to $g^4$ in $\gamma_{WH}$.}
\begin{align} 
\gamma_{Bq}  = & \, \frac{g^{\prime\,2}}{(4\pi)^2}(4Q_q) \ , \\
\gamma_{Wq}  = & \, 0 \ , \\
\gamma_{BH}  = & \, -\frac{6}{(4\pi)^2}\left(g^{\prime\,4}+g^2 g^{\prime\,2}\right) \ , \label{gammaBH} \\
\gamma_{WH}  = & \, - \frac{6}{(4\pi)^2}\left(3 g^{4}+g^2 g^{\prime\,2}\right) \ . \label{gammaWH}
\end{align}
where $Q_q$ denotes the quark charge in units of $|e|$ (e.g. $Q_t = +2/3$). 

At the scale $m_S$ we integrate out the scalar resonance and work with the effective Lagrangian
\beq
\begin{split}
\mathcal{L}_{\rm EFT}^{m_t < \mu < m_S}  = & \, \mathcal{C}_{WW}\, \bar \chi \chi\, W^{I\,\mu\nu} W^I_{\mu\nu} +
\mathcal{C}_{BB}\, \bar \chi \chi\, B^{\mu\nu} B_{\mu\nu}  + \ , \\  &
\sum_{q = u, d, s,c,b,t} \mathcal{C}_{yq} y_q \, \bar \chi \chi \,\left(\overline{q}_L H q_R + \mathrm{h.c.} \right) + \mathcal C_H\,(H^\dagger H)^2\,\bar \chi \chi \ .
\end{split}
\eeq
As done for the QCD running, we set our boundary conditions at the scale $\mu = m_S$
\begin{align}
 \mathcal C_{yg}(m_S) \simeq & \, 0 \ ,  \\
 %\frac{c_{\chi S}}{\Lambda m_S^2} c_{yq}(m_S) = \frac{c_{\chi S}}{\Lambda m_S^2}\gamma_{Bq} \log \left(\frac{m_S}{\Lambda}\right)c_{BB}(\Lambda) \ ,\\
 \mathcal C_H(m_S) \simeq & \,  0 \ ,  \\
 % \frac{c_{\chi S}}{\Lambda m_S^2} c_H(\Lambda)=\frac{c_{\chi S}}{\Lambda m_S^2}\left[\gamma_{Bh}c_{BB}(\Lambda)+\gamma_{Wh} c_{WW}(\Lambda) \right] \, \log \left(\frac{m_S}{\Lambda}\right)\ , \\
 \mathcal{C}_{BB}(m_S) = & \, \frac{c_{\chi S}}{\Lambda m_S^2} c_{BB}(m_S) \ ,\\
  \mathcal{C}_{WW}(m_S) = & \, \frac{c_{\chi S}}{\Lambda m_S^2} c_{WW}(m_S) \ ,
\end{align}
%\begin{eqnarray}
% \mathcal C_q(m_S) &=& \frac{c_{\chi S}}{\Lambda m_S^2} c_q(m_S) =  \frac{g^{\prime\,2}}{(4\pi)^2}(4Q_q) \, \log \left(\frac{m_S}{\Lambda}\right)\,C_B(\Lambda) \ , \nn\\
% \mathcal C_h(m_S) &=&   \frac{c_{\chi S}}{\Lambda m_S^2} c_h(\Lambda)\nn\\
%&=&-\frac{6}{(4\pi)^2}\left[\left(g^{\prime\,4}+g^2 g^{\prime\,2}\right)C_B(\Lambda)+\left(3 g^{4}+g^2 g^{\prime\,2}\right)C_W(\Lambda) \right] \, \log \left(\frac{m_S}{\Lambda}\right)\ ,\nn\\
% \mathcal{C}_{BB}&=& \frac{c_{\chi S}}{\Lambda m_S^2} c_{BB}(\Lambda)\ ,\nn\\
%  \mathcal{C}_{WW}&=& \frac{c_{\chi S}}{\Lambda m_S^2} c_{WW}(\Lambda)\, .
%\end{eqnarray}
and do not account for the negligible running from $\Lambda$ to $m_S$. We do evolve the couplings from $m_S$ to the scale of electroweak symmetry breaking. As before, we integrate out the heavy SM degrees of freedom at the common scale $m_t$ and then match to the effective Lagrangian 
\begin{equation}
\mathcal{L}_{\rm EFT}^{m_b<\mu < m_t}  = \sum_{q = u, d, s,c,b} \mathcal{C}_{q} m_q \, \bar \chi \chi \,\overline{q} q +
 \mathcal{C}_{GG} \, \bar \chi \chi\, G^{A\,\mu\nu} G^A_{\mu\nu}+
 \mathcal{C}_{FF} \, \bar \chi \chi\, F^{\mu\nu} F_{\mu\nu}\ .
\end{equation}
We find the following Wilson coefficients
\begin{align}
\mathcal C_q (m_t)  = & \, \mathcal C_{yq} (m_t^+)-\frac{v^2}{m_h^2}\mathcal C_H(m_t^+) \ ,\nn \\
= & \,  \gamma_{Bq} \, \log \left(\frac{m_t}{m_S}\right)\,\mathcal C_{BB}(m_S)-\frac{v^2}{m_h^2} \left[\gamma_{Bh}\mathcal C_{BB}(m_S)+\gamma_{Wh} \mathcal C_{WW}(m_S) \right] \, \log \left(\frac{m_t}{m_S}\right)\ , \\
\mathcal {C}_{GG}(m_t) = & \,  -\frac{\alpha_s(m_t)}{12\pi}\mathcal C_{yt}(m_t^+)\ , \\
 \mathcal{C}_{FF}(m_t) = & \, c_w^2\, \mathcal{C}_{BB}(m_S)+ s_w^2\, \mathcal{C}_{WW}(m_S) \ .
\end{align}
where $m_t^+$ denotes a scale slightly above the top quark mass and $v=246$ GeV. Note that here we neglected an $\mathcal O(\alpha_{\mathrm{em}}^2)$ threshold correction to $\mathcal{C}_{FF}$ from integrating out the top quark.

We can evolve this set of operators to lower energies. As $\mathcal {C}_{GG}(m_t)$ is only induced at the two-loop level, we neglect additional mixing from $\mathcal C_{GG}$ into $\mathcal C_q$. The only mixing we then need to consider is the mixing between $ \mathcal{C}_{FF}$ and $\mathcal C_q$ described by the anomalous dimension \cite{Frandsen:2012db}
\begin{equation}
\gamma_{Fq} = \frac{\alpha_{\mathrm{em}}}{4\pi}(24 Q_q^2)\ . 
\end{equation}
At the scale $\mu_N$ we then obtain for $q=\{u,d,s\}$
\begin{align}
\mathcal C_q (\mu_N) = & \,  \mathcal C_q(m_t)+\gamma_{Fq} \, \log \left(\frac{\mu_N}{m_t}\right)\,\mathcal C_{FF}(m_t)\ , \\
\mathcal {C}_{GG}(\mu_N) = & \,  -\frac{\alpha_s(m_t)}{12\pi}\mathcal C_t(m_t) -\frac{\alpha_s(m_b)}{12\pi}\mathcal C_b(m_b) -\frac{\alpha_s(m_c)}{12\pi}\mathcal C_c(m_c)\ , \\
 \mathcal{C}_{FF}(\mu_N)  = & \,  \mathcal{C}_{FF}(m_t) \ .
\end{align}
Using the numerical value~\cite{Agashe:2014kda} $g^2/(4\pi) \simeq 0.034$, $ g^{\prime\,2}/(4\pi)  \simeq  0.010$ and $s_w^2 \simeq 0.23$ we obtain the results in Eqs.(\ref{eq:RGEWFINALa})-(\ref{eq:RGEWFINALd}) in Section~\ref{sec:RGE}.

\section{Relic Density Calculation}
\label{app:relic}

We compute the DM relic density by numerically solving the Boltzmann equation 
\beq
\frac{d n_\chi}{d t} + 3 H n_\chi = -\langle \sigma v_{{\rm rel}}\rangle \left[n_\chi^2 - n_\chi^{{\rm eq}\,2}\right] \ ,
\label{eq:BoltzFO}
\eeq
where $H$ is the Hubble parameter. The thermally averaged cross section as a function of the temperature $T$ for a specific annihilation channel is computed as described in Ref~\cite{Gondolo:1990dk}
\beq
\langle \sigma v_{{\rm rel}}\rangle_{\chi \chi \rightarrow ij} = \frac{1}{8 \, m_\chi^4 \, T \, K_2^2[m_\chi / T]} 
\int_{4m_\chi^2}^{\infty} ds \,( s -4 m_\chi^2) \, \sqrt{s} \, \sigma_{\chi \chi \rightarrow ij}(s) \,K_1\left[\frac{\sqrt{s}}{T}\right] \ .
\eeq
Here, the total cross sections as a function of the CM energy can be found in Appendix~\ref{app:DecayAndXS}. 

We rewrite the Boltzmann equation in terms of the comoving density $Y_\chi = n_\chi / s$, with $s$ the entropy density, and by using the quantity $ x = m_\chi / T$ as the time variable. The Boltzmann equation in its final form reads
\beq
\frac{d Y_\chi}{dx} = - \langle \sigma v_{{\rm rel}}\rangle \, \frac{s}{H \, x} \left( 1 - \frac{1}{3} \frac{d \ln g_{* s}}{d \ln x} \right) \left[Y_\chi^2 - Y_\chi^{{\rm eq}\,2}\right] \ ,
\eeq
where $g_*(x)$ is the effective number of relativistic degrees of freedom as a function of the temperature. We solve the above equation by imposing the boundary condition at $ x_0 = 1$
\beq
Y_\chi(x_0) = Y^{\rm eq}_\chi(x_0)  = \frac{2}{g_{* s}(x_0)} \frac{45}{4 \pi^4} \, x_0^2 \, K_2[x_0] \ .
\eeq
The numerical solution provides us with the asymptotic value $Y_\chi^\infty$ of the comoving number density. The number and mass density today are
\begin{align}
\label{eq:nchi0} n^\infty_\chi  = & \, 2 \times Y^\infty_\chi s_0  \ , \\
\rho^\infty_\chi  = & \,  m_\chi n^\infty_\chi \ ,
\end{align}
where we have for the current entropy density~\cite{Agashe:2014kda} 
\beq
 s_0 = 2891.2 \, {\rm cm}^{-3} \ .
\eeq
The factor of $2$ in \Eq{eq:nchi0} is because we deal with a Dirac fermion and we add the contribution of the antiparticles. Finally, we compute the DM contribution to the $\Omega$ parameter
\beq
\Omega_\chi = \frac{\rho_\chi}{\rho_{{\rm cr}}} \ ,
\eeq
where for the critical density we have~\cite{Agashe:2014kda} 
\beq
\rho_{\rm cr} / h^2 = 1.05375 \times 10^{-5} \,  \GeV \, {\rm cm}^{-3} \ .
\eeq

%****************************Biblio****************************

\end{document}